\documentclass[aps,prd,floats,floatfix,twocolumn,amssymb,amsmath,
superscriptaddress,nofootinbib,showpacs,longbibliography]{revtex4-2}

\usepackage[T1]{fontenc}
\usepackage[utf8]{inputenc}
\usepackage{txfonts} % match Phys. Rev. font
\usepackage[normalem]{ulem} %for \sout

\usepackage{graphicx}
\usepackage[linktocpage,breaklinks]{hyperref}
\usepackage[capitalize]{cleveref}
\usepackage[usenames,dvipsnames]{xcolor}
\hypersetup{colorlinks=true,citecolor=NavyBlue,linkcolor=NavyBlue,urlcolor=NavyBlue}

\usepackage{multirow,array,booktabs}
\DeclareMathAlphabet{\pazocal}{OMS}{zplm}{m}{n}
\usepackage{microtype}
\usepackage{subfigure}
\usepackage{journals}
\newcommand{\nn}{\nonumber\\}
\usepackage{enumitem} %for enumerate as i, ii, iii, ...

\definecolor{BerlinU1}{HTML}{62AD2D}

\begin{document}
%%%%%%%%%%%%%%%%%%%%%%%%%%%%%%%%%%
% % %
\title{The error budget of binary neutron star merger simulations for configurations with high spin}

\date{\today}
%%  Add Abbreviations for Journals

\author{Hao-Jui Kuan}
\email{hao-jui.kuan@aei.mpg.de}
\affiliation{Max Planck Institute for Gravitational Physics (Albert Einstein Institute), 14476 Potsdam, Germany}

\author{Ivan Markin}
\affiliation{Institute for Physics and Astronomy, University of Potsdam, Haus 28, Karl-Liebknecht-Str. 24/25, 14476, Potsdam, Germany}

\author{Maximiliano Ujevic}
\affiliation{Centro de Ci$\hat{e}$ncias Naturais e Humanas, Universidade Federal do ABC, 09210-170, Santo Andr{\'e}, 9210-170, SP, Brazil}

\author{Tim Dietrich}
\affiliation{Institute for Physics and Astronomy, University of Potsdam, Haus 28, Karl-Liebknecht-Str. 24/25, 14476, Potsdam, Germany}
\affiliation{Max Planck Institute for Gravitational Physics (Albert Einstein Institute), 14476 Potsdam, Germany}

\author{Kenta Kiuchi}
\affiliation{Max Planck Institute for Gravitational Physics (Albert Einstein Institute), 14476 Potsdam, Germany}
\affiliation{Center of Gravitational Physics and Quantum Information, Yukawa Institute for Theoretical Physics, Kyoto University, Kyoto, 606-8502, Japan} 

\author{Masaru Shibata}
\affiliation{Max Planck Institute for Gravitational Physics (Albert Einstein Institute), 14476 Potsdam, Germany}
\affiliation{Center of Gravitational Physics and Quantum Information, Yukawa Institute for Theoretical Physics, Kyoto University, Kyoto, 606-8502, Japan} 

\author{Wolfgang Tichy}
\affiliation{Department of Physics, Florida Atlantic University, Boca Raton, FL 33431}

\begin{abstract}
Numerical-relativity simulations offer a unique approach to investigating the dynamics of binary neutron star mergers and provide the most accurate predictions of waveforms in the late inspiral phase. However, the numerical predictions are prone to systematic biases originating from the construction of initial quasi-circular binary configurations, the numerical methods used to evolve them, and to extract gravitational signals.
To assess uncertainties arising from these aspects, we analyze mergers of highly spinning neutron stars with dimensionless spin parameter $\chi=0.5$.
The initial data are prepared by two solvers, \textsc{FUKA} and \textsc{SGRID}, which are then evolved by two independent codes, \textsc{SACRA} and \textsc{BAM}.
We assess the impact of numerical discretizations, finite extraction radii, and differences in numerical frameworks on the resulting gravitational waveforms.
Our analysis reveals that the primary source of uncertainty in numerical waveforms is the evolution code, while the initial data solver has {a smaller} impact.
We also compare our numerical-relativity waveforms with state-of-the-art analytical models, finding that the discrepancies between them exceed the estimated numerical uncertainties.
Few suggestions are offered: (i) the analytic waveform becomes an inadequate approximation after the two neutron stars come into contact and the binary enters the essentially-one-body phase, (ii) the analytical models may not capture finite-size effects beyond quadrupole moment, and (iii) the inconsistent use of the binary black hole baseline in the analytical models may also be contributing to these discrepancies.
The presented results benchmark the error budget for numerical waveforms of binary neutron star mergers, and provide information for the analytic models to explore further the high spin parameter space of binary neutron star mergers.
\end{abstract}
\maketitle

\section{Introduction}

Coalescing binary neutron stars (BNSs) are promising candidates for studying the state of matter at supranuclear density and probing the phase diagram of quantum chromodynamics, either at low temperatures during the late inspiral (cf.~\cite{Dietrich:2020eud,Chatziioannou:2020pqz,Radice:2020ddv,Suvorov:2024cff} for recent reviews), or at finite temperature during the postmerger (e.g.,~\cite{Figura:2020fkj,Raithel:2023gct,Fields:2023bhs,Miravet-Tenes:2024vba}).
Current gravitational wave (GW) detectors, such as Advanced LIGO~\cite{LIGOScientific:2014pky} and Advanced Virgo~\cite{VIRGO:2014yos}, and next-generation observatories like the Einstein Telescope~\cite{Hild:2010id,Punturo:2010zz,Branchesi:2023mws,Abac:2025saz} and Cosmic Explorer~\cite{LIGOScientific:2016wof,Reitze:2019iox,Evans:2021gyd}, will primarily reveal equation-of-state (EOS) information during the inspiral through tidal interactions between the neutron stars~\cite{Branchesi:2023mws}.
These tidal interactions affect the GW phase evolution and can be extracted from the observed data through matched filtering.
The tidal effects of an NS are characterized by a response function sensitive to its internal structure, which is itself determined by the nuclear EOS and the spin of the star.
The leading-order term, proportional to the tidal deformability \cite{Flanagan:2007ix,Hinderer:2007mb,Hinderer:2009ca,Damour:2012yf,Read:2013zra}, has been measured, though with large uncertainties, from the first BNS event GW170817 \cite{LIGOScientific:2017vwq}, yielding certain constraints on the EOS via the GW signal \cite{LIGOScientific:2018cki,LIGOScientific:2018hze,Annala:2017llu}.
These GW-based constraints can be positioned in a broader multimessenger framework \cite{Dietrich:2020efo,Lattimer:2021emm,Pang:2021jta,Ascenzi:2024wws}. 
Notably, the EOS constraints from GW170817 could be further refined when combined with observations of electromagnetic counterpart emissions  \cite{LIGOScientific:2017ync,Radice:2017lry,Coughlin:2018fis,Kiuchi:2019lls}. 

Expanding the tidal response in a series of the frequency, the zeroth order contribution comprises static tides.
These can be classified into the polar (gravitoelectric Love number) and axial (gravitomagnetic Love number) deformations of metric, corresponding to the tidally-induced mass and current multipole moments of the star, respectively \cite{Binnington:2009bb,Damour:2009vw}. 
The tidal deformability originates from the former class \cite{Hinderer:2007mb,Hinderer:2009ca}, which enters in the GW phasing effectively at 5th post-Newtonian (PN) order $\left[\propto (v/c)^{10}\right]$ while the latter class is at least of 6PN effect \cite{Yagi:2013sva,Banihashemi:2018xfb,Henry:2020ski,Narikawa:2023deu}.
Static tides adequately describe the early inspiral, but the frequency-dependent part of the tidal response becomes progressively more important when the system approaches the merger.
This dynamic response includes dissipative effects at the linear level, characterized by an imaginary component of Love number ~\cite{Ripley:2023qxo,HegadeKR:2024agt,Saketh:2024juq}, which may induce measurable GW phase shifts in high signal-to-noise ratio events \cite{Ripley:2023lsq}.
At the quadratic level, oscillation modes of neutron stars contribute to dynamical tides~\cite{Andersson:2019ahb,Gupta:2020lnv,Pitre:2023xsr,HegadeKR:2024agt}.
As the case for the static tides, the dynamical ones can be divided into an axial and a polar sector \cite{Gupta:2023oyy,Mandal:2023lgy} with the dominant mode in each sector being the Rossby mode (r--mode, \cite{Flanagan:2006sb,Lai:2006pr,Ma:2020oni}) and the fundamental mode (f--mode; \cite{Shibata:1993qc,Kokkotas:1995xe,Lai:1993di,Ho:1998hq,Andersson:2019ahb,Ma:2020rak,Kuan:2022etu,Gamba:2022mgx,Kuan:2023qxo}), respectively. 
It has also been shown that odd-parity dynamical tides play a subdominant role~\cite{Ho:1998hq,Andersson:2019ahb}.
Incorporating dynamical tides into the Love number formalism introduces an effective tidal Love number for late-inspiral evolution \cite{Steinhoff:2016rfi,Hinderer:2016eia,Andersson:2019ahb,Andersson:2019dwg,HegadeKR:2024agt}, which effectively capture the frequency-dependent nature of the tidal response.

Tidal effects deliver important information about the internal structure of NSs through the connections between GW observations to the coefficients of tidal response function that are determined by the underlying EOS.
For interpreting GW signals, accurate waveform models are essential to precisely infer source parameters.
In this regard, significant efforts have been made to develop such models.
One important family of models leverages PN information and numerical-relativity (NR) simulations to construct phenomenological waveform models, e.g.,~\cite{Dietrich:2017aum,Kawaguchi:2018gvj,Dietrich:2018uni,Dietrich:2019kaq,Abac:2023ujg}.
Another focuses on tidal effective-one-body (EOB) models, branching out into two families (see, e.g., \cite{Dietrich:2017feu,Rettegno:2019tzh} for a comparison): 
(i) \textsc{TEOBResumS}, incorporating tidal effects inspired by gravitational self-force computations ~\cite{Damour:2009wj,Bernuzzi:2014owa,Nagar:2018zoe,Akcay:2018yyh,Nagar:2018plt,Gamba:2023mww}, and 
(ii) \textsc{SEOBNRv*T}, which extends point-particle baselines with PN tidal effects, including dynamical tides~\cite{Hinderer:2016eia,Steinhoff:2016rfi,Steinhoff:2021dsn,Haberland:2025luz}. 

Tidal effects in spinning BNS systems are more complex than in non-spinning cases due to additional spin-tidal Love numbers \cite{Abdelsalhin:2018reg,Castro:2022mpw}, 
potential resonances between the redshifted branch of f--modes \cite{Lai:1997wh,Kruger:2019zuz,Kruger:2021zta} and tidal pushing force \cite{Ma:2020rak,Gupta:2020lnv,Dewberry:2022,Kuan:2024jnw}, and non-linear tides may be revealed \cite{Yu:2022fzw,Kuan:2024jnw,Yu:2025ptm}.
Furthermore, the multipole moments of rapidly spinning NSs are significantly different from those of a black hole (BH) with an identical spin \cite{Ryan:1995wh}.
Therefore, accurately incorporating higher-order multipole moments such as the mass quadrupole, mass octupole, and current quadrupole into the binary dynamics can be crucial for predicting the emitted GWs.
Analytical descriptions of spinning BNS systems have been developed in the PN formalism~\cite{Steinhoff:2021dsn,Kuan:2022etu,Kuan:2023qxo,Mandal:2022nty,Mandal:2022ufb,Yu:2024uxt}, and in the EOB framework~\cite{Steinhoff:2021dsn,Schmidt:2019wrl}.
The analytic attempts, however, have limited power in handling the late-time tidal response of NSs. 
For the last $<0.1$~s of inspiral, NR simulations are required to model the merging process and to provide accurate waveforms~\cite{Baiotti:2011am,Bernuzzi:2012ci,Hotokezaka:2013mm,Dietrich:2017feu}.

Acquiring NR waveforms from quasi-circular BNS consists of two separate ingredients: 
(i) preparation of binary initial data in a quasi-equilibrium state and 
(ii) a robust numerical scheme to evolve the system forward in time, tracking the inspiral dynamics. 
Due to the indispensability of NR simulations, several collaborations have invested to develop extensive databases, e.g., the Simulating eXtreme Spacetimes (SXS) collaboration\footnote{\url{https://data.black-holes.org/waveforms/index.html}}~\cite{Mroue:2013xna,Boyle:2019kee}, the Computational Relativity (CoRe) collaboration\footnote{\url{http://www.computational-relativity.org}}~\cite{Dietrich:2018phi,Gonzalez:2022mgo}, and the \textsc{SACRA} data bank\footnote{\url{http://www2.yukawa.kyoto-u.ac.jp/~nr_kyoto/SACRA_PUB/catalog.html}} \cite{Kiuchi:2017pte,Kiuchi:2019kzt}.
It is generally of interest to quantify the error budget for numerical waveforms constructed by different initial data (ID) solvers and evolution codes, especially since the state-of-the-art EOB and phenomenological waveform models rely on numerical data.
Hence, comparing results across codes is essential to clarify systematic errors in initial data construction and/or due to differences in evolution schemes for the spacetime and hydrodynamic sectors \cite{Bernuzzi:2012ci,Radice:2013hxh,Foucart:2018lhe}, even though simulations of the same physical system should converge to a consistent continuum solution.

For BNSs, Ref.~\cite{Baiotti:2010ka} have compared numerical results obtained in the evolution codes \textsc{SACRA} and \textsc{Whiskey} by using the same initial data from \textsc{LORENE}~\cite{Gourgoulhon:2000nn,Taniguchi:2001qv}, where non-spinning NSs merge in $\sim6$ orbits; Ref.~\cite{Nagar:2018zoe} has shown the consistency in the numerical waveforms of \textsc{BAM} and \textsc{THC} for some selected non-spinning BNS models; and, recently, Ref.~\cite{Hamilton:2024ziw} presented a comparison survey more inclusive to the evolution codes as well as initial data solvers.
In addition, there are some long-term ($>15$ orbits) NR simulations of aligned and/or anti-aligned spinning BNS with a moderate magnitude of dimensionless spin of $|\chi|<0.2$~\cite{Dietrich:2015pxa,Dietrich:2016lyp,Dietrich:2018uni,Dietrich:2018upm,Foucart:2018lhe} and a larger magnitude of $|\chi|\gtrsim0.3$~\cite{Tacik:2015tja,Most:2019pac,Tichy:2019ouu,Dudi:2021wcf,Rosswog:2023rqa}.
A subset of the results have been compared with the analytic models~\cite{Steinhoff:2021dsn,Gamba:2022mgx}.
The error budget from initial data has been pointed out as well (e.g.,~\cite{Bernuzzi:2013rza,Tsokaros:2016eik}).
Agreement in numerical data of BNS configurations with high spins across different codes is yet-to-be confirmed.

It is our purpose here to extend the comparison studies and to provide estimates of other numerical errors by considering more challenging binary parameters and longer simulations. 
The estimation will be based on long-term simulations covering the last $\sim15$--18 orbits of inspiral for rapidly-spinning BNSs with both components having a large dimensionless spin of $\chi_1=\chi_2=0.5$ where the positive sign of $\chi$ denotes a spin aligned with the orbital angular momentum. 
With such a large value of $\chi$ the spin period of each NS becomes $\approx 1.2$--$1.6$\,ms (see Table~\ref{tab:bulkprop}).
We first prepare ID and then evolve it with two codes to analyze the systematic difference employing different evolution schemes along with numerical errors due to finite resolution and extraction radius of GWs.
We then turn to focus on the uncertainties adhered to the difference preparation of ID. 
We use a single evolution code to evolve initial data generated by two initial data solvers to estimate the associated uncertainties.

Since the performed simulations are the first attempt to explore the parameter space of spins as high as $\chi=0.5$, we next carry out a detailed comparison to some state-of-the-art analytic waveform models to scrutinize their performance.
We show that the accuracy of the current waveform models becomes poor when the binary enters the essential-one-body regime after the two NS participants come into contact.
Before that, the dephasing between NR and analytic waveforms seems to arise from not-yet sufficiently accurate treatment of finite-size effects of spin-induced multipole moments of NSs.
Our results emphasize the importance of higher-moment (the current-octupole, the mass-hexadecapole etc.) effects for rapidly-spinning BNSs.
We also show that the dephasing behaves as if it were a $\sim2.5$ PN order term in very late times.
This brings us to speculate that the deviation at late time might partially come from the fact that the horizon absorption effects -- relevant only for binary-black-hole (BBH) systems but not for BNS mergers -- contaminates the BNS waveform models through the use of a BBH baseline in the construction.

Finally, we summarize the numerical uncertainties that can stem from discretizations of evolution codes, numerical extraction of waveforms, and the different ID preparations, whereby we conclude that errors linked to ID solvers are mostly negligible, and the primary source for numerical uncertainties attributes to the different evolution codes.
With that being said, the established numerical error budget is unambiguously less than the deviation from current analytic waveform models to the NR predictions.

The remainder of this paper presents the details of the results discussed above. 
We begin by introducing the setups of the initial data solvers (\cref{ID.setup}) and evolution codes (\cref{sacra.setup,bam.setup}), followed by a detailed analysis of the numerical results in \cref{secIII}. There, we quantify the error budgets in GW phasing coming from the finite extraction radius of the waveform, finite resolutions, different initial data solvers, and evolution codes. We conclude our findings in \cref{secIV}. Throughout this paper, unless explicitly stated otherwise, the geometric units $G=c=M_\odot=1$ are assumed, where $G$ is the gravitational constant, $c$ is the speed of light, and $M_\odot$ is the solar mass.

\section{Numerical Setup}\label{secII}

\subsection{Initial data}\label{ID.setup}

We focus on binaries consisting of two identical neutron stars (NSs), each with a gravitational mass of $1.35M_\odot$ and a dimensionless spin of $\chi_{1,2} = 0.5$ aligned with the orbital angular momentum. 
This configuration yields an effective spin of $\chi_{\rm eff} \equiv (m_1\chi_1 + m_2\chi_2)/M - 38(\chi_1 + \chi_2)m_1m_2/(113M^2) \simeq 0.416$, where $m_1$ and $m_2$ are the Arnowitt–Deser–Misner (ADM) masses of the individual NSs, and $M = m_1 + m_2$ is the total mass.
We employ the piecewise-polytropic approximants \cite{Read:2008iy} of H4~\cite{Lackey:2005tk} and SLy~\cite{Douchin:2001sv} for the cold part of the EOS with a remark that these two EOSs are consistent with the current astrophysical observational constraints \cite{LIGOScientific:2018cki,LIGOScientific:2018hze,Annala:2017llu,Shibata:2017xdx,Rezzolla:2017aly,Margalit:2017dij,De:2018uhw,Most:2018hfd,Tews:2018iwm,Shibata:2019ctb,Capano:2019eae,Dietrich:2020efo,Landry:2020vaw,Koehn:2024set}.
The bulk properties of individual stars in the considered binary are summarized in \cref{tab:bulkprop}. 
The quadrupole ($\Lambda_2$) and octupole ($\Lambda_3$) Love numbers are computed for a non-spinning NS with the same rest mass ($M_0$) as the spinning NSs of interest.
These values will be used to generate tidal EOB waveforms in \cref{secIV}.

\begin{table}
\centering
\caption{Bulk properties of the individual NS comprising the studied binaries. The first to the last columns contain the rest mass, the quadrupole and octupole Love numbers, and the spin frequency, respectively.
}
{\renewcommand{\arraystretch}{1.7}
\setlength{\tabcolsep}{2.5pt}
\begin{tabular}{p{0.05\textwidth}
                >{\centering}p{0.1\textwidth}
				>{\centering}p{0.08\textwidth}
				>{\centering}p{0.08\textwidth}
				>{\centering\arraybackslash}p{0.08\textwidth}}
\toprule
EOS   & $M_0$ ($M_\odot$) & $\Lambda_2$ & $\Lambda_3$ & $f_s$ (Hz)\\
\midrule
SLy & $1.478956$ &  $414.411$ & $767.045$ & $832.827$ \\
H4  & $1.457339$ & $1156.860$ & $3170.426$ & $629.331$ \\
\bottomrule

\end{tabular}
}
\label{tab:bulkprop}
\end{table}

\begin{table*}%[htp]
    \centering
    \caption{Initial data properties for different configurations and different initial data solvers. The columns represent the model name, the initial data solver, ADM mass of the binary ($M_{\mathrm{ADM}}$), the angular momentum of the binary ($J$), the initial orbital angular frequency scaled by the total mass of NSs ($M\Omega_{\rm orb,ini}$), and the residual eccentricity, respectively.
    }
    {\renewcommand{\arraystretch}{1.7}
    \setlength{\tabcolsep}{2.5pt}
    \begin{tabular}{p{0.05\textwidth}
                    >{\centering}p{0.07\textwidth}
                    >{\centering}p{0.1\textwidth}
                    >{\centering}p{0.1\textwidth}
                    >{\centering}p{0.15\textwidth}
                    >{\centering\arraybackslash}p{0.1\textwidth}}
    \toprule
    Model   & ID & $M_{\mathrm{ADM}}~(M_\odot)$ & $J~(M_\odot^2)$ & $M\Omega_{\rm orb,ini}\,(\times10^{-2})$ & $e\,(\times10^{-4})$ \\
    \midrule
    SLy$_{++}$ & FUKA  &  2.68149 & 9.85632 & 1.49661 & $\lesssim6.2$ \\
    SLy$_{++}$ & SGRID &  2.68230 & 9.87132 & 1.49859 & $\lesssim8.5$ \\
    H4$_{++}$  & FUKA  &  2.68123 & 9.86442 & 1.49715 & $\lesssim6.5$ \\
    H4$_{++}$  & SGRID &  2.68142 & 9.89536 & 1.49731 & $\lesssim4.8$ \\
    \bottomrule
    \end{tabular}
    }
    \label{tab:fuka_sgrid_params}
\end{table*}

We use the publicly available solvers \textsc{FUKA}~\cite{Papenfort:2021hod,Grandclement:2009ju} and \textsc{SGRID}~\cite{Tichy:2009yr,Tichy:2011gw,Tichy:2012rp,Tichy:2016vmv,Dietrich:2015pxa,Tichy:2019ouu} to generate binaries in quasi-circular equilibrium as ID.
Both codes solve the constraint equations in the extended conformal thin-sandwich formalism~\cite{Pfeiffer:2002iy,Pfeiffer:2005jf} using pseudo-spectral methods.
\textsc{FUKA} utilizes the \textsc{KADATH} library \cite{Grandclement:2009ju} as its spectral solver.
A resolution $d=13$ (using the first 13 Chebyshev coefficients) is sufficient for our purposes, as will be evidenced by the $\gtrsim3$-order convergence achieved in the computations.
An eccentricity-reducing scheme based on the 3.5 PN estimation is implemented \cite{Papenfort:2021hod} in \textsc{FUKA}, which can yield configurations with a low eccentricity ($e$) of $\mathcal{O}(10^{-3})$. 
Building upon this configuration, we further remove the residual eccentricity to the extent of $e\lesssim10^{-3}$ through the iterative procedure proposed in~\cite{Pfeiffer:2007yz,Boyle:2007ft,Buonanno:2010yk} and proven valid for BNS systems in~\cite{Kyutoku:2014yba,Moldenhauer:2014yaa}.
For the same BNS configurations, we also prepare the ID using \textsc{SGRID}. 
Similarly to \textsc{FUKA}, we apply the iterative scheme to reduce residual eccentricity to $< 10^{-3}$.
The eccentricity here is estimated from the first few orbits simulated with the two evolution codes that will be introduced shortly in \cref{evo.setup}.
The values quoted in the last column of~\cref{tab:fuka_sgrid_params} are approximately estimated ones from the simulations by the two evolution codes for the \textsc{FUKA} ID, while those quoted for the \textsc{SGRID} ID are obtained by the evolution done in one of the evolution codes (\textsc{BAM}; see below for descriptions of evolution codes).

\subsection{Evolution}\label{evo.setup}

All simulations presented in this work are performed with the codes \textsc{SACRA-MPI}~\cite{Yamamoto:2008js,Kiuchi:2017pte} and \textsc{BAM}~\cite{Bruegmann:2006ulg,Thierfelder:2011yi,Dietrich:2015iva}. \textsc{SACRA-MPI} uses the Baumgarte-Shapiro-Shibata-Nakamura formalism \cite{Shibata:1995we,Baumgarte:1998te} with the moving puncture gauge \cite{Campanelli:2005dd,Baker:2005vv} and employs a Z4c-type constraint propagation prescription~\cite{Hilditch:2012fp}. 
\textsc{BAM} uses the Z4c evolution scheme directly, combined with the moving puncture gauge~\cite{Alcubierre:2002kk,Baker:2005vv,Campanelli:2005dd}.
To incorporate thermal effects, we augment the cold EOS with a thermal law of ideal gas \cite{Shibata:2005ss} with an adiabatic exponent $\Gamma_{\rm th}=1.67$ in \textsc{SACRA} and $\Gamma_{\rm th}=1.75$ in \textsc{BAM}; cf.~\cite{Bauswein:2010dn} for more details about the prescription of the thermal effects. 
Note that the thermal effect plays essentially no role in the inspiral phase, and hence, the choice of this parameter does not play any major role in the context of this paper.
For both codes, the grid is structured in an adaptive moving mesh algorithm to evolve the late-time inspiral dynamics of BNSs.
The specific numerical scheme in discretizations and grid configurations used in our simulations are detailed separately for each code below, while the verbose list of numerical setup parameters are listed in \cref{appendixA}.

\subsubsection{SACRA-MPI}\label{sacra.setup}

In this study, we employ a box-in-box grid with 10 refinement levels (from level 0 to 9) of increasing spatial resolution. 
The first six levels each contain a fixed box centered on the binary’s center of mass at ($x,y,z$)=(0,0,0), while the remaining four levels contain 2 boxes centered around the two NSs, respectively. 
A plane symmetry is applied to the grid, and thus the computational domain of the box in the $(9-n)$--th level spans $[-2^{n}L,\,2^{n}L]$ in the $x$- and $y$-directions, while the domain covers $[0,\,2^{n} L]$ along the $z$-axis. 
In this work, we set $L\approx15$~km irrespective of EOSs, and denote the grid spacing at the finest mesh by $\Delta=L/N$ with $(2N+1)\times(2N+1)\times(N+1)$ the number of non-staggered grid points employed in the computational domain.
We perform simulations for four grid resolutions characterized by $N\in\{78,94,118,158\}$, which are equivalent to $\Delta\simeq$190\,m, 160\,m, 125\,m, and 93\,m, respectively.
In what follows, we refer to these resolutions as R1--R4, from lowest to highest.
For spatial discretizations, we adopt a fourth-order finite difference scheme, while time integration is handled via a fourth-order Runge-Kutta method.
For the high-resolution shock-capturing scheme, we use the approximate Riemann solver HLLE \cite{Kurganov:2000ovy,Amiram:2006zjz}.

\subsubsection{BAM}
\label{bam.setup}
\textsc{BAM} has nested Cartesian grids with 7 refinement levels, two of which are non-moving outermost boxes. The grid origin is set to the center of mass of the binary system. For each finer refinement level, the spatial resolution is twice as high. The finest leaf boxes of the level tree follow both of the punctures and are selected in a way, so that they fully contain the NSs within a margin of around 15\%. This way, the spacing depends on the radius of the NS and, in turn, on the stiffness of the EOS. We perform simulations with the resolutions of $\Delta = 184~\mathrm{m},163~\mathrm{m},123~\mathrm{m},92~\mathrm{m},$ for SLy EOS, and $\Delta = 235~\mathrm{m}, 209~\mathrm{m},157~\mathrm{m},117~\mathrm{m},$ for H4 EOS. Throughout this article, we refer to these resolutions as R1--R4, respectively.

For the spacetime, we use a fourth-order finite differencing scheme. For the evolution of matter, we use fifth-order WENO-Z reconstruction~\cite{BORGES20083191}, MC2 slope limiter, and high-order Local Lax-Friedrichs (HO-LLF) Riemann solver~\cite{Bernuzzi:2016pie}. For time integration, we employ a fourth-order Runge-Kutta method.

\section{Error budgets of numerical gravitational waves}\label{secIII}

We derive the waveform by extracting the outgoing component of the complex Weyl scalar $\Psi_4$. It can be decomposed in the spin-weighted spherical harmonics by
\begin{align}
    \Psi_4(t,r,\iota,\psi) =\sum_{\ell m} \Psi_4^{\ell m}(t,r)\,\, {}_{-2}Y_{\ell m}(\iota,\psi),
\end{align}
where $\iota$ and $\psi$ are the polar and azimuthal angles, respectively.
The GW strain for each mode is obtained by double time integration of $\Psi_4(t,r,\iota,\psi)$ as,
\begin{align}\label{eq:psi2h}
    (h_+^{\ell m}-ih_\times^{\ell m})(t,r,\iota,\psi) = -\int^{t} {\rm d}t' \int^{t'} {\rm d}t'' \Psi_4(t'',r,\iota,\psi).
\end{align}
The retarded time at which the (2,2)-mode of strain $h^{\ell m}$ reaches its maximum is defined as the merger time, $t_{\rm mrg}$. 
Note, however, that by the time of merger, the two NSs have already come into contact, meaning the actual onset of merging occurred earlier.
Roughly speaking, the contact sets in when the tidal interaction overcomes the internal gravity of NSs. To quantify the contact, there have been different approaches in the literature, e.g., \cite{Bernuzzi:2013rza} estimated the contact by checking when particular contour density lines of the two stars start touching, another option to determine the contact is based on the mass shedding limit as discussed in ~\cite{Taniguchi:2010kj}.
Overall, the separation when the stars come into contact depends on the masses, spins and the EOS of BNSs [e.g., \cite{Lai:1993pa,Taniguchi:1996bu,Shibata:1997xn}]. Independent of the exact criterion used for determining the contact, it is about $a_{\rm contact} = 2\,\text{--}\,4\,R_1$ with $R_1$ being the circumferential radius of the first star.
Assuming $a_{\rm contact}=3\,R_1$, it is found as $a_{\rm contact}=36$ and 42 km for the SLy$_{++}$ and H4$_{++}$, respectively.

We perform the following bottom-up analysis of the waveform quality to provide a comprehensive error measure of numerical waveforms. First, we study the errors arising from the finite extraction radii in \cref{ext_err} inside each simulation run. Then, we assess the errors at different grid resolutions in \cref{res_err}. After the code error budgets are quantified, we compare the waveform systematics produced by different evolution codes with the same initial data in \cref{evo_err}, and by different initial data solvers with the same evolution code in \cref{id_err}. 

We note that violations in the rest mass of the BNS can also lead to inaccuracies~\cite{Dietrich:2017feu,Kiuchi:2017pte}.
However, in our simulations, the rest mass is conserved within $<10^{-5}\%$ until the last 2--3~ms, and the conservation is maintained as $\simeq10^{-4}\%$ at the merger time for the simulations.
The associated phase error is therefore $\mathcal{O}(10^{-4})$ radians estimated by Equation (B1) of \cite{Kiuchi:2017pte}.
This is orders of magnitude less than other errors that will be discussed in this Section and thus we will ignore the phase error due to the violation of the rest-mass conservation hereafter.

\subsection{Finite extraction radii within individual evolution codes}
\label{ext_err}

Due to the finite computation domain, one cannot extract GWs at future null infinity but has to evaluate them in the local wave zone where the radius of the extraction sphere is comparable to the wavelengths of GWs.
Such extraction introduces certain phase errors \cite{Bernuzzi:2011aq}, making it potentially challenging to ensure consistency across waveforms obtained at different radii.
For example, Refs.~\cite{Hotokezaka:2015xka, Bernuzzi:2016pie} reported that waveforms extracted at larger radii tend to exhibit faster phase evolution.
It is, therefore, necessary to appropriately extrapolate the waveforms extracted at finite radii to obtain the gauge-independent asymptotic waveform at the future null infinity. 

One way is to approximate both the phase and the amplitude of the waveform by a polynomial relation~\cite{Boyle:2007ft,Boyle:2009vi,Pollney:2009ut,Iozzo:2020jcu},
\begin{align}\label{eq:polyfit}
    f(t_{\rm ret};r_{A,j})=f(t_{\rm ret})_\infty + \sum_{k=1}^{K} a_k(t_{\rm ret}) r_{A,j}^{-k}\,\,\,\,\text{for}\,\,j=0,...,N-1,
\end{align}
where $r_{A,j}$ is the areal radius of the $j$-th out of $N$ extraction spheres and $f(t_{\rm ret};r_{A,j})$ is either the phase or the amplitude of the waveforms computed at $r_{A,j}$ while the extrapolated waveform is denoted by $f(t_{\rm ret})_\infty$, and $K < N$ is the extrapolation order.
On the right, the polynomial fitting coefficients $a_k(t_{\rm ret})$ are functions of the retarded time $t_{\rm ret}$, defined by
\begin{eqnarray}
    \label{eq:retarded_time_full}
    t_{\rm ret} &=& \int_0^t \frac{\langle \alpha \rangle}{\left[1-\displaystyle\frac{2M_{\mathrm{ADM}}}{r_{A,j}}\right]^{\frac{1}{2}}} {\rm d}t' \nonumber \\
    &&- \left[ r_{A,j} + 2M_{\mathrm{ADM}}\ln\left( \frac{r_{A,j}}{2M_{\mathrm{ADM}}}-1 \right) \right]\,,
\end{eqnarray}
where $M_{\mathrm{ADM}}$ ($\ne M$) is the initial ADM mass of the system and $\langle \alpha \rangle$ is the average lapse over the associated extraction sphere \cite{Iozzo:2020jcu}.

Evaluating \eqref{eq:retarded_time_full} requires temporal data of average lapse and areal radius from the extraction spheres.
At the time these simulations were performed, neither of the employed evolution codes provided output for these metrics, and thus we have to resort to reasonable approximations for these quantities. 
As the first-order approximation, we assume the Schwarzschild spacetime in isotropic coordinates in the far zone with the source mass being the initial ADM mass of the system. 
Under this assumption, the areal radius is given by $r_A\simeq r_j [1+M_{\mathrm{ADM}}/(2r_j)]^2$ for $r_j \gg M_\mathrm{ADM}$. 
This also simplifies \eqref{eq:retarded_time_full} to
\begin{align}
    t_{\rm ret} = t - \left[ r_A + 2M_{\mathrm{ADM}}\ln\left( \frac{r_A}{2M_{\mathrm{ADM}}}-1 \right) \right].
\end{align}
We note that, in reality, $r_A$ and the average lapse evolve according to the gauge condition and deviate slightly from the isotropic coordinates. 
However, in our experience with similar simulations and the \textsc{BAM-SGRID} configuration discussed below, this difference is small for BNS waveforms in 1+log slicing: around $10^{-3}$ difference in the areal radius, and $5 \times 10^{-3}$ difference in average lapse for a large enough radius $r\simeq 1000\,M_\odot$.

\begin{figure}%[htp]
    \includegraphics[width=\columnwidth]{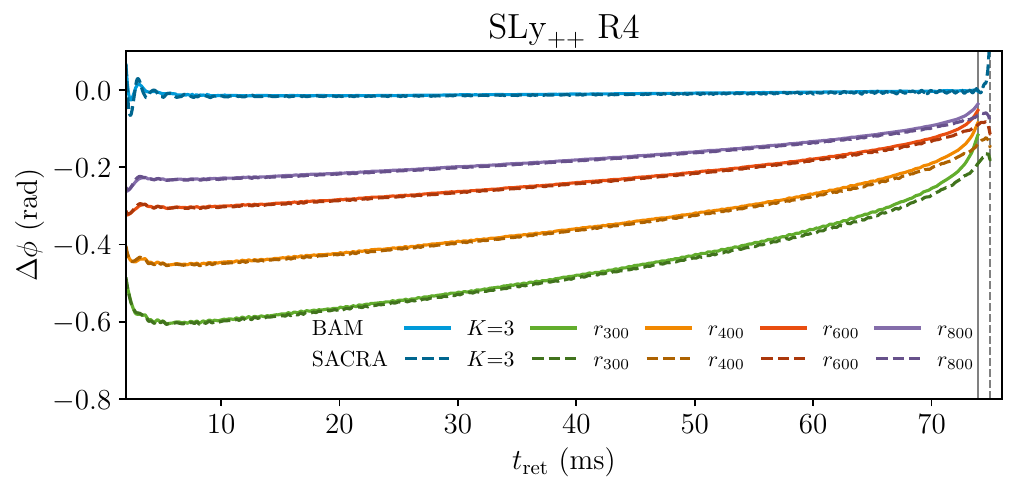}
    \includegraphics[width=\columnwidth]{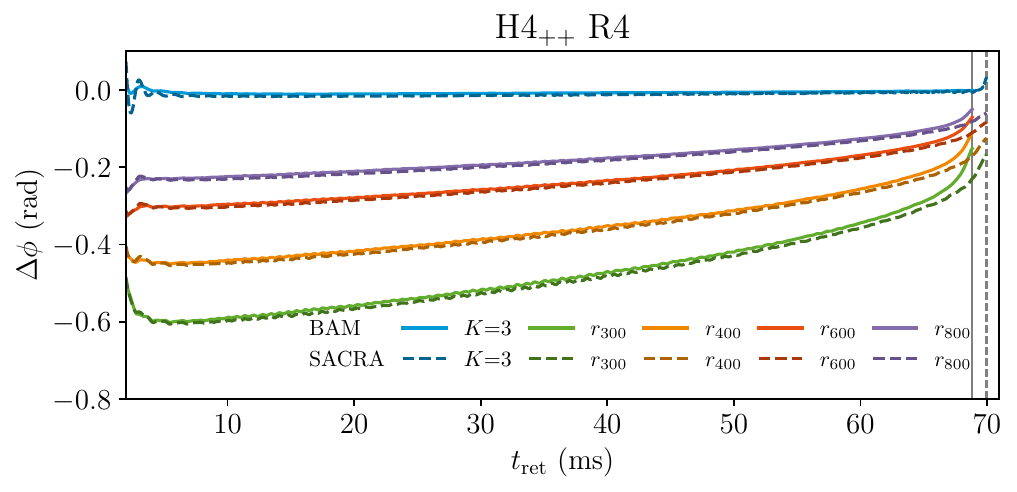}
  \caption{Evolution of the finite-radius extraction phase uncertainty ($\Delta \phi$) for both codes at the highest resolution (R4) at matching extraction coordinate radii. The solid curves correspond to the \textsc{BAM} code, and the dashed curves to the \textsc{SACRA} code. 
  For each code, the curves labeled with $r$ illustrate the phase difference between the waveform extracted at the corresponding coordinate radii and the second-order ($K{=}2$) extrapolated waveform.
  The $K{=}3$ curves depict the phase differences between $K{=}3$ and $K{=}2$ extrapolated waveforms.
The data is terminated at the corresponding merger times for each code marked as vertical gray lines.} 
  \label{fig:finite_radii}
\end{figure}

We leverage the perturbative method of \textsc{scri} package~\cite{scri} to obtain extrapolated waveforms.
The uncertainty due to finite-radii extraction is then estimated by the phase difference between the extrapolated waveforms and the ones extracted at coordinate spheres.
The results are shown in Fig.~\ref{fig:finite_radii}, where we see that the phase uncertainty is the highest in the early inspiral and decreases with frequency, as also seen in previous works, e.g. \cite{Hotokezaka:2015xka, Bernuzzi:2016pie}. 
For the outermost extraction radius present in both codes, $r{=}800M_\odot$, the phase uncertainty starts at the level of $\sim -0.25$~rad and steadily decreases throughout the inspiral.
For most of the duration of the waveform, the evolution of the phase uncertainty is remarkably similar between \textsc{BAM} and \textsc{SACRA}.
Differences appear mainly during the merger phase, which begins at $\sim67$~ms for the SLy$_{++}$ model and $\sim58$~ms for H4$_{++}$.
In this phase, \textsc{BAM} tends to show slightly lower uncertainty. 
At the merger time and extraction radius $r{=}800M_\odot$ (purple), the phase uncertainty is $\sim -0.05$~rad for the \textsc{BAM} result and $\sim -0.06$~rad for the \textsc{SACRA} one in the H4$_{++}$ model; 
for SLy$_{++}$, the uncertainties are $\sim -0.04$~rad and $\sim -0.08$~rad, respectively.

Beside the extraction radii matching with those in \textsc{SACRA}, the \textsc{BAM} simulations had additional, larger ones: $r\in\{900M_\odot, 1000M_\odot, 1100M_\odot, 1200M_\odot\}$ for the H4 configuration, and $r{=}900M_\odot$ for the SLy configuration. 
As the error of extraction at the larger radii is lower, we use the waveforms extrapolated from the largest radii available in the following analysis. 

On top of the described method for obtaining extrapolated waveforms at infinity, Refs.~\cite{Lousto:2010qx,Nakano:2011pb} proposed an analytic extrapolation based on the next-to-leading order asymptotic behavior of the complex Weyl scalar, which we refer to as Nakano's extrapolation method following \cite{Hotokezaka:2015xka,Hotokezaka:2016bzh}. 
In this method, we first obtain the extrapolated $(\ell,m)$ component of $\Psi_4$ from the data extracted at a given distance $r_j$ as
\begin{align}\label{eq:nakano}
    \Psi_4^{\ell m,\infty}(t_{\rm ret}) &= \Big(1-\frac{2M}{r_{A,j}}\Big) \Bigg[ \Psi_4^{\ell m}(t_{\rm ret};r_j) \nonumber\\
    &-\frac{(\ell-1)(\ell+2)}{2r_{A,j}}\int^{t_{\rm ret}}\Psi_4^{\ell m}(t';r_j) dt'
    \Bigg].
\end{align}
The waveform at infinity is then derived by integrating \cref{eq:nakano} twice [cf.~Eq.~\eqref{eq:psi2h}], for which we use the fixed frequency integration proposed by \cite{Reisswig:2010di}.
The extrapolated waveforms are depicted in the top panels of \cref{fig:sacra_wv,fig:bam_wv}, and will be used to estimate the various numerical uncertainties below.

\subsection{Waveform convergence within individual evolution codes}\label{res_err}

\begin{figure}
    \centering
    \includegraphics[width=\columnwidth]{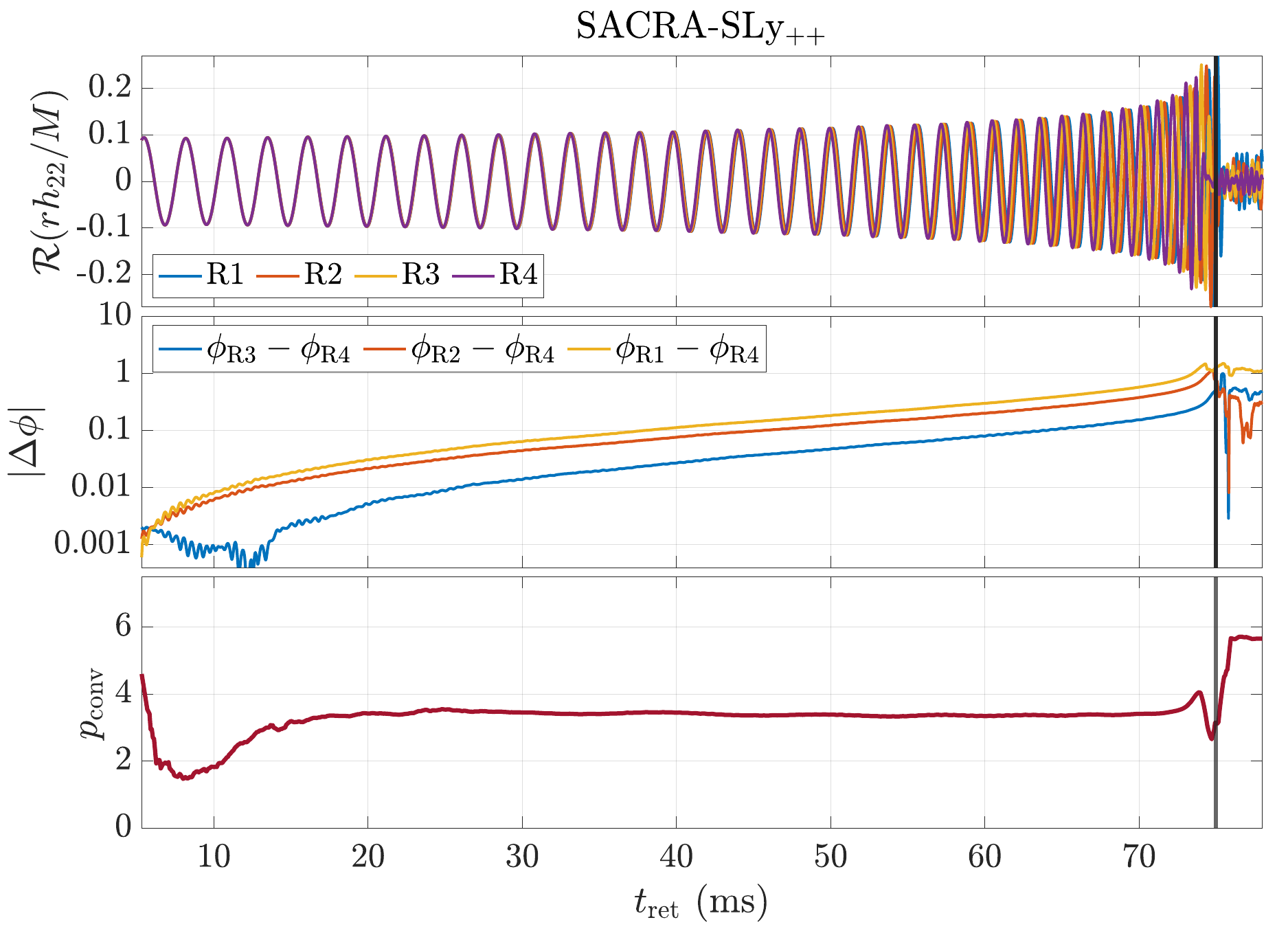}
    \includegraphics[width=\columnwidth]{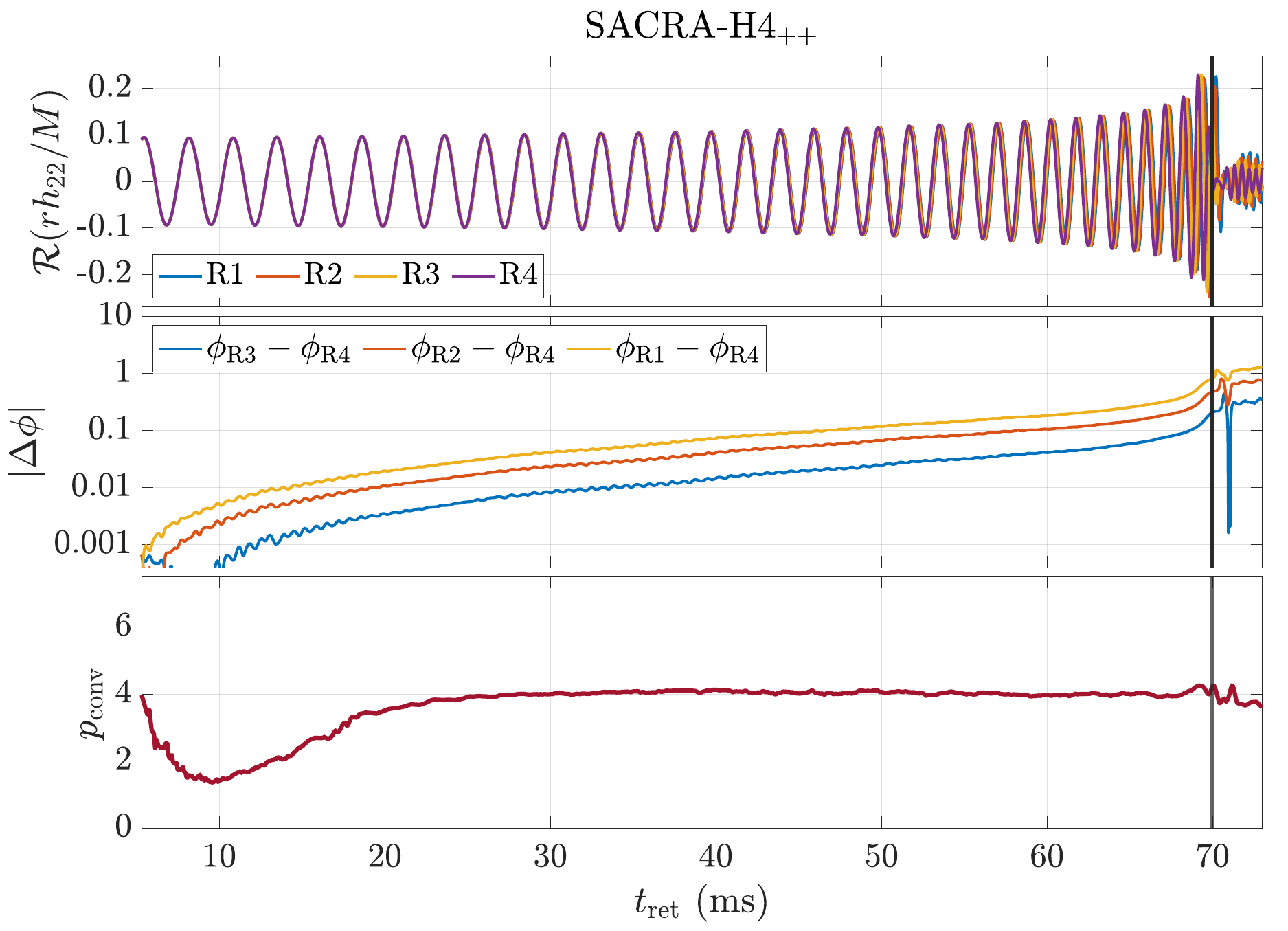}
    \caption{Real part of $(2,2)$ mode of the computed waveform for the four resolutions considered (first); phase shift between resolutions as functions of time (second); and the convergence power estimated through \cref{eq:pconv} (third).
    The vertical black line marks the merger time of the R4 waveform, i.e., when the amplitude of $h_{22}$ peaks.
    The waveforms are generated by \textsc{SACRA-MPI} for models SLy$_{++}$ and H4$_{++}$. 
    The numerical extraction radius is $r=800\,M_\odot$, and Nakano's method [\cref{eq:nakano}] is applied to extrapolate the waveform to spatial infinity.
    }
    \label{fig:sacra_wv}
\end{figure}
\begin{figure}
    \centering
    \includegraphics[width=\columnwidth]{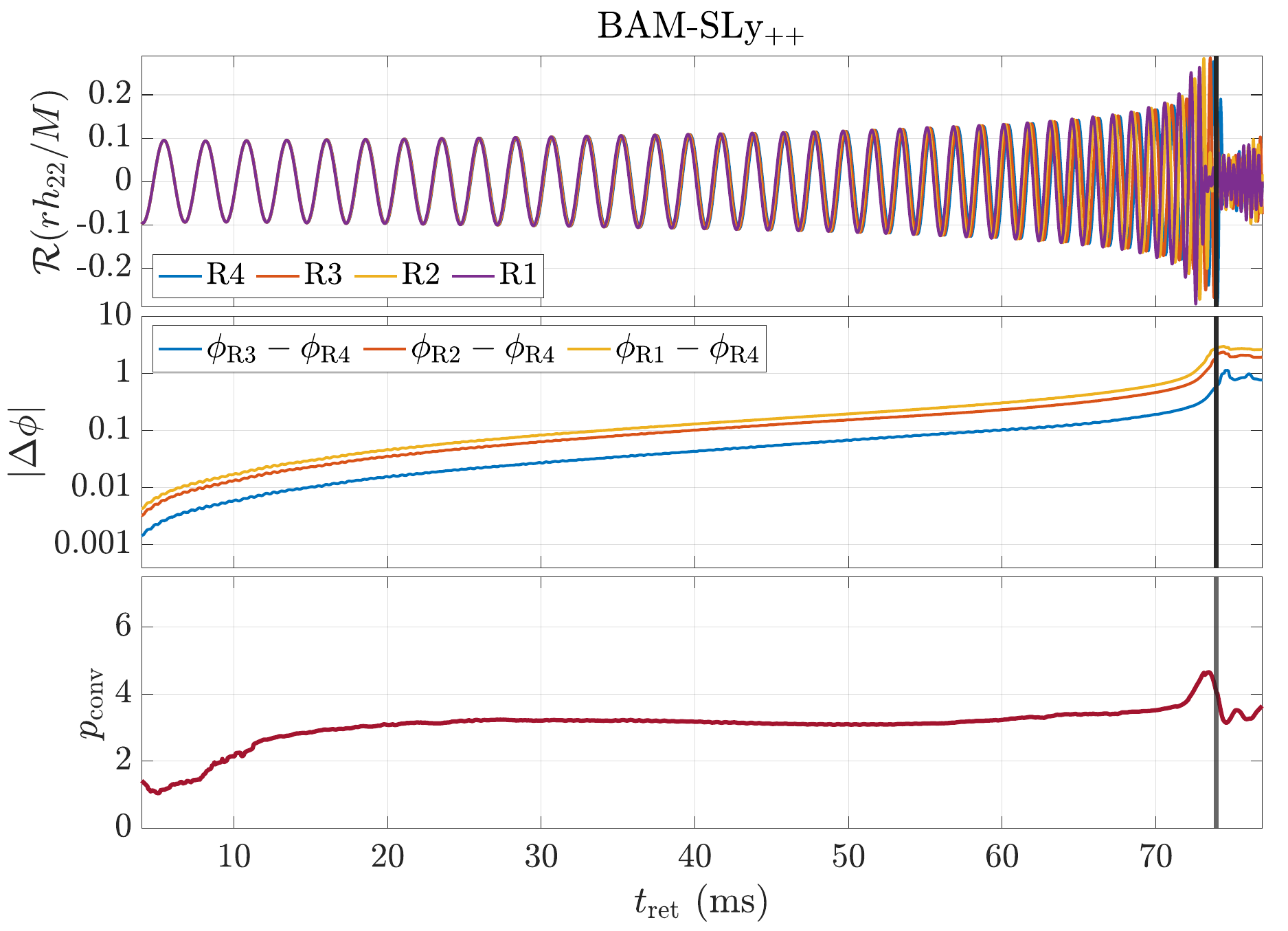}
    \includegraphics[width=\columnwidth]{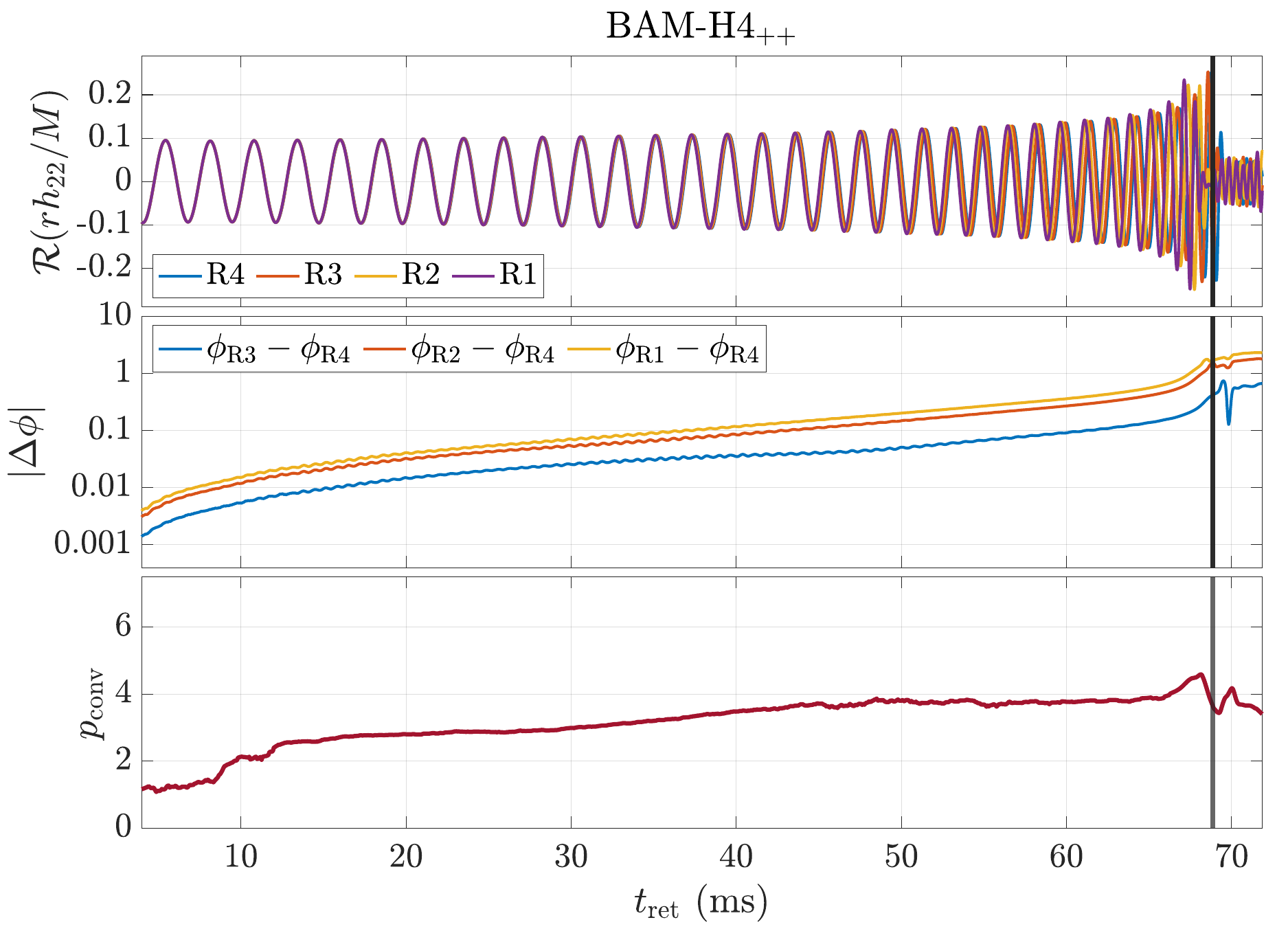}
    \caption{Same as Fig.~\ref{fig:sacra_wv} but for waveforms from \textsc{BAM}, where the numerical extraction radii for the waveforms are set at $900\,M_\odot$ and $1200\,M_\odot$ for SLy$_{++}$ and H4$_{++}$, respectively.
    }
    \label{fig:bam_wv}
\end{figure}

For the extrapolated waveforms based on \cref{eq:nakano}, the dephasing $\Delta\phi=\phi_{R_i}-\phi_{R4}$ between two grid resolutions $R_i$ and $R4$ are shown in the middle panels of \cref{fig:sacra_wv,fig:bam_wv} for simulations of \textsc{SACRA-MPI} and \textsc{BAM}, respectively.
Both codes demonstrate that the binary merges earlier at lower grid resolutions due to stronger numerical dissipation. 
Taking the merger time (indicated by the black vertical lines) from the R4 simulations, it occurs at $\sim75$~ms and $\sim70$~ms for the SLy$_{++}$ and H4$_{++}$ models, respectively, after completing $\alt18.7$ and $\alt15.6$ orbits in simulations.
The difference in the phasing of R3 and R4 remains below one radian until shortly before the merger time for all the shown cases.

To assess how results at different grid resolutions approach the continuum limit, one has to estimate the convergence order of the system. Because of the complexity of NR codes, in which several different error terms converge with different convergence orders, one can not expect to find a clean convergence order that stays constant over time.
As a first approach, one can approximate a convergence order $\hat{p}_{\rm conv}(t)$ by 
\begin{align}\label{eq:pconv_0}
    \phi(t;R)=\phi(t;\infty)+\hat{a}(t) \Delta_R^{ \hat{p}_{\rm conv}(t)}\,.
\end{align}
where $\phi(t;\infty)$ denotes the approximation of the continuum solution with infinite resolution.
This allows estimation using waveforms from any three of the adopted resolutions.
While this convergence order could be used to determine the quality of the simulation, it may not be the most optimal method to capture the multiple, competing convergence orders that might be more dominant throughout the different times of the simulation.
Therefore, as a second method, we introduce another approach resembling but different from \cref{eq:pconv_0}.
This one uses all the considered resolutions to determine a convergence order by a least-squares fit of
\begin{align}
    \label{eq:pconv}
    \phi(t;R_i)-\phi(t;R4) = \tilde{a}(t)\left( \frac{\Delta_{R_i}}{\Delta_{R4}}\right)^{ p_{\rm conv}(t) }
\end{align}
across R1--R4. 
This approach has the advantage that we can access the convergence order in a cleaner way, though it also comes with the risk that unresolved issues or non-convergent contributions could be missed.
The analysis below is based on the four-resolution estimation \cref{eq:pconv}, while the estimation using three resolutions can be found in \cref{appendixB}.

The convergence order as functions of time is shown in the bottom panels of \cref{fig:sacra_wv,fig:bam_wv}.
Notably, the consistent convergence behavior with an approximately constant value of $p_\mathrm{conv}$ only reveals after $\sim20$~ms for all the simulations.
This delay suggests that the initial data require some time to relax into a state compatible with the full Einstein equations and the gauge condition in the evolution code.
At later times, the estimates of $p_{\rm conv}$ made for \textsc{SACRA} and \textsc{BAM} waveforms coincide for the SLy$_{++}$ model, and is found to be $p_{\rm conv}\simeq3.4$.
For the H4$_{++}$ model, the \textsc{SACRA} run shows a relatively constant convergence order of $p_{\rm conv}\lesssim 4$, while the \textsc{BAM} run exhibits a gradual increase from 3 to 4 during 30--40~ms, which then maintains at $\lesssim 4$ up to the merger. Therefore, irrespective of the codes and EOSs, we find a reasonable convergence order of 3--4.

\begin{figure}
    \centering
    \includegraphics[width=\columnwidth]{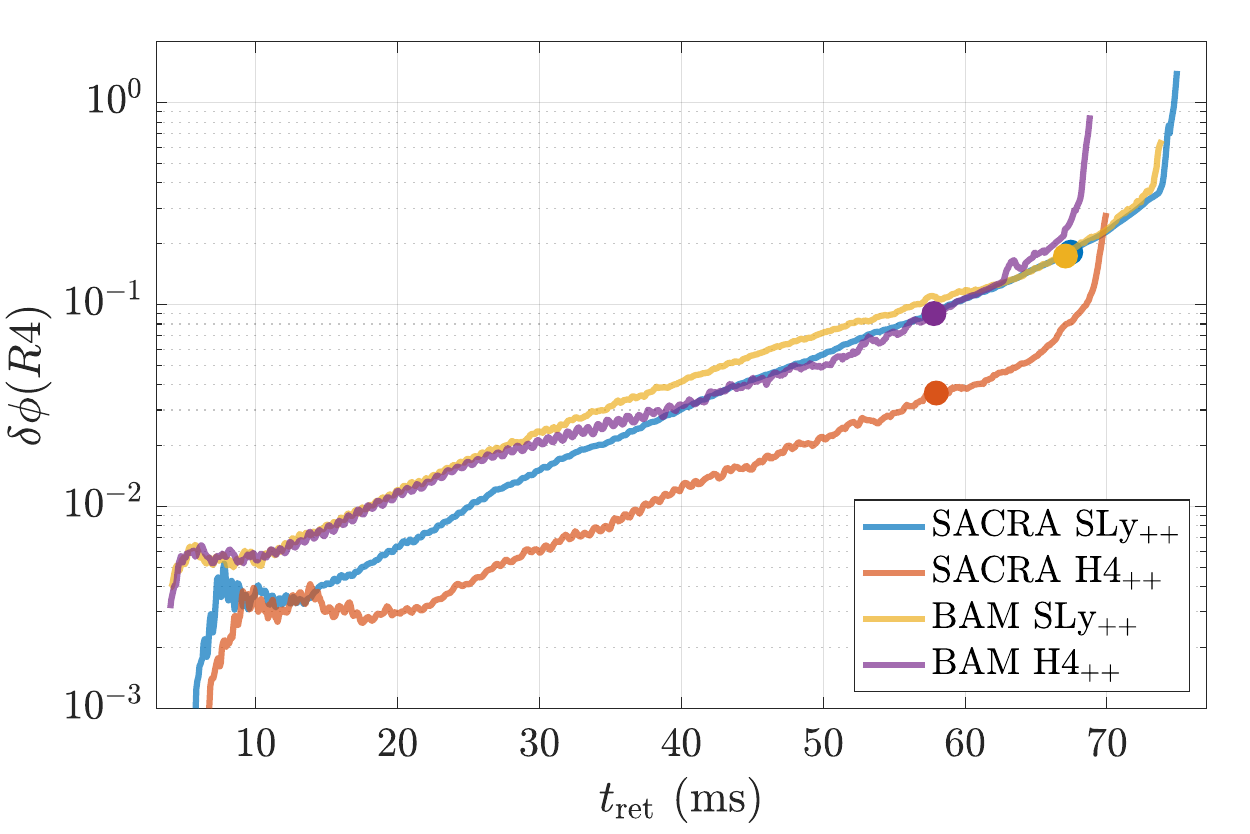}
    \caption{Estimate of the phase error due to finite grid resolution for the highest resolution run and employing the convergence behavior of merger times [cf.~\cref{eq:poly_R4}].
    The end times of the curves are set by the merger time of the simulations with R4 in the two codes, while the moment of contact of two NSs are represented by the filled circles.
    }
    \label{fig:poly_res}
\end{figure}

As the evolution of convergence power dictates how the numerical waveforms approach the continuum limit at each time step, one way to approximate this limit is to employ \cref{eq:pconv}.
In particular, the last term captures the deviation from the waveform at a given resolution to the continuum limit based on the information provided by all resolutions adopted.
Among the datasets, the results with R4 should most closely approximate the continuum solution.
We therefore use the quantity,
\begin{align}\label{eq:poly_R4}
    \delta\phi(t,R4):=\phi(t;R4)-\phi(t;\infty) \simeq a(t) \Delta_{R4}^{p_{\rm conv}(t)}\,,
\end{align}
to estimate the phase error due to finite grid resolutions.
We focus on the inspiral phase of GWs, and the error estimates are truncated at the merger time of the corresponding R4 simulation.
The results are presented in \cref{fig:poly_res}. 
This shows a monotonic growth of the deviation following the early stage of the evolution.
The phase error remains less than one radian at the merger time of R4 except for the SACRA run of the SLy$_{++}$ model, which exhibits a slightly larger deviation of $1.2$~rad.

\subsection{Dephasing between the evolution codes}
\label{evo_err}

The grid resolutions adopted here are shown to be high enough that the phase at merger ($\phi_{t_{\rm mrg}}$) approximately converges to the true solution, i.e., $\phi_{t_{\rm mrg}}$ computed with the simulation of $\Delta^{-1}\rightarrow\infty$.
However, a direct comparison of this limit deducted from two codes is not possible because the same model undergoes different relaxation in each code during the initial phase, which will lead to undesired contamination into the analysis.
The phase evolution can nevertheless be analyzed in more detail. 

For the comparison of the waveforms from each code, we first align 
\footnote{We find that despite the usage of the same initial data, an early time alignment of the data is necessary. While this could come from subtle differences in the numerical setups, such as grid structure and outermost boundary treatments. Also, differences in the initial gauge conditions could introduce such visible differences, in particular, in \textsc{SACRA}, the shift vector is reset to zero, and the lapse function is computed as $\psi^{-2}$ with $\psi$ representing the conformal factor. The simulations performed with \textsc{BAM} instead directly use the lapse function and shift vector solved in \textsc{FUKA}.}
the waveforms of the resolution R4 by minimizing the integral
\begin{align}\label{eq:I_phase}
    I_{\rm phase}=\int_{t_i}^{t_f} \lvert [\phi_1(t+t_b)+\phi_d]-\phi_2(t) \rvert^2{\rm d}t
\end{align}
over a time and phase offsets $t_b$ and $\phi_d$, where $\phi_2$ is the phase of the target waveform to which the phase $\phi_1$ is aligned.
The alignment window is set to $t_i-t_{\rm mrg}=-60$~ms and $t_f-t_{\rm mrg}=-40$~ms, while we note that our results remain essentially unchanged when using alternative alignment intervals.

After time alignment, the phase difference is computed as
\begin{equation}
    \Delta \phi = \phi_1(t+t_b) + \phi_d - \phi_2(t).
\end{equation}
We use the \textsc{SACRA} waveforms to determine the merger time since the \textsc{BAM} waveforms merge earlier by $\simeq0.3$~ms and $\simeq0.16$~ms for SLy$_{++}$ and H4$_{++}$, respectively.

\cref{fig:aligned} presents the aligned waveforms, where the merging phase (defined as the interval between the moment of contact and the merger time) is magnified in the right panels.
We find that the phase difference remains at sub-radian level even into the merger phase, up to the last $\alt1.2$ and $\alt0.5$~GW cycles for the SLy$_{++}$ and H4$_{++}$ models, respectively.
Beyond this point, the difference accumulates more rapidly, and reaches to $\sim2.15$ and $\sim1.55$~radians at the merger time for the SLy$_{++}$ and H4$_{++}$ models.
It should be noted that variations in $\Gamma_\mathrm{th}$ may influence the dynamics during the merging phase.
The extent to which this affects the numerical waveforms is not thoroughly investigated here.
In addition, the alignment was performed for waveforms with slightly different grid resolutions, and hence, the estimated phase error also contains uncertainty due to the finite-resolution and not just due to differences in the code.

\subsubsection{Artificial time-stretching for cross-code comparison}

From the results shown in \cref{fig:sacra_wv,fig:bam_wv}, it can be noticed that the dephasing between the waveform with a lower resolution and that of R4 steadily accumulates over time.
This suggests that the acceleration of coalescence due to numerical dissipation is {\it monotonic in time} as is found in previous works (e.g., \cite{Hotokezaka:2015xka, Hotokezaka:2016bzh}).
Therefore, the artifact due to numerical dissipation might be eliminated by stretching the timescale of waveforms by a certain factor, thus hypothetically achieving the waveform with asymptotically infinite spatial resolution as first proposed by Hotokezaka~\cite{Hotokezaka:2013mm}.
We emphasize that although the time-stretching procedure is an \textit{ad-hoc} way to estimate the continuum limit, i.e., the method does not stem from a rigorous derivation, the convergence property of our numerical waveforms supports employing this method.
Based on this observation, we will employ this method as a second approach to access the numerical uncertainty in evolution codes; notably, we will seek the time-dilation factors for waveforms of each code so as to obtain hypothetical zero-spacing grid resolutions (i.e., $\Delta\rightarrow0$), which are then compared with each other.

\begin{figure}
    \centering
    \includegraphics[width=\columnwidth]{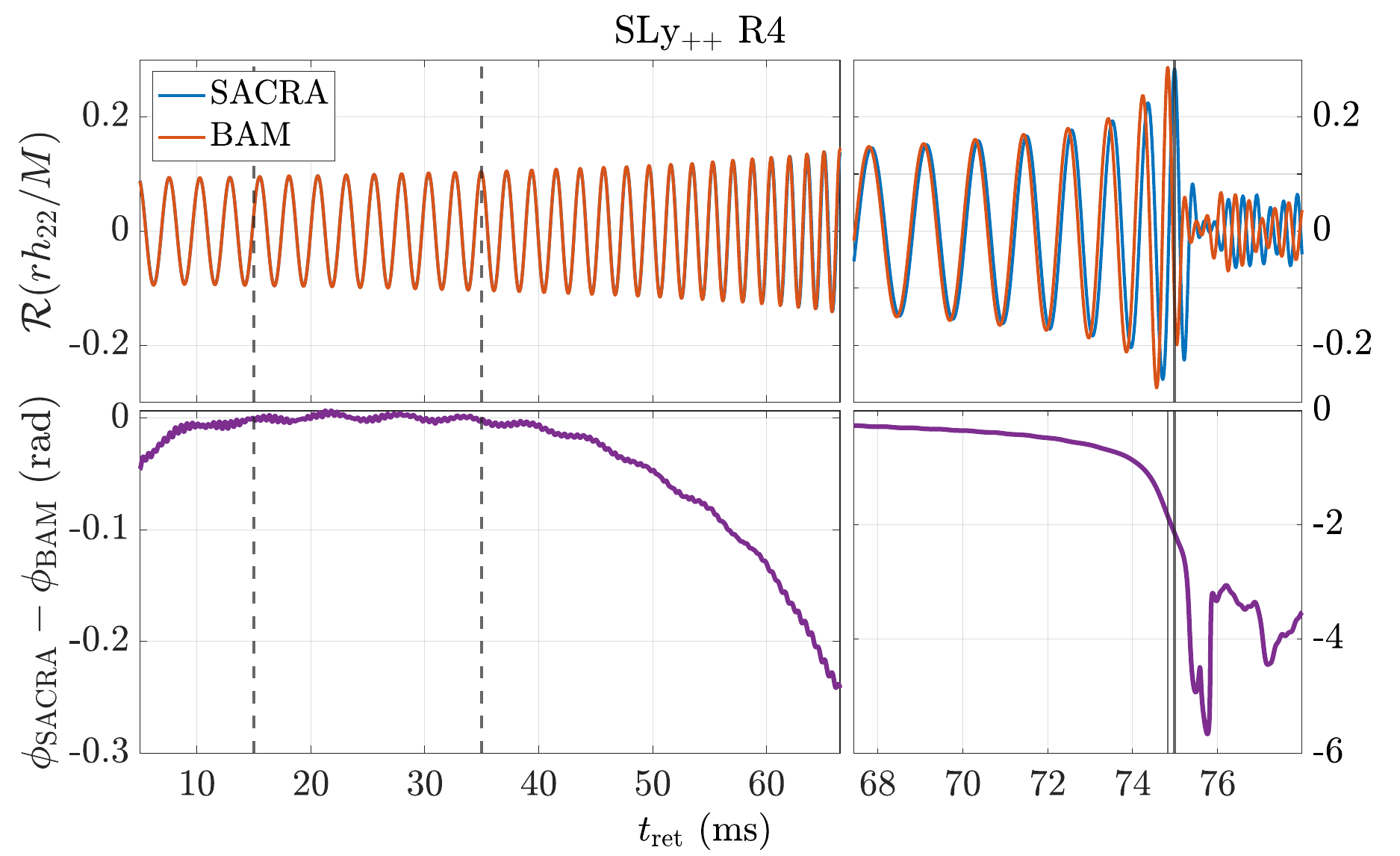}
    \includegraphics[width=\columnwidth]{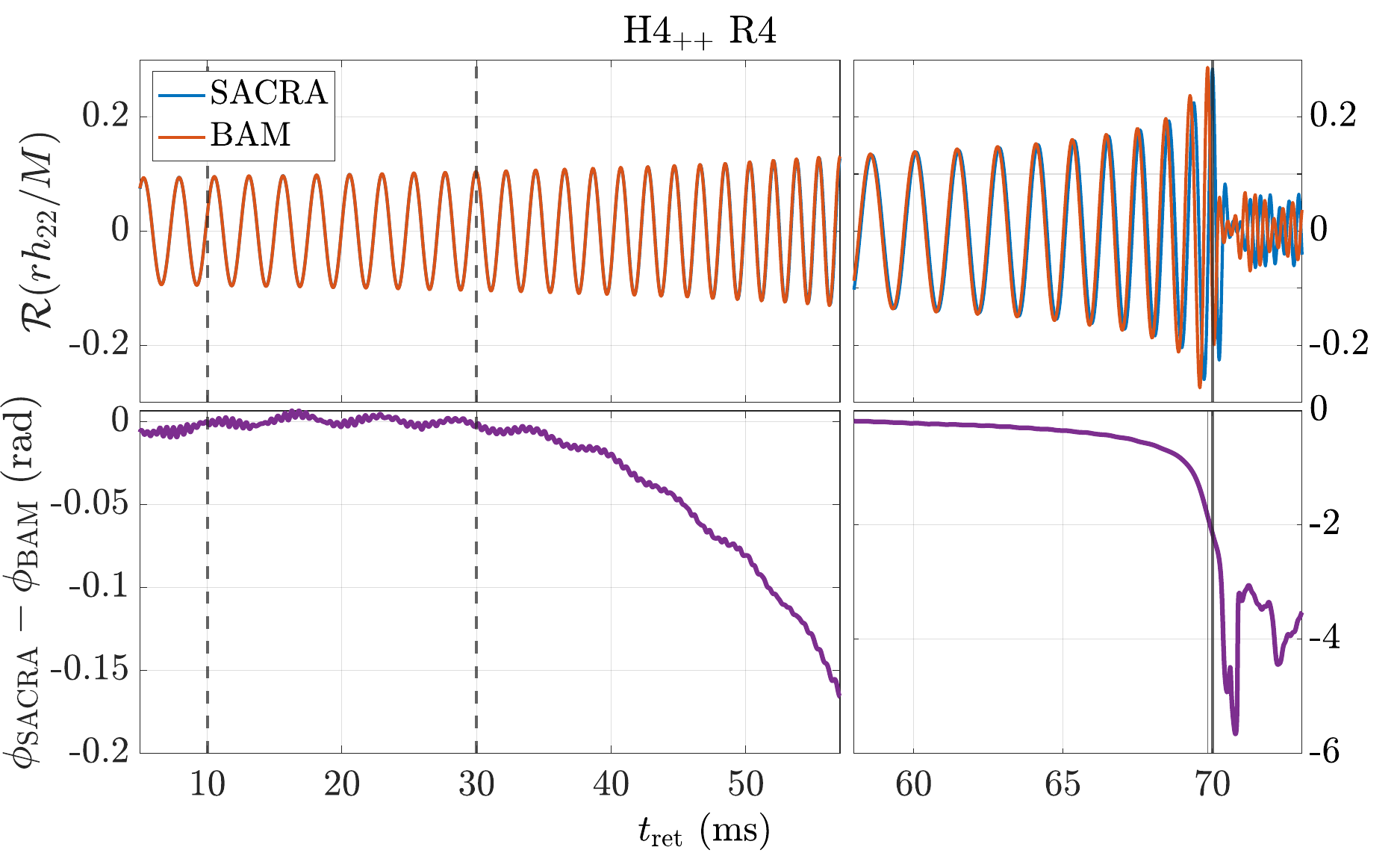}
    \caption{Waveform comparison in the time domain between the waveforms of the resolution R4 from the two codes (see the legends) along with the associated dephasing (bottom panels) for models SLy$_{++}$ and H4$_{++}$.
    The waveforms are aligned in phase and time over the window: $[-60,\,-40]$~ms before the merger time as indicated by the vertical dashed lines.
    The part of waveform after the onset of contact is magnified in the right panels, where the merger time, determined from the \textsc{SACRA} waveforms, is marked by the solid vertical line.
    }
    \label{fig:aligned}
\end{figure}

\begin{figure}
    \centering
    \includegraphics[width=\columnwidth]{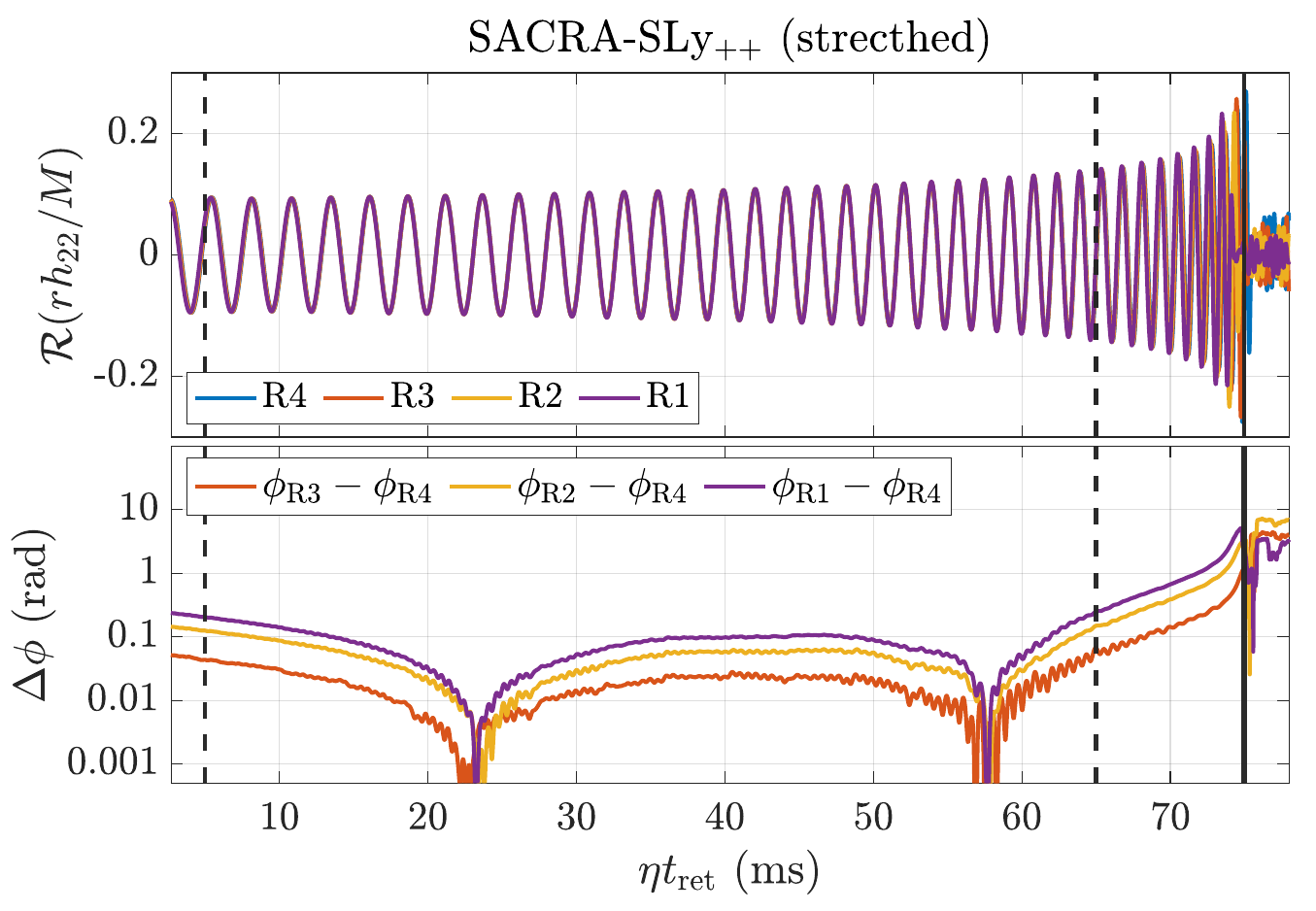}
    \includegraphics[width=\columnwidth]{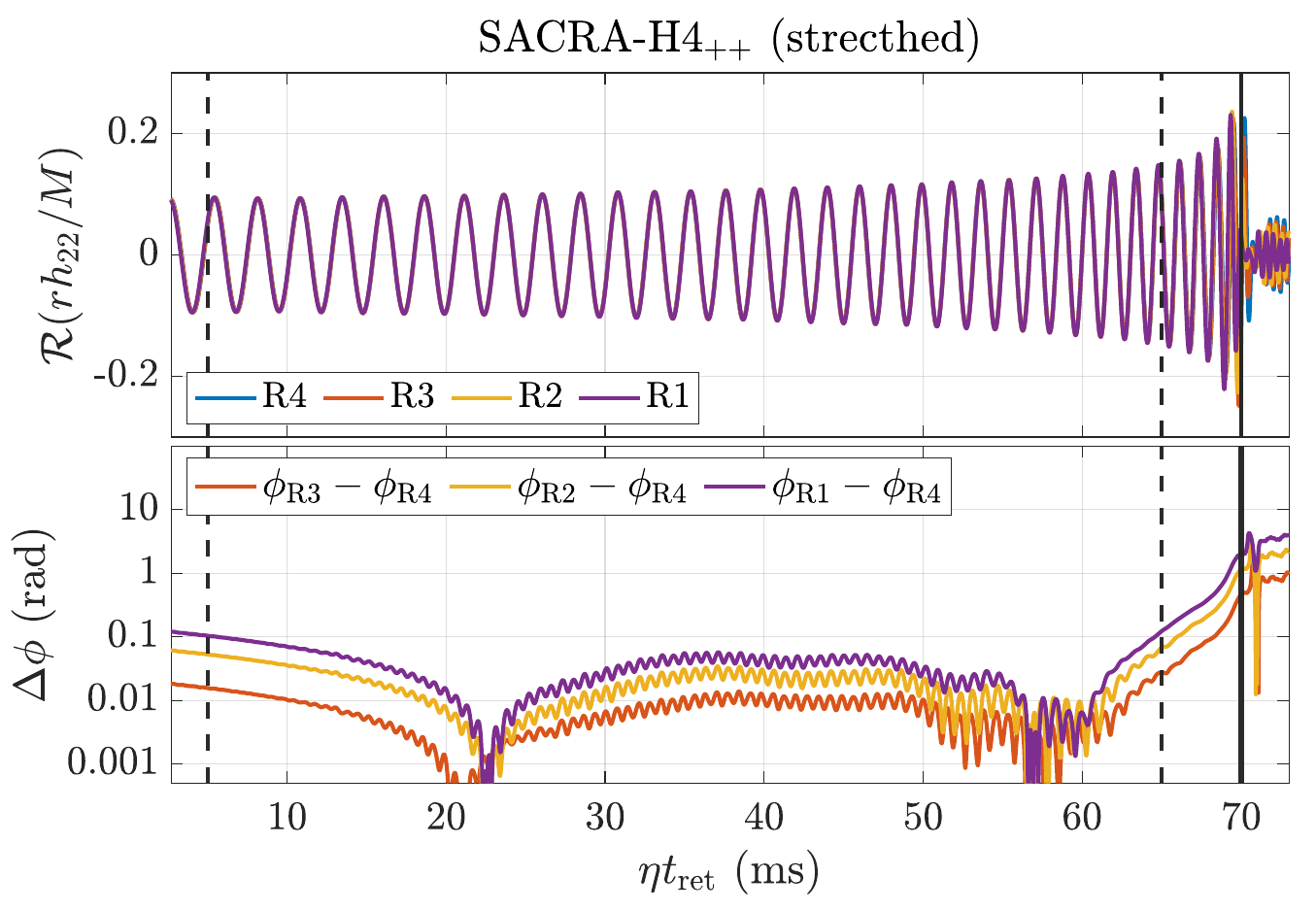}
    \caption{Waveforms at different resolutions as functions of the \emph{stretched time} (see the main text for details) for the SLy$_{++}$ (top) and H4$_{++}$ (bottom) models.
    The merger time is indicated by the vertical solid line, and the alignment window is shown between the two vertical dashed lines.
    The shown data are for \textsc{SACRA} results, while a similar alignment can be achieved for the time-stretched \textsc{BAM} data.
    }
    \label{fig:inf_res}
\end{figure}

The time-stretching scheme is detailed in \cite{Hotokezaka:2013mm,Hotokezaka:2015xka,Hotokezaka:2016bzh}, for which (at least) four grid resolutions are required. 
In short, a stretching factor ($\eta\gtrsim1$) is introduced to minimize the following integral
\begin{align}\label{eq:I_ex}
I_{\rm ex}=\int_{t_i}^{t_f}\lvert A_1(\eta t)e^{i(\phi_1(\eta t)+\phi_d)}-A_2(t)e^{i\phi_2(t)}\rvert^2{\rm d}t,
\end{align}
which aligns the self-similar waveforms obtained with different resolutions.
Here $A_2e^{i\phi_2}$ represents the waveform with which $A_1e^{i\phi_1}$ is geared to align. 
Taking the waveform of R4 as target, we can obtain $\eta\,(>1)$ for a lower resolution by setting the R4 waveform as $A_2$ and $\phi_2$ and, and that of the lower resolution is temporally dilated by a factor that minimizes the integral \eqref{eq:I_ex}.
In the present study, we use the interval window of $[t_i,\,t_f]=[5,60]$~ms for the time stretching, while we have examined that the results are not sensitive to the choice of the time interval.

\begin{figure}
    \centering
    \includegraphics[width=\columnwidth]{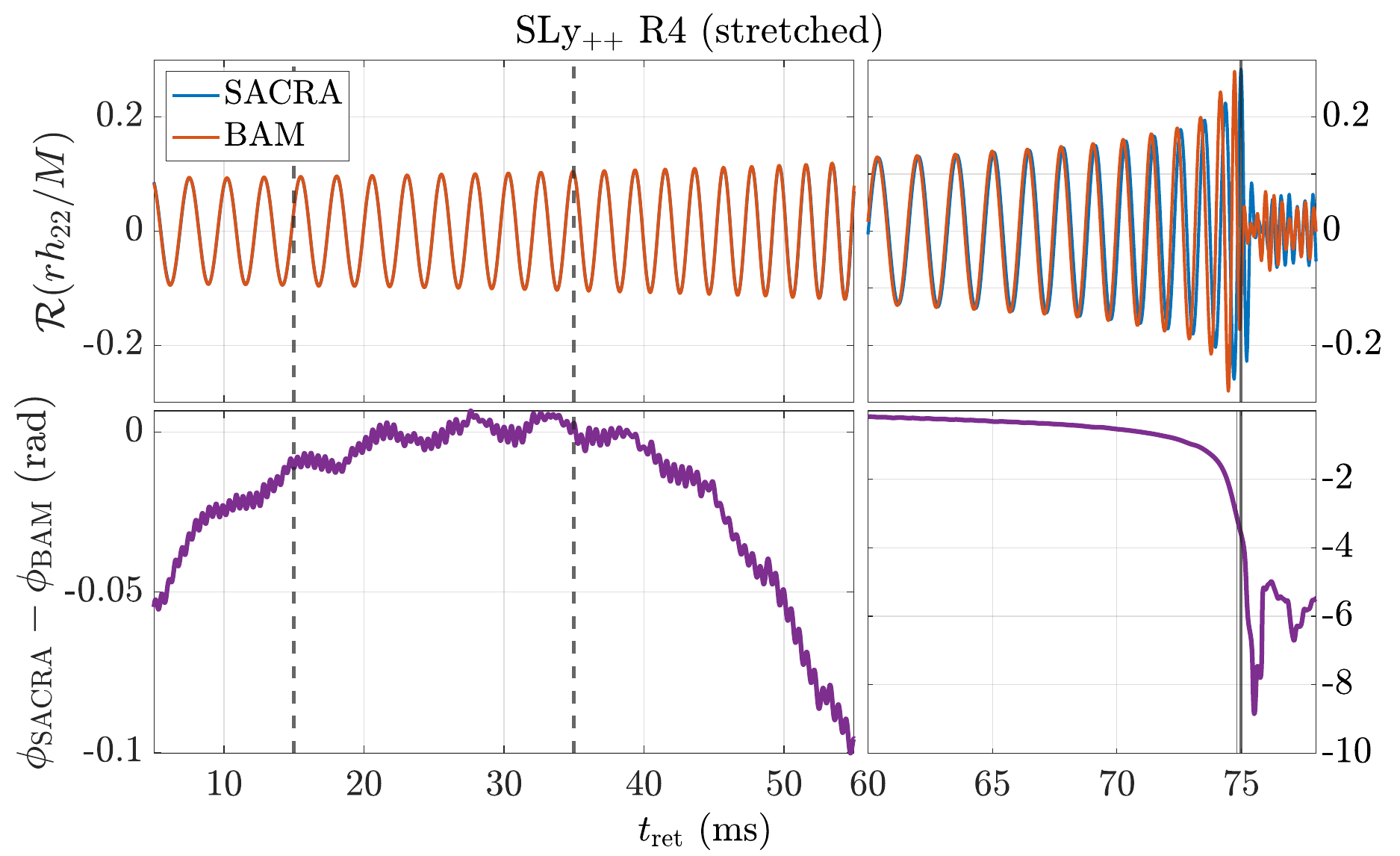}
    \includegraphics[width=\columnwidth]{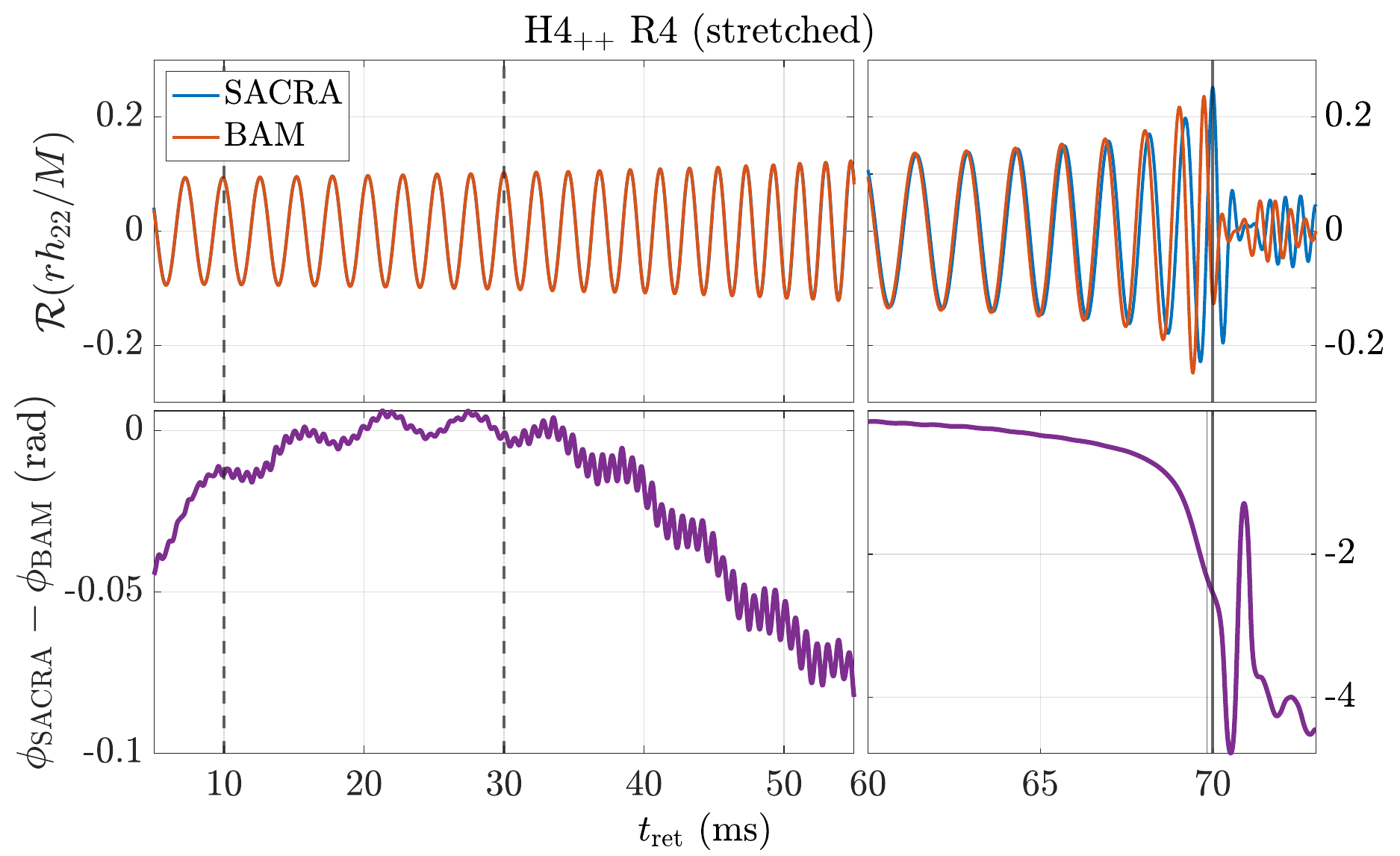}
    \caption{Numerical waveforms, expressed in terms of stretched time and aligned between two codes, are presented for models SLy$_{++}$ and H4$_{++}$ as indicated in the plot title.
    }
    \label{fig:aligned_streched}
\end{figure}

The time-stretched waveforms of \textsc{SACRA} results are shown in \cref{fig:inf_res}.
The phase difference between resolutions is suppressed to approximately $\mathcal{O}(10^{-2})$ radians for most of the simulation while rising to $\sim1$~radian when approaching the merger. 
Although, after stretched in time, those with lower resolutions still evolve faster in phase and merge earlier, the dephasing is rather constant and the difference is substantially reduced.
The stretching factor that optimizes the match between a lower resolution waveform with that of R4 approaches unity from R1 to R3, suggesting the expected convergent behavior.
The order of the convergence then offers an estimate on the stretching factor to be performed on the R4 waveform to achieve the hypothetically infinite-resolution waveform.
\cref{fig:aligned_streched} shows the comparisons of the stretched results of two codes after aligned based on the integral \eqref{eq:I_phase}.
The results are similar to what have been found by comparing the R4 waveforms of two codes (cf.~\cref{fig:aligned}) while an increase by $\lesssim0.5$~rad is observed in the comparison of stretched waveforms.

\subsection{Influence of the initial data code}
\label{id_err}

Another component that can influence waveform accuracy is the initial data and the way it is constructed. 
To assess the uncertainties arising from initial data, we perform additional simulations of the same configurations with initial data produced by the \textsc{SGRID} code.
We ensure that the baryonic masses of the NSes in \textsc{SGRID} match those of the solutions obtained with \textsc{FUKA}.

In contrast to \textsc{FUKA}, \textsc{SGRID} does not have an automatic iterative process to obtain the target value of the dimensionless spin. That means that the velocity potential for the NS matter must be set, and the dimensionless spin parameter $\chi$ can only be calculated after the solution is complete and the angular momentum of the star is known. One can empirically derive fitting formulae to estimate the velocity potential~\cite{Dietrich2016BinaryNS}. However, in our case, the fitting formula does not provide an accurate solution, which we attribute to the high spins that lie outside the fitting and, thus, the validity region of the relation. We resort to a manual root-finding procedure to obtain a solution with the required $\chi$ value. 
In addition, \textsc{SGRID} does not employ automatic eccentricity reduction using 3.5PN estimates and yields the initial data with residual eccentricities up to $\mathcal{O}(10^{-2})$.
We use the code supplied as part of the \textsc{SGRID} source to obtain the eccentricity reduction parameters from the proper distances between two NSs. 
As with \textsc{FUKA}, we terminate the iteration process when the eccentricity is reduced to $e < 10^{-3}$. We quote the resulting values for the eccentricity in \cref{tab:fuka_sgrid_params}.

\begin{figure}%[htbp]
    \centering
    \includegraphics[width=\columnwidth]{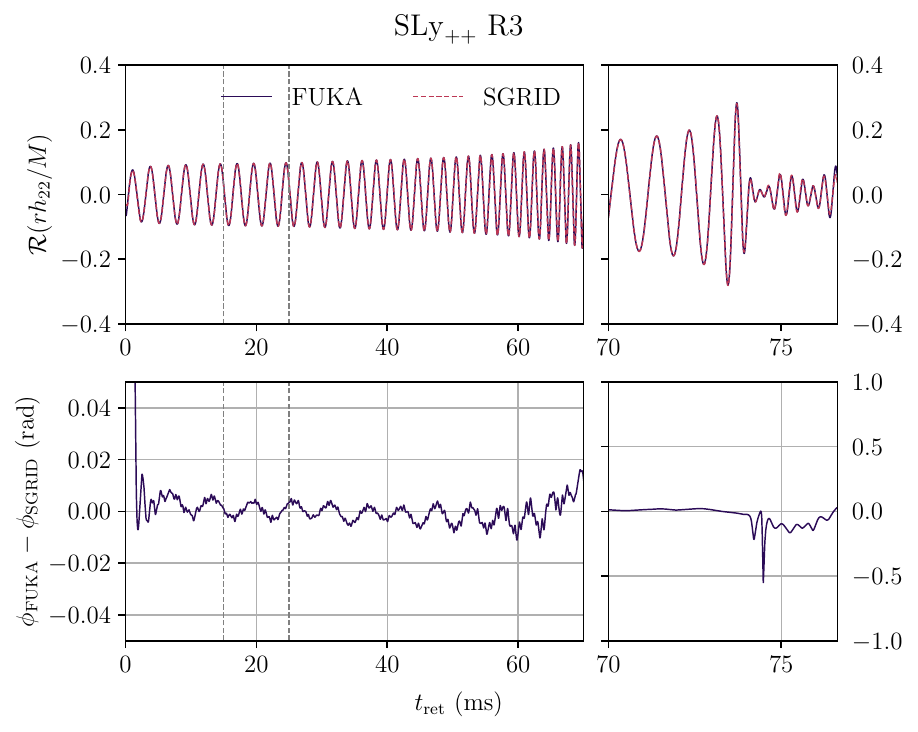}
    \includegraphics[width=\columnwidth]{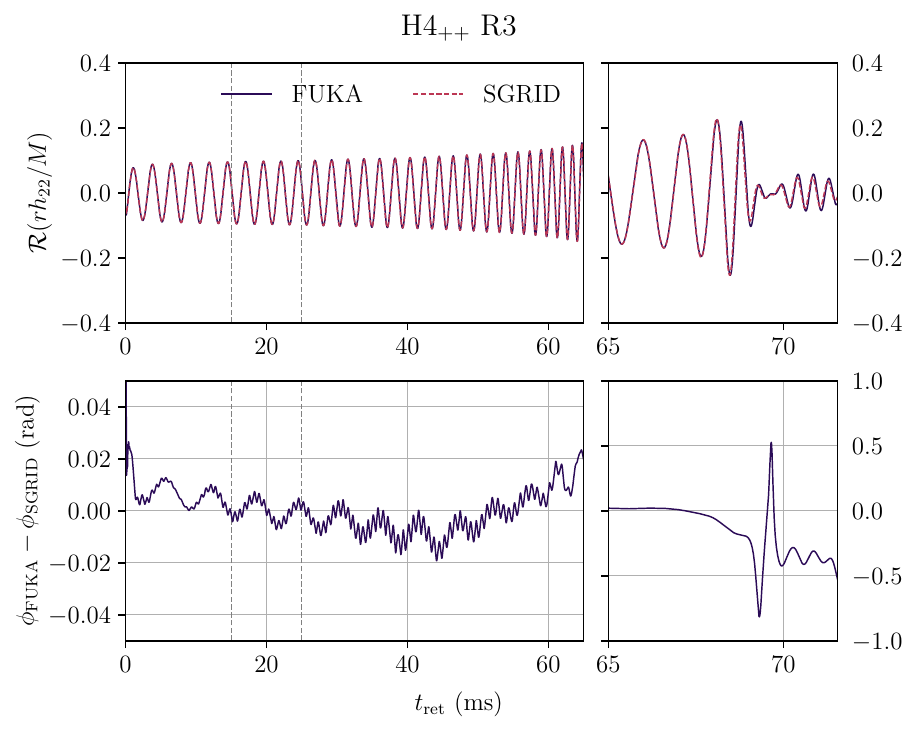}
    \caption{Waveforms produced in simulations by the \textsc{BAM} code with different initial data solvers, \textsc{FUKA} and \textsc{SGRID} (top panels), and the corresponding phase differences (bottom panels) at resolution R3 for both EOSs at coordinate extraction radius $r=900M_\odot$ as a function of retarded time. The vertical dashed lines designate the alignment window.}
    \label{fig:fuka_vs_sgrid}
\end{figure}

The evolution of the \textsc{SGRID} data was performed using the same grid and evolution configuration as earlier employed for \textsc{FUKA}, but only for a single resolution, namely R3. The resulting waveforms are compared to the ones obtained in the \textsc{FUKA} counterparts in \cref{fig:fuka_vs_sgrid}. 
The waveforms display excellent agreement, with the dephasing remaining within $\pm 0.02$~rad during the inspiral, generally oscillating around zero. We suggest that these oscillations are caused by the differences in the residual eccentricities, as their frequency is similar to the initial orbital frequency. The waveforms show a high level of agreement even after the merger, with typical values for the dephasing of $0.2$~rad for SLy and $0.4$~rad for H4, excluding the short spikes of dephasing in the post-merger phase; cf.~\cite{Bernuzzi:2013rza}. The residual eccentricity does not appear to have any noticeable effect on the waveforms during postmerger -- the \textsc{SGRID} waveform for SLy has higher eccentricity than for H4, yet the postmerger dephasing is higher in the H4 case.

To conclude the study of the initial data error, we want to highlight the presence of high-frequency central density oscillations at 2.273~kHz in case of $\rm{H4}_{++}$, and 2.720~kHz in case of $\rm{SLy}_{++}$.
These oscillations are present in both evolution codes and have the same frequencies regardless of the choice of the initial data solver. 
These oscillations are significant enough to influence the gravitational waveform, modulating the inspiral frequency. 
We have examined the pattern of the oscillation across the star and identified them as a $(2,0)$ density mode.
Judging by the frequency, it could be an f-mode, but a detailed eigenfunction analysis is required to clarify this further.
The exact origin of their excitations is also unknown, but we suggest that it arises from approximations employed in derivation of the equations solved for the initial data construction, such as neglecting higher-order spin terms.

\section{Comparison to analytic waveform models}\label{secIV}

\begin{figure}
    \centering
    \includegraphics[width=\columnwidth]{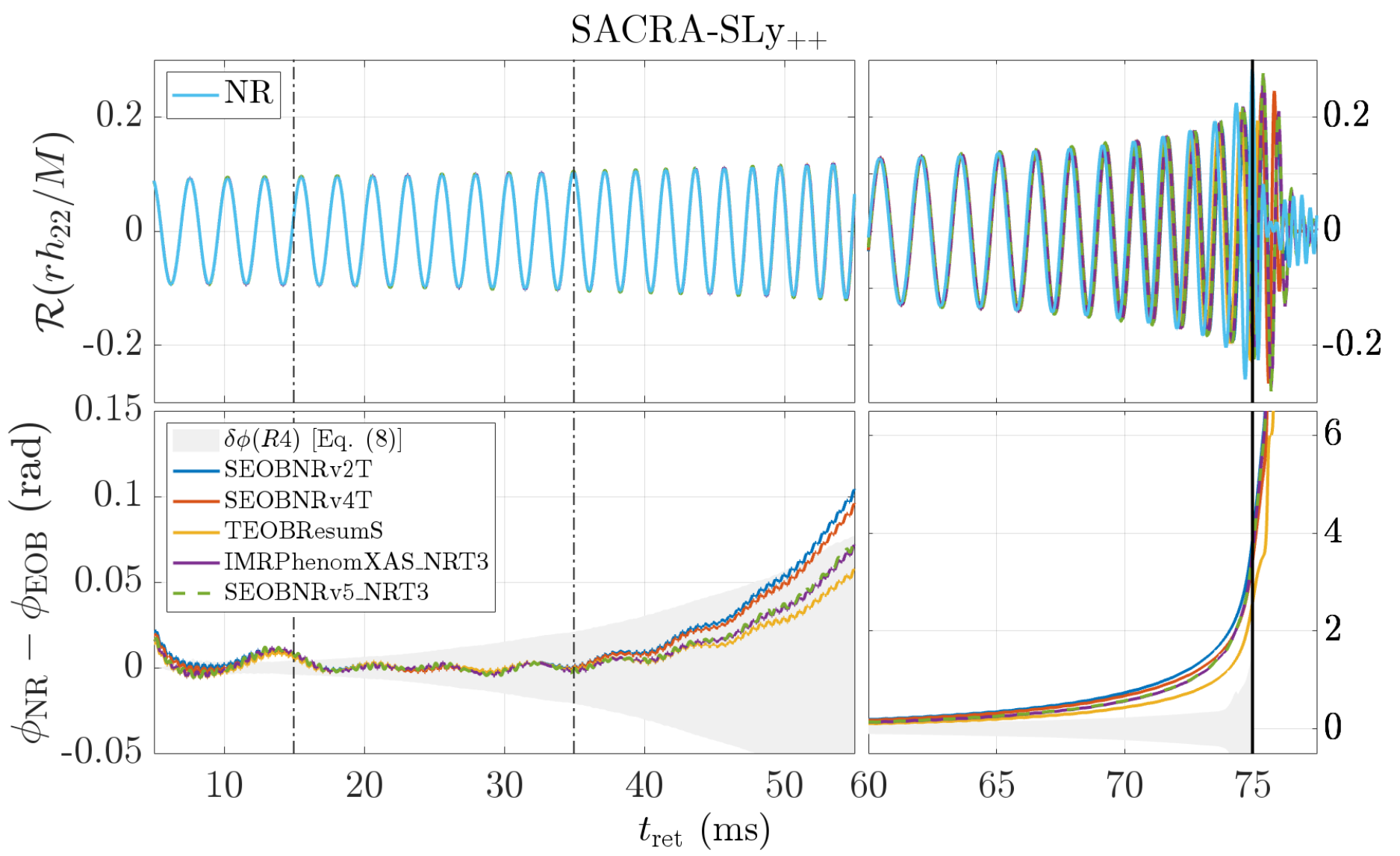}
    \includegraphics[width=\columnwidth]{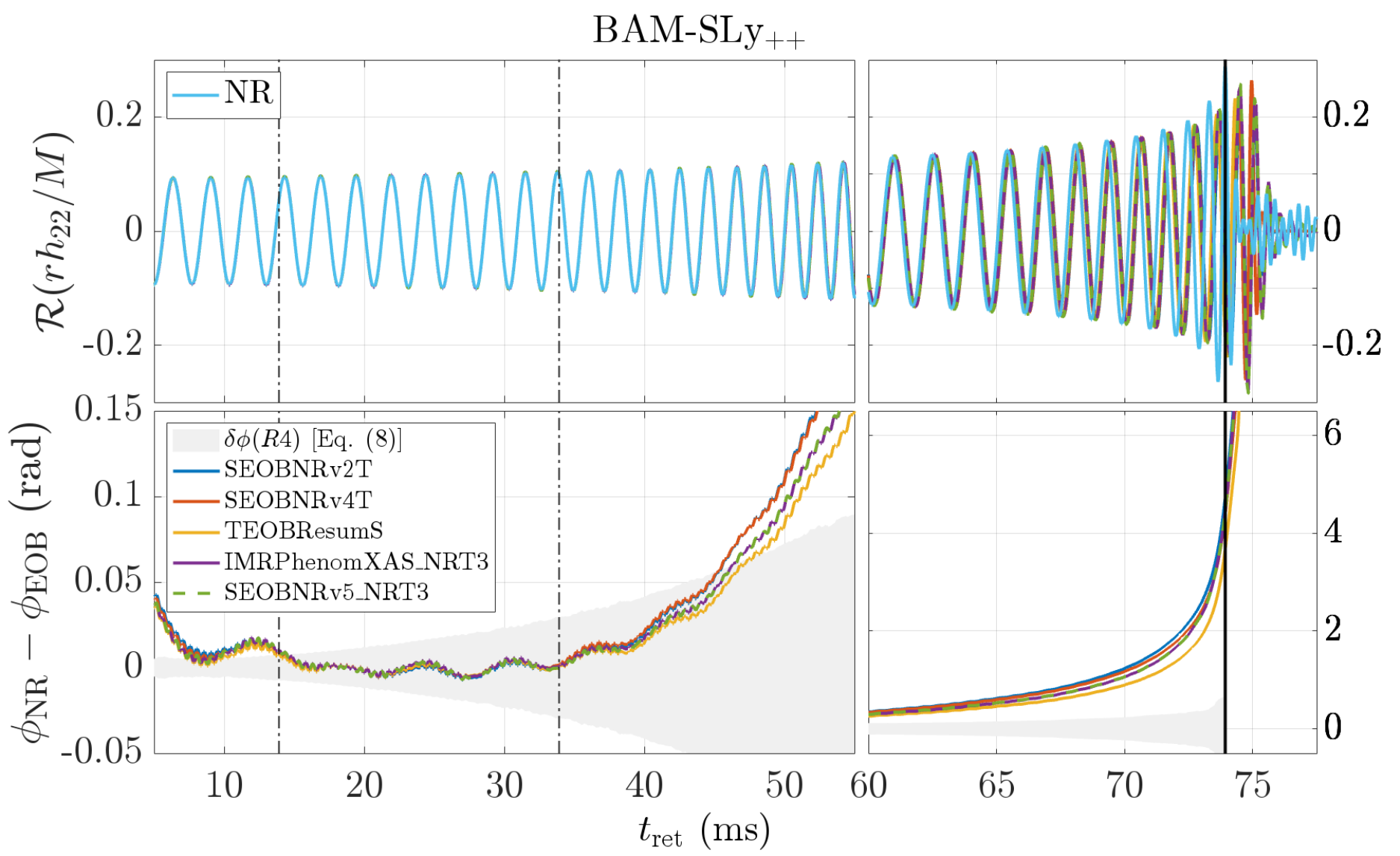}
    \caption{Comparison of NR waveforms with the selected EOB and phenomenological waveform models for the SLy$_{++}$ configuration. 
    The R4 numerical waveforms by \textsc{SACRA} (top panel) and \textsc{BAM} (bottom panel) are used for the analysis.
    The analytic waveforms are aligned with the NR one in the time interval of $t_{\rm ret}-t_{\rm mrg}=[-60, -40]$~ms relative to the merger time (between vertical dashed lines).
    The curves of \textsc{IMRPhenomXAS\_NRT3} and \textsc{SEOBNRv5\_NRT3} almost overlap with each other, making one of them barely visible.
    The merger time on the plot is determined from the NR waveforms and shown as the solid vertical lines.
    }
    \label{fig:vs_eob_sly}
\end{figure}
\begin{figure}
    \centering
    \includegraphics[width=\columnwidth]{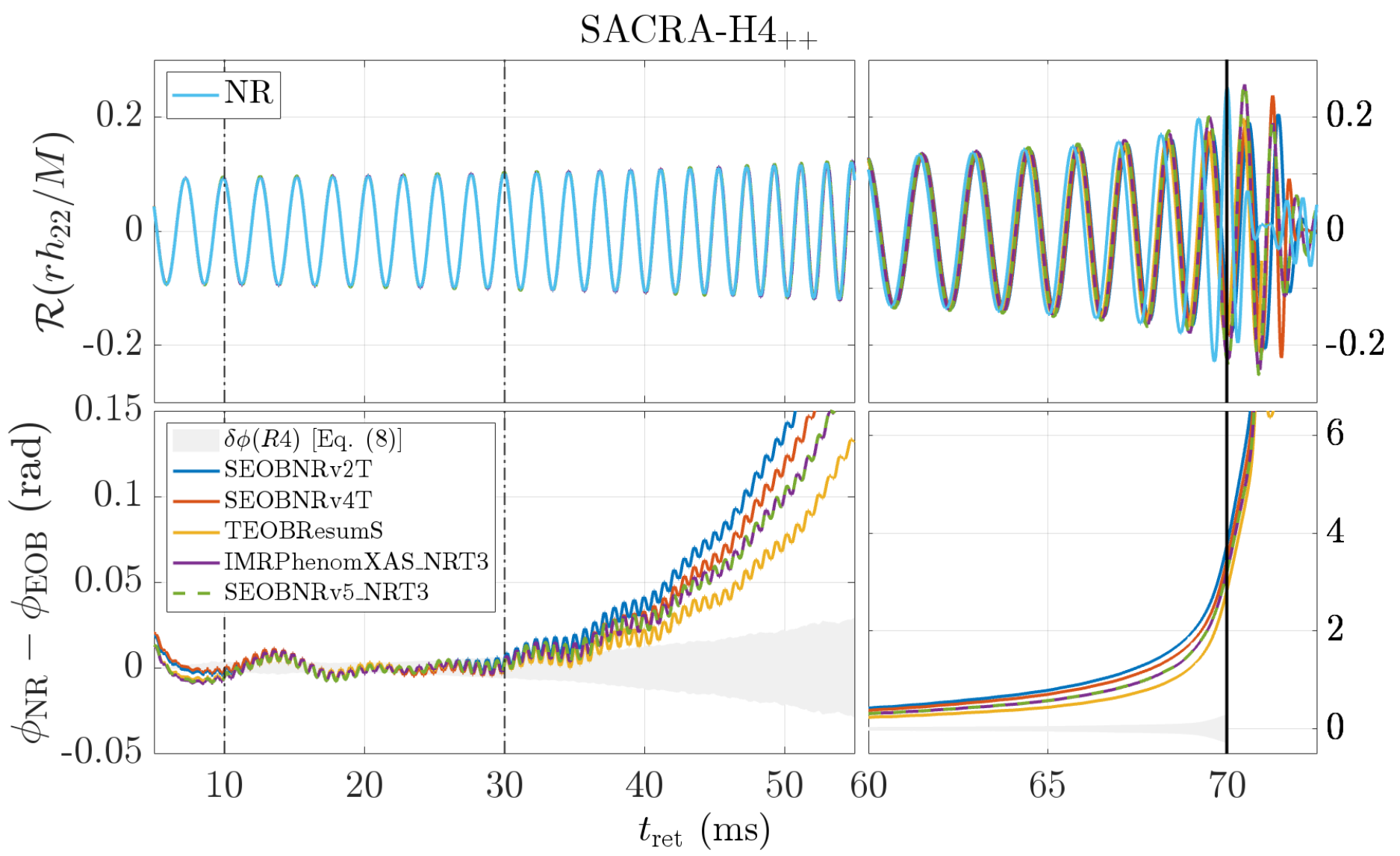}
    \includegraphics[width=\columnwidth]{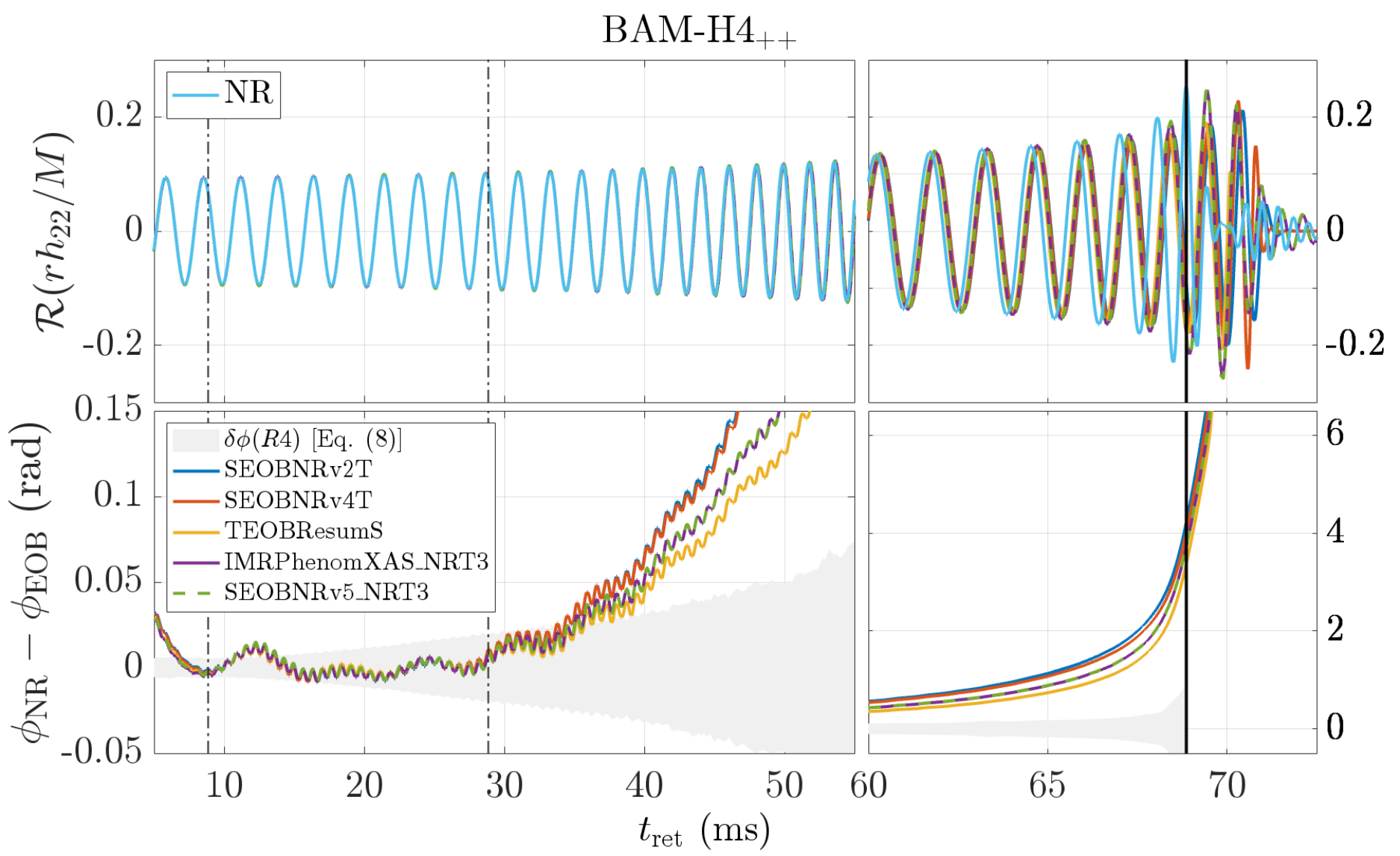}
    \caption{%\uj{Please shift down the x-caption.} 
    Same as \cref{fig:vs_eob_sly} while for model H4$_{++}$.
    }
    \label{fig:vs_eob_h4}
\end{figure}

\subsection{Time-domain Comparison}

The numerical waveforms of rapidly spinning BNSs performed here are outside the parameter space that have been covered in the literature to date, and thus the accuracy of analytic waveform models that rely fully on calibration with NR results remains to be examined.
In this section, we explore this issue for some of the latest EOB models -- \textsc{SEOBNRv2T} and \textsc{SEOBNRv4T}~\cite{Hild:2010id,Steinhoff:2016rfi,Steinhoff:2021dsn,Lackey:2018zvw}, and \textsc{TEOBResumS}~\cite{Damour:2009wj,Bernuzzi:2014owa,Nagar:2018zoe,Nagar:2018plt,Akcay:2018yyh} -- and phenomenological models \textsc{IMRPhenomXAS$\_$NRTidalv3} and \textsc{ SEOBNRv5$\_$ROM$\_$NRTidalv3}~\cite{Abac:2023ujg}.
Analytic waveforms are obtained from LALSuite~\cite{lalsuite} via PyCBC~\cite{alex_nitz_2024_10473621}.

We align the numerical results of the highest resolution models R4 with the selected waveforms by minimizing $I_{\rm phase}$ of Eq.~\eqref{eq:I_phase}.
The aligned waveforms for SLy$_{++}$ and H4$_{++}$ are shown in~\cref{fig:vs_eob_sly,fig:vs_eob_h4}, respectively, overplotted with the phase uncertainties due to finite grid resolution [shaded area; \cref{eq:poly_R4}].
In the early part of the waveform, the agreement between NR data and the analytic waveforms is within the numerical error due to finite resolution.
However, a sizeable phase difference between the waveform approximants and the NR data accumulates within $\lesssim 20$~ms before the merger, reaching $\lesssim4$ radians at the merger time (black vertical line) for both \textsc{SACRA} and \textsc{BAM} results.
For EOB models, the peak GW amplitude occurs by $\sim1.5$~ms later relative to the NR merger time for \textsc{SEOBNRv2/4T} and by $\sim1$~ms later for \textsc{TEOBResumS} in the case of H4$_{++}$.
The SLy$_{++}$ model shows smaller delays: $\lesssim1$~ms for \textsc{SEOBNRv2/4T} and within $\pm0.1$~ms for \textsc{TEOBResumS}. 
Overall, \textsc{TEOBResumS} aligns slightly better with NR waveforms in the phase shift at merger and the delay of coalescence for the configurations considered here.

It is also critical to quantify the uncertainty of the BBH sector of these analytic models, as they could, in principle, be an important source of deviation rather than tidal dephasing.
This uncertainty is estimated by comparing the waveforms of the same binary parameters in the SXS catalog with these EOB models (see \cref{appendixC} for details), and is found to be at least twice as small as those for BNS waveforms.
Hence, the discrepancies of modeling tidal effects noticeably exceed the uncertainties involved in modeling the BBH baselines.
The much smaller error seen in the BBH sector suggests that the primary source of error in the analytic BNS waveform models may be (i) the finite size effects associated with the multipole moments of NSs, or (ii) effects that are present in BBH system while irrelevant to the BNS binaries such as horizon absorption \cite{Tagoshi:1997jy,Alvi:2001mx,Porto:2007qi,Saketh:2022xjb}.
To further investigate the influence of these two possibilities, it would be beneficial to understand how the analytic error behaves in terms of GW frequency $\omega$ since different effects are of different PN order, and thus they scale distinctly with $\omega$.

\subsection{Frequency-domain Comparison}

The raw numerical data of $\omega$ contains high frequency noise, and the oscillatory behavior prevents a straightforward $\Delta\phi$-$\omega$ analysis.
This issue can be hurdled by the scheme detailed in \cite{Baiotti:2010xh,Baiotti:2011am} to smooth out this quantity for NR waveforms, which is recapped as follows.
The raw data of GW phasing is first cleaned by fitting it to an analytic PN expansion.
The expression of the latter expansion is given as
\begin{align}
    \phi=\phi_0-\frac{2M^2}{m_1m_2}x^{-5}\left(1+p_2x^2+p_3x^3+p_4x^4\right),
\end{align}
where we introduced $x=\left[m_1m_2(t_{\rm mrg}-t)/5M^2\right]^{-1/8}$ and the fitting coefficients $p_2$, $p_3$, and $p_4$.
To stave off the potential overfitting problem, we have confirmed that the deviation between the raw data ($\phi^{\rm NR}$) and the clean phase ($\phi^{\rm fit}$) is $\vert\phi^{\rm fit}/\phi^{\rm NR}-1\vert<10^{-4}$ throughout the inspiral up to the merger time.
Even with the "cleaned" phase, the fitting will be deteriorated by including the initial signal and that in the very last moment before the merger.
We thus cut the first 1--2 ms and the last 0.1--0.2 ms of the simulated inspiral waveforms in this work to keep as much numerical data as possible while seeking a reasonable quality of the fitting procedure.
The time derivative of the cleaned phase then produces a smooth $\omega$.

\begin{figure}
    \centering
    \includegraphics[width=\columnwidth]{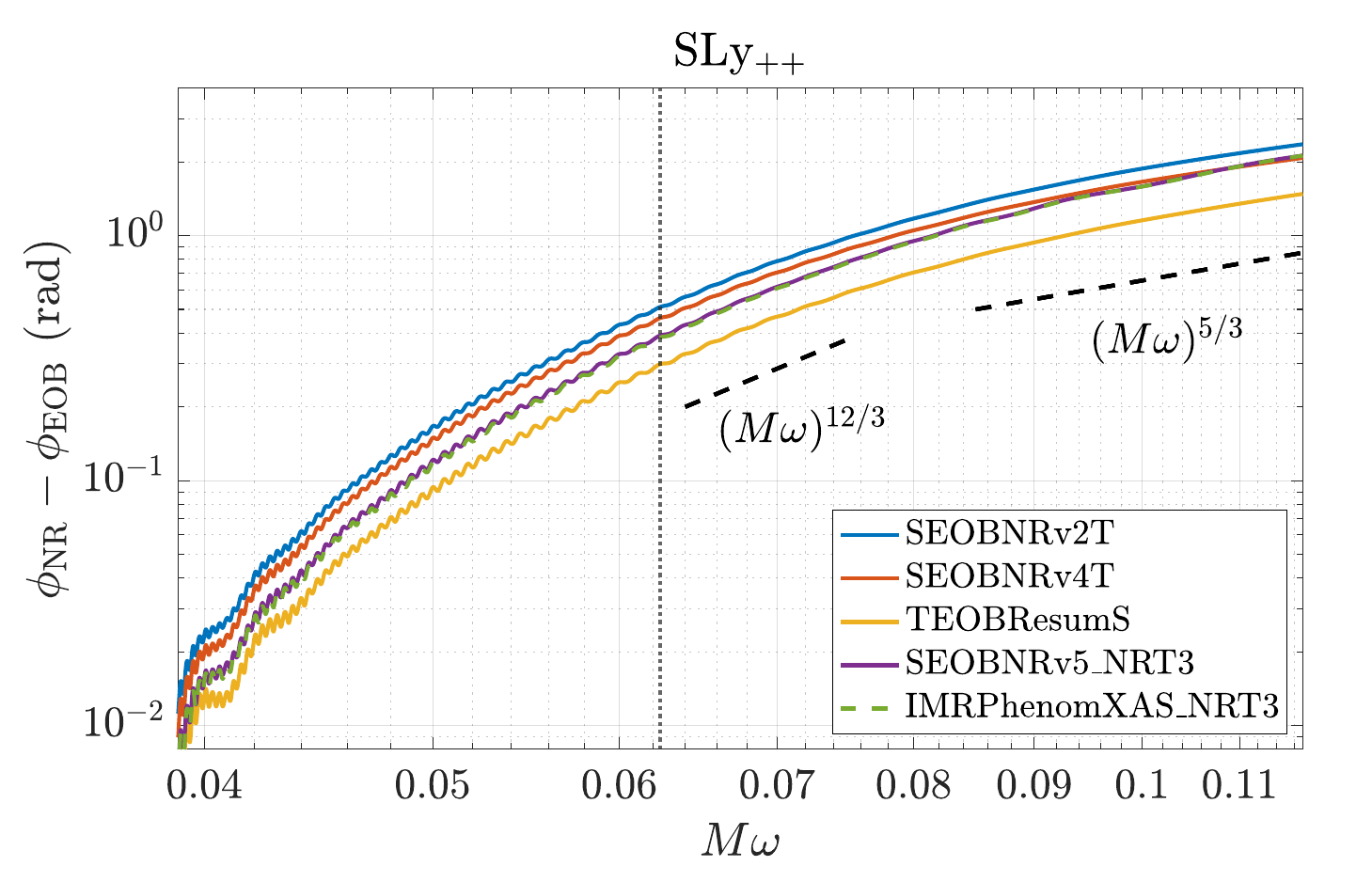}
    \includegraphics[width=\columnwidth]{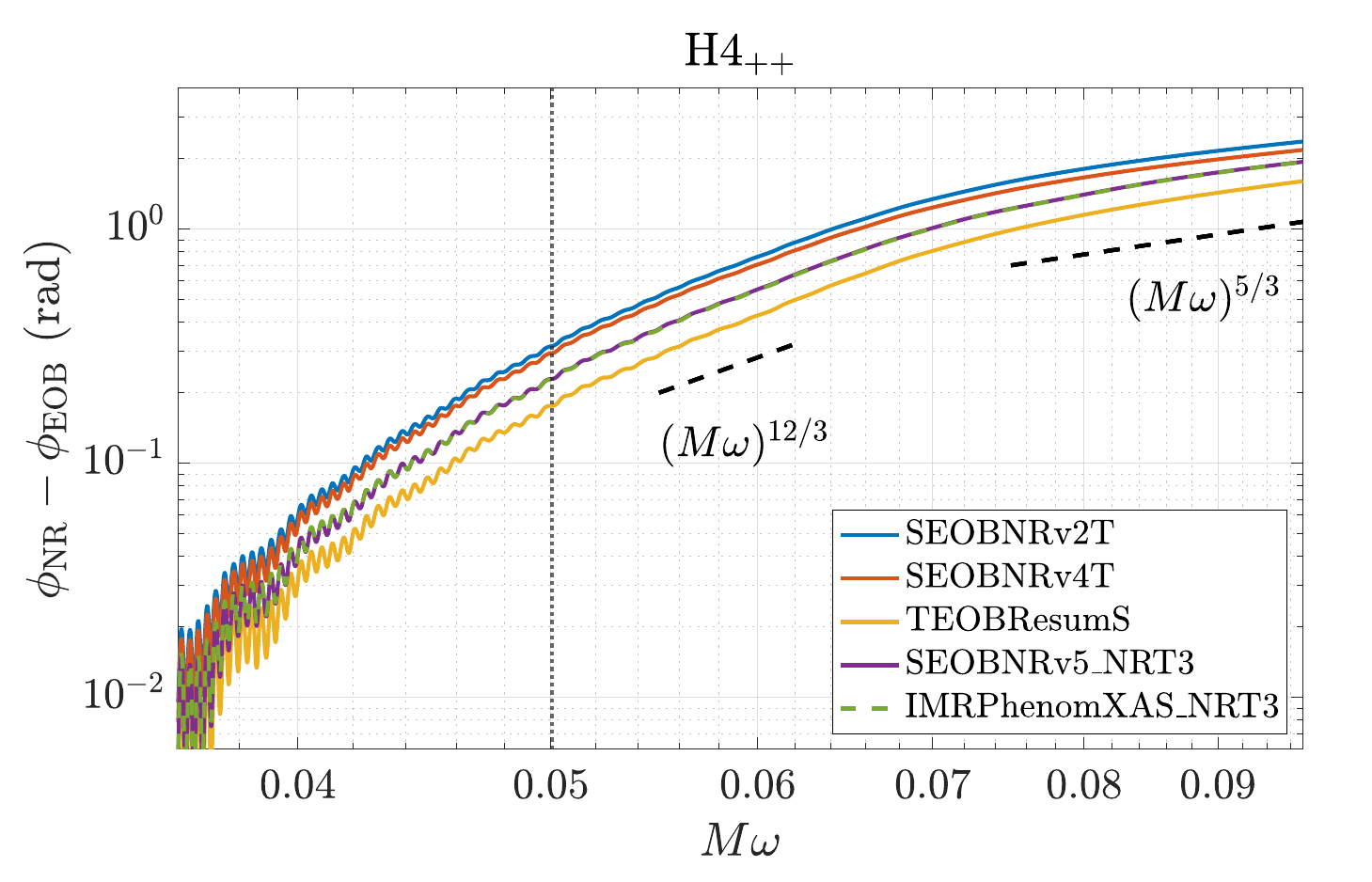}
    \caption{Phase difference between the \textsc{SACRA} waveform and the considered analytic waveform models as functions of $M\omega$.
    The dash slope lines depict a tendency of 2.5 PN and 6 PN orders, and the frequency at the contact of two NSs are represented as the vertical dotted line.
    }
    \label{fig:dphase_ome}
\end{figure}

We have tested our findings and the dephasing with respect to individual approximants is independent of the time matching window and the evolution codes.
For better visibility of this behavior.
However, for better visibility, we only plot the dephasing for the match window of $[-60,\,-40]$~ms before merger for the \textsc{SACRA} waveforms; cf.~\cref{fig:dphase_ome}.
Denoting the PN expansion parameter as $v=(M\Omega)^{2/3}$, a trend of $\propto v^{2.5}$ and $v^6$ suggests that $\Delta\phi$ scales as 2.5 and 6 PN terms at late times, respectively.
This varying behavior indicates that the discrepancies could arise from several possible sources, including contributions of multipole moments of NSs \cite{Thorne:1980ru, Ryan:1995wh}, and/or the horizon absorption effect \cite{Tagoshi:1997jy,Mino:1997bx,Alvi:2001mx,Porto:2007qi,Saketh:2022xjb}. 
We examine these possibilities in detail in the following discussion.

\subsubsection{Effects of spin-induced moments}

Mass ($M,M_2,M_4,\dots$) and current moments ($J,S_3,S_5,\dots$) of NSs contribute to the radiative moments of spacetime, thereby affecting the energy spectrum and flux of GWs \cite{Ryan:1995wh}.
For slowly spinning binaries, the dominant finite-size effect comes from the spin-induced mass quadrupole, $M_2 = -Q_2 M^3$, while higher moments are typically subleading.
However, for the spin as high as $\chi=0.5$ considered here, the effects due to the current octupole $S_3=q_3M^4$ and mass $2^4$--pole moments $M_4=Q_4M^5$ can also contribute significantly as $q_3$ and $Q_4$ become comparable to $Q_2$ \cite{Shibata:1998xw,Pappas:2013naa,Yagi:2014bxa}.
In fact, these coefficients of moments are found as $\{Q_2,\,q_3,\,Q4\}=\{1.288,\,1.604,\,4.047\}$ and $\{1.782,\,2.656,\,7.854\}$ for the spinning NSs in the SLy$_{++}$ and H4$_{++}$ models, respectively.
The $M_2$-related effects on conservative dynamics have been computed up to 5PN order (3PN relative to leading order) \cite{Levi:2022rrq,Bautista:2024agp}, and are included in the \textsc{TEOBResumS} model up to 4PN \cite{Nagar:2018plt}.
By contrast, the \textsc{SEOBNRv4T} model only incorporates them at leading (2PN) order \cite{Lackey:2018zvw}.
In addition to the EOB models, the $M_2$ influence in wavform's phase has been formulated into the \textsc{TaylorF2} model and can be found in Sec.~III.~C of \cite{Nagar:2018plt}.

The effects of the current octupole and mass hexadecapole moments on the conservative dynamics have been derived to next-to-leading order (4.5PN and 5PN, respectively) \cite{Levi:2014gsa,Levi:2022rrq,Bautista:2024agp}.
\textsc{TEOBResumS} includes leading-order contributions from both, though the octupole is only treated phenomenologically \cite{Nagar:2018plt}.
On the other hand, these effects were not included in \textsc{SEOBv2/4T} and only present in the latest version of \textsc{SEOBv5THM} \cite{Haberland:2025luz}.
The omission of these effects in \textsc{SEOBv2/4T} could perhaps explain the lower deviation from NR waveforms observed for \textsc{TEOBResumS}.
The leading-order effects of $S_5,M_6,\dots$ enter at $\ge5.5$~PN order, and have not yet been completely computed in the PN framework.
On top of the aforementioned effects, tidal effects are no longer well-described by the tidal deformability alone for rapidly spinning BNSs.
In fact, spin-induced multipolar deformations sizeably enhance tidal effects, and the corrections also start at 5PN order \cite{Kim:2021rfj,Levi:2022rrq,Kim:2022bwv,Mandal:2022ufb}.

We expect that the comparison of NR to analytic waveforms should become spurious once the two NSs come into contact.
Acquiescing the validity of analytic models after NSs touch on each other could thus lead to nonphysical predictions especially since these models have not been calibrated against NR simulations for such high-spin configurations.
Prior to contact, the dephasing led by finite size effects is more prominent for stiffer EOS, which is indeed seen in \cref{fig:dphase_ome} up to the onset of contact of the H4$_{++}$ model at $M\omega\simeq 0.05$.

To close, we provide an order of magnitude estimates for some of the aforementioned effects.
Since the influence in phasing resulting from hexadecapole and beyond has not been written as a closed form, we only estimate the effects of $M_2$ and $S_3$ moments in below.
For the equal-mass, equal-spin BNS considered here, the GW's phase due to quadrupole-monopole effect in the \textsc{TaylorF2} model is given by [Eqs.~(44)--(47) of \cite{Nagar:2018zoe}]
\begin{align}
    \phi_{\rm QM} &= \frac{75Q_2}{128\nu}\left(\frac{M\omega}{2}\right)^{-1/3}
    -\left(\frac{45}{16}+\frac{15635}{896\nu}\right)\frac{Q_2}{2}\left(\frac{M\omega}{2}\right)^{1/3}
    \nn
    &+\frac{75}{16\nu}\pi Q_2
    \left(\frac{M\omega}{2}\right)^{2/3}\,,
\end{align}
where $\nu=m_1m_2/M^2=0.25$\footnote{The connection between the notations of the spin-induced mass quadrupole coefficient here and that in \cite{Nagar:2018zoe} is $Q_2/2=-\tilde{a}^2C_Q$. }.
The octupole contribution is quoted from the \textsc{TaylorT2} model and reads \cite{Marsat:2014xea}
\begin{align}
    \phi_{\rm oct} = \frac{55}{16}q_3x\,.
\end{align}
Between $M\omega = 0.04$ and contact, the accumulated phases are $\{\phi_{\rm QM},\,\phi_{\rm oct}\}=\{-1.611,0.138\}$ and $\{-1.318,0.118\}$~rad for the SLy$_{++}$ and H4$_{++}$ models, respectively.
The obtained values of $\phi_{\rm oct}$ are roughly consistent with the differences between the \textsc{SEOBNRv2/4T} and \textsc{TEOBResumS} waveforms, which aligns with the fact that the octupole effect is not included in the former model.
The observed $\sim 6$~PN scaling in \cref{fig:dphase_ome} indicates that finite-size effects of higher PN orders may also be important for the considered spin parameters. 
However, estimating these higher PN order effects is beyond the reach of current analytic knowledge.

\subsubsection{Astray horizon absorption effects}
\label{Sec:astrayhorizon}

While it is evident that (semi-)analytic models that treat the two objects as being separated even after their contact will have intrinsic errors in their modelling, it might till be important to understand the exact reason behind the observed tendency of $\sim 2.5$ PN order at late times for the dephasing.
Adding the fact that the observed dephasing is independent of EOS, this directs us to consider that the dephasing in late phase could be due to the inconsistent inclusion of horizon absorption effects because this effect appears from 2.5 PN order for spinning BHs~\cite{Galtsov:1982hwm, Mino:1997bx}.
Despite their irrelevance to BNSs, these effects are inherited in all the analytic models adopted here since these models rely on certain BBH baseline, which includes these effects.

The phase corrections due to horizon absorption of BHs are given by [Eqs.~(5.10)--(5.15) of \cite{Saketh:2022xjb}]
\begin{align}
    \phi_{\rm HA} &=-\frac{5}{192}\chi(1+3\chi^2)
    -\frac{5}{96}\chi(1+3\chi^2)\log\left( \frac{M\omega}{2}\right)\nn
    &+\frac{15\chi}{5376}\left[
        \frac{105}{2}(1+3\chi^2)
        -\frac{4707\chi^2+1779}{4}
    \right] \left( \frac{M\omega}{2} \right)
\end{align}
up to the next-to-leading order at 3.5PN.
The accumulated effect from $M\omega=0.04$ to the merger amounts to $-0.145$ and $-0.112$~rad for the SLy$_{++}$ and H4$_{++}$ models, respectively.
The inclusion of $\phi_{\rm HA}$ in the BBH baseline will then underestimate the phase for BNS cases. 
While this potential problem could be eliminated if spinning BNS waveforms would be employed for the calibration of phenomenological BNS models or effenctive-one-body models describing the BNS coalescence, none of the existing waveform models employed spinning BNS systems during the calibration. This highlights the needs for further tests and comparison on a larger parameter space region to validate our observation.

\subsection{Phase acceleration}\label{sec:Qw}
In above, comparing phase errors requires us to align the waveforms, and the alignment itself could introduce some systematics in the measurement of error budget.
In this section, we perform another sort of comparison, making use of the dimensionless quantity,
\begin{align}
    Q_\omega=\frac{{\rm d}\phi}{{\rm d}\ln\omega}=\frac{\omega^2}{{\rm d}\omega/{\rm d}t}\,.
\end{align}
This quantity effectively estimates the number of GW cycles spent at a given logarithm GW frequency $\omega$ \cite{Damour:2000gg}. 
In addition, its inverse measures the validity of the stationary phase approximation, often assumed when deriving frequency-domain phasing from time-domain waveforms.
With this quantity, the comparison is conducted in the frequency domain, and it thus helps to avoid the potential issue of alignment.

\begin{figure}
    \centering
    \includegraphics[width=\columnwidth]{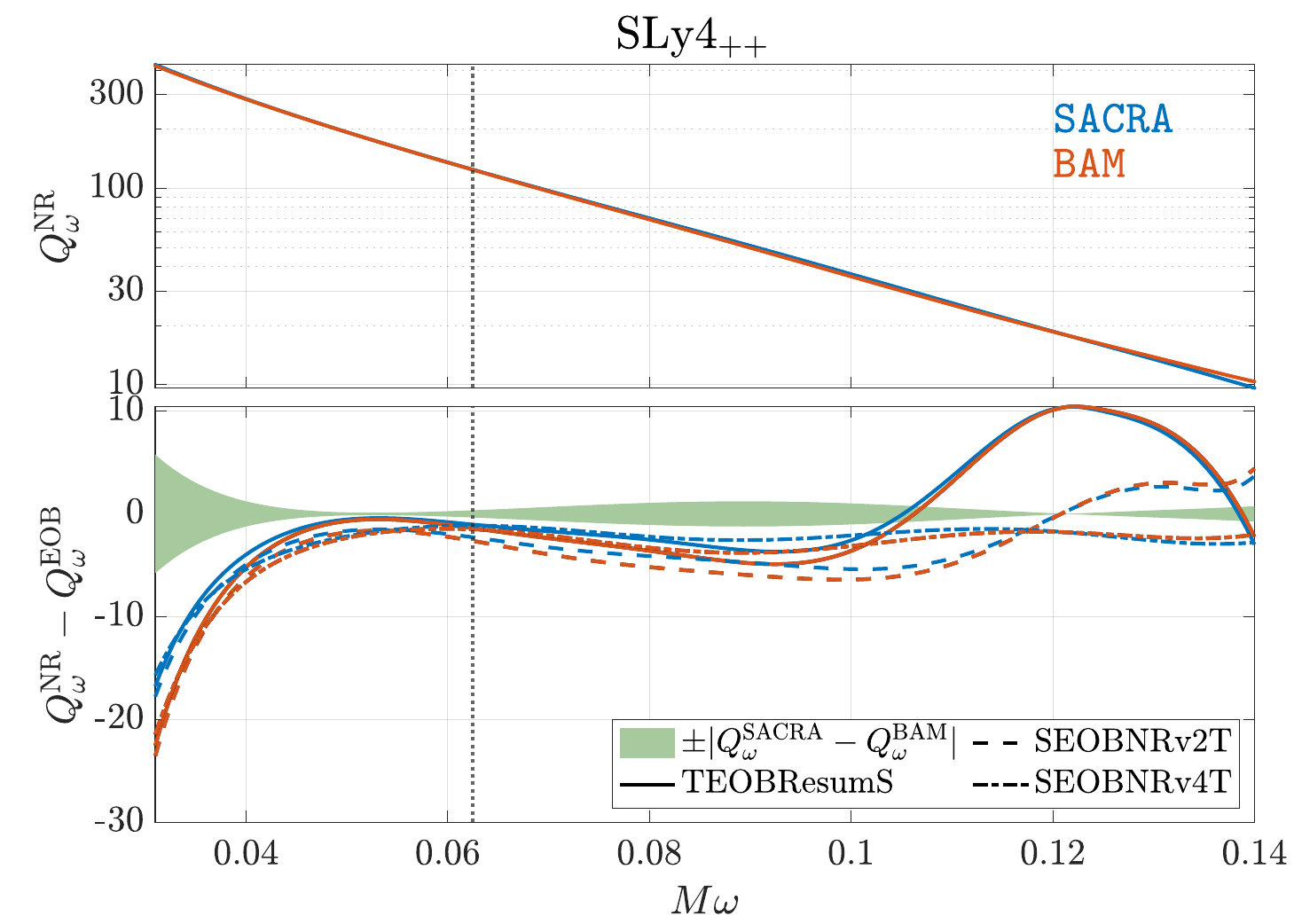}
    \includegraphics[width=\columnwidth]{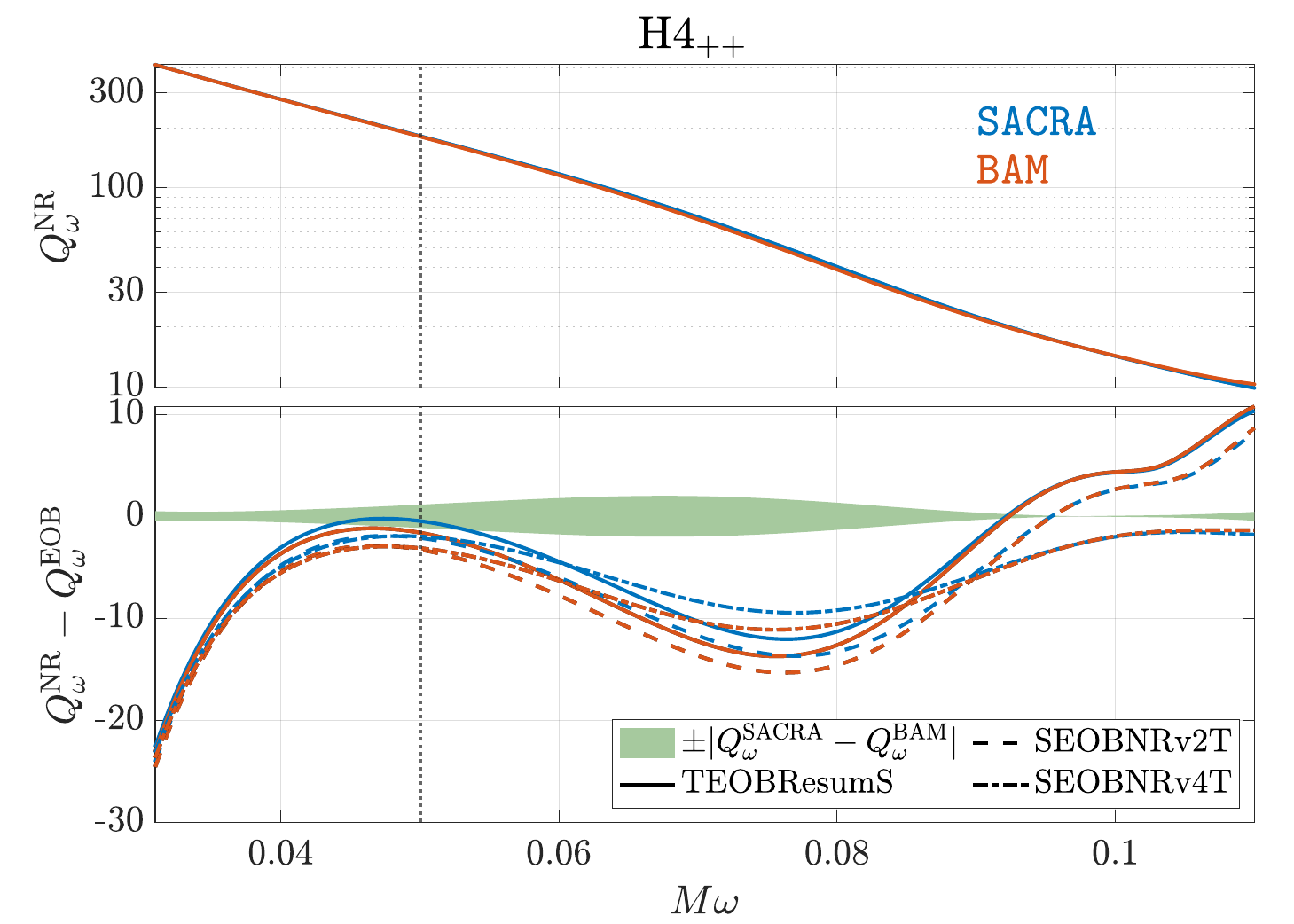}
    \caption{Deviation in the quality factor $Q_\omega$ between NR waveforms of \textsc{SACRA} (blue) and \textsc{BAM} (red) to the considered EOB approximants for models SLy$_{++}$ (top) and H4$_{++}$ (bottom). 
    The shaded region represents the discrepancy between the two numerical codes, with its boundaries defined by $\pm |Q_\omega^{\rm SACRA}-Q_\omega^{\rm BAM}|$.
    The frequency at the contact of two NSs are represented as the vertical dotted line.
    The analysis was not conducted for data up to merger time (see the main text), which happens at $M\omega\simeq0.16$ and 0.12 for models SLy$_{++}$ and H4$_{++}$, respectively. 
    }
    \label{fig:Qw}
\end{figure}

We compare the R4 NR waveforms and EOB models in terms of this quality factor as a function of mass-scaled frequency $M\omega$ in~\cref{fig:Qw}. 
Several observations can be made:
\begin{enumerate}[label=(\roman*)]
    \item The difference $\lvert Q_\omega^{\rm SACRA}-Q_\omega^{\rm BAM}\rvert$ reflects numerical uncertainties and is represented by the shaded area in \cref{fig:Qw}.
    The percentage mismatch between them become most prominent after two NSs come into contact (dashed vertical line).
    At peak, the percentage difference grows to $<5\%$ and then damps out until the final rise shortly before the merger.
    The waveforms for two codes coincide at $M\omega\simeq0.12$ and 0.1 for models SLy$_{++}$ and H4$_{++}$, respectively.
    This transient agreement also appears (though at different frequencies) when comparing NR and EOB models (see below).
    In addition, an early-time deviation of $\lesssim 2\%$ is noticeable in the SLy$_{++}$ model, while it is much smaller in the H4$_{++}$ model.
    \item The deviation between EOB models is comparable to the NR-EOB difference, aligning with the results shown in \cref{fig:dphase_ome}.
    However, the deviations of analytic waveform models from simulation data (curves in the bottom panels) generally exceeds the mismatch between two numerical waveforms as the curves most reside outside the the shaded area.
    \item Same as the deviation between the two NR waveforms, the deviation of $Q_\omega$ here is not monotonic.
    This fluctuant behavior does not allow for a discussion in term of PN effects.
    For the SLy$_{++}$ model, the deviation stays at a comparable level throughout the inspiral for \textsc{SEOBNRv2/4T} models except for the early signal at $M\omega<0.04$.
    One of them (\textsc{SEOBNRv2T}) flips the sign at the moment that the NR waveforms coincide, while the other retains the sign up to the merger.
    The behavior of NR-\textsc{SEOBNRv2/4T} differences is similar for the H4$_{++}$ model, including the \textsc{SEOBNRv2T}'s flipping sign when NR waveforms agree with each other.
    The transient agreement with NR of \textsc{SEOBNRv2T} there and also that of \textsc{TEOBResumS} at slightly lower frequencies is in line with the transient consistency between NR and EOB models reported in the literature (e.g., \cite{Bernuzzi:2014owa,Albertini:2024rrs}).
    \item Numerical uncertainty between the two codes is generally smaller than the NR--EOB deviation.
    Also, the $Q_\omega$-analysis suggests that the \textsc{SEOBNRv2T} models match better the NR results while the above analyses of dephasing and merger-time delay suggest otherwise: the NR-EOB phase difference is lower for \textsc{TEOBResumS} (cf.~\cref{fig:vs_eob_sly,fig:vs_eob_h4}).
    More comprehensive investigation of rapidly spinning BNS inspiral is needed to further distinguish the validity of the waveform models over wider parameter space.
\end{enumerate}

\section{Summary and Conclusion}\label{secV}

We have analyzed the numerical uncertainties in the NR waveform of various types using BNS inspirals with large aligned spins $\chi_1=\chi_2=0.5$.
A soft and a stiff EOS, SLy4 and H4, were employed to ensure that our conclusions were not biased by the EOS stiffness.
We performed simulations that covered the last $\approx30$ and 35 GW cycles before merger time for the SLy$_{++}$ and H4$_{++}$ models, respectively.
These simulations were the longest to date of highly spinning BNS systems.
In light of the novelty of the conducted simulations, we monitor the convergence order of the numerical data before making estimates of the waveform uncertainties. 
This is examined in \cref{res_err}.

We found that the ID underwent a relaxation phase as it adapted to the evolution code.
As a result, the convergence order estimated during this initial phase differs noticeably from that obtained at later times.
After the relaxation phase, the waveform of the SLy$_{++}$ model exhibits a convergence order of $p_{\rm conv}\sim3.4$ in both codes.
On the other hand, the behavior of $p_{\rm conv}$ after relaxation differs slightly between two codes for the H4$_{++}$ model: the \textsc{SACRA} simulation shows a convergence order of $\lesssim 4$, while the \textsc{BAM} simulation features $p_{\rm conv} \gtrsim 3$ shortly after relaxation, which subsequently increases to and remains at $\lesssim 4$ after 40~ms of the simulation.
In addition, the duration and manner of the relaxation phase depend on the evolution code, making it challenging to compare the initial phases of evolution between two codes.
It is for this reason that a direct comparison of waveforms from two codes is unplausible, and waveforms should be aligned before conducting detailed comparison of them.

\begin{figure}
    \centering
    \includegraphics[width=\columnwidth]{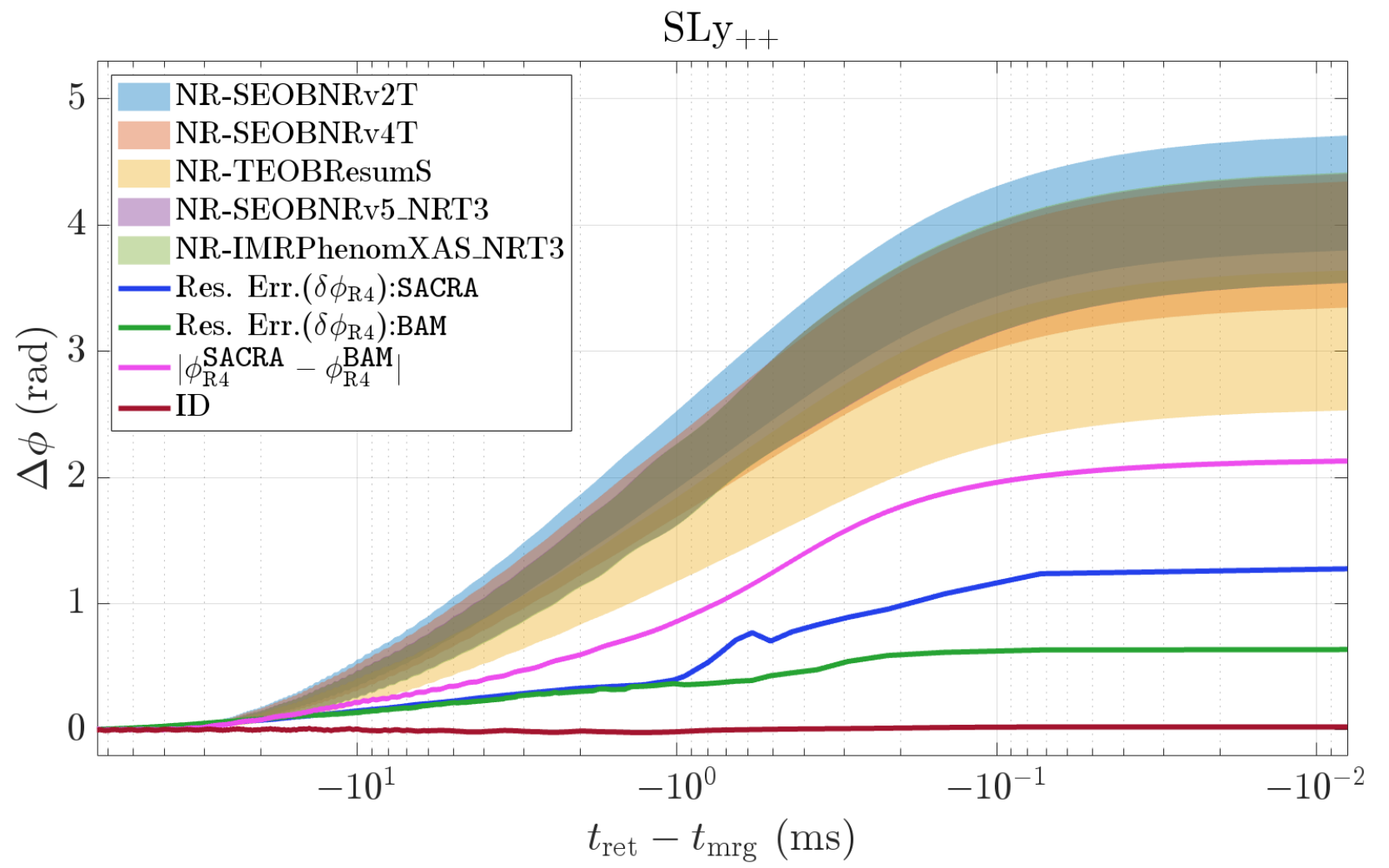}
    \includegraphics[width=\columnwidth]{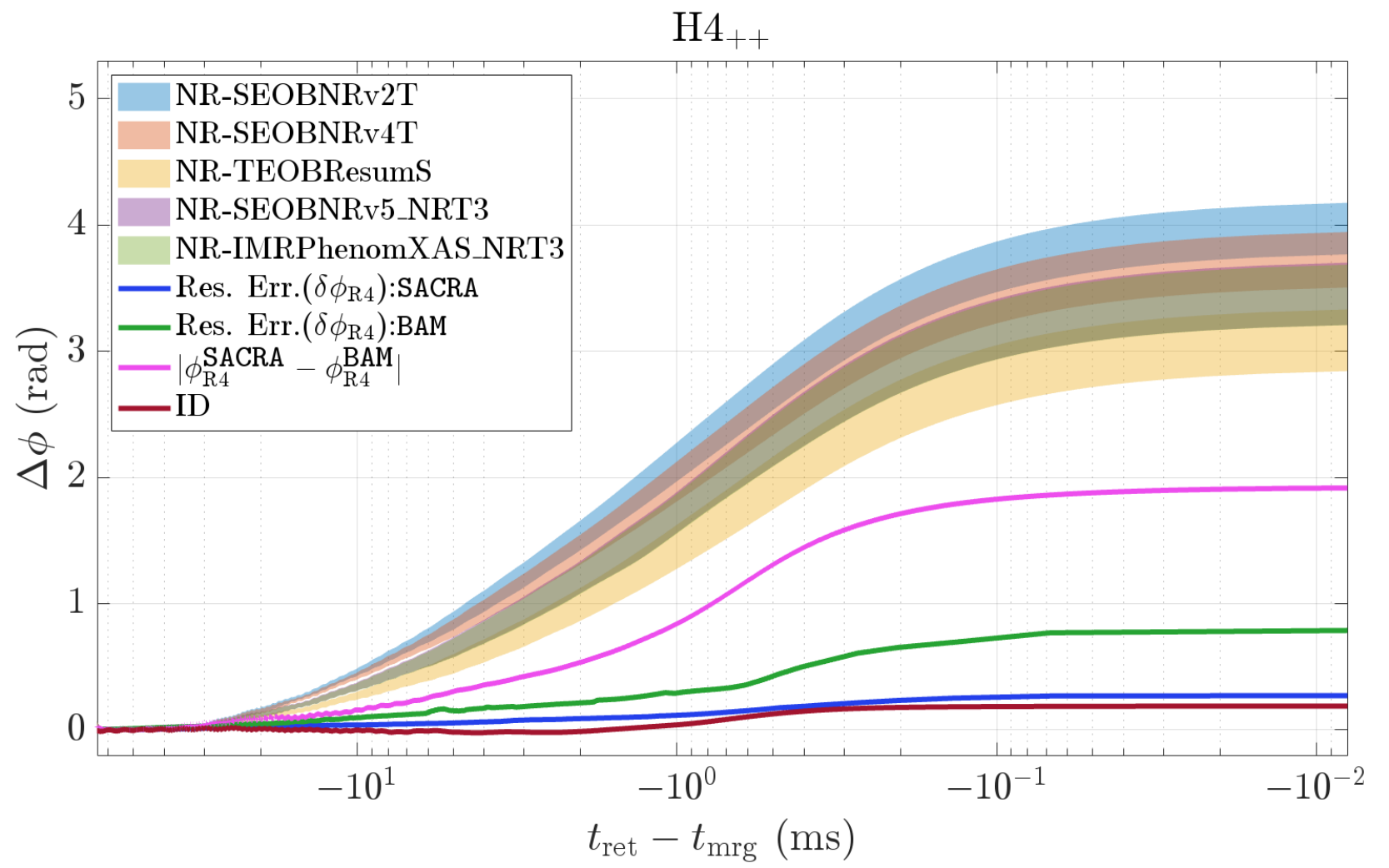}
    \caption{Numerical error budgets studied in this article together with the deviation from analytic waveform models to NR waveforms.
    The deviation from analytic models to NR waveforms with the R4 resolution spreads a finite width that accounts for varying time matching windows and comparisons to the two evolution codes.
    It can be noticed that the results of SEOBNRv5$\_$NRT3 and IMRPhenomXAS$\_$NRT3 almost overlap.
    }
    \label{fig:summary}
\end{figure}

With the waveforms aligned in time and phase and extrpolated to future infinity, we went through a list of comparison between two codes to measure the following error budget:
\begin{itemize}
    \item \emph{Evaluation uncertainties} due to finite grid resolution, which is to be understood as the difference of the best-resolved NR waveform with the hypothetical continuum solution (\cref{res_err});
    \item \emph{Code uncertainties} due to evolution codes (\cref{evo_err}) and different initial data solvers (\cref{id_err});
    \item \emph{Analytical uncertainties} quantified by the discrepancies between NR waveforms and some state-of-the-art analytical waveform models (\cref{secIV}).
\end{itemize}
This classification should provide an indicator of which factors contribute significantly to the numerical errors, while one should keep in mind that there is no clear boundary between different types of uncertainties. 
For example, bias introduced by finite extraction radii inevitably affects the assessment of \emph{code uncertainties}.
The estimated uncertainties are summarized in \cref{fig:summary}.
In brief, the deviation between the two codes is the primary source of numerical errors, which, however, is still clearly below the level of discrepancies between NR and analytic waveforms.
Generally, the uncertainties due to finite grid resolution are a factor of a few less than the code deviation, while the errors associated with different ID solvers contribute the least to the numerical errors.

It should be noted that the total numerical uncertainty cannot be obtained by simply summing up the individual error budgets.
If the uncertainty \cref{eq:poly_R4} accurately reflects the phase shift from the R4 waveforms to the continuum solution, then the difference between the R4 waveforms produced by the two codes (\emph{code uncertainties}) should reflect the difference in their respective distances to the continuum limit (\emph{evaluation uncertainties}). 
In other words, the absolute difference between the two code's R4 waveforms should be approximately equal to the absolute difference between their respective deviations to the continuum:
\begin{align}
    \lvert \phi_{\rm R4}^{\textsc{SACRA}}- \phi_{\rm R4}^{\textsc{BAM}} \rvert\simeq\lvert \delta\phi_{\rm R4}^{\textsc{SACRA}}- \delta\phi_{\rm R4}^{\textsc{BAM}} \rvert\,.
\end{align}
Reading from \cref{fig:summary}, the latter error is smallerthan the former.
This discrepancy suggests that the different error estimation methods are not independent of each other, and naively summing up the curves in \cref{fig:summary} will overestimate the numerical error.

The \emph{analytical uncertainties} were analyzed for some state-of-the-art waveform models, including three that incorporate tidal effects with the theoretical framework and two that introduce the tidal dephasing to a BBH baseline model.
The results of the latter two models (i.e., \textsc{SEOBNRv5}/\textsc{IMRPhenomXAS}\_\textsc{NRT3}) are remarkably close to each other.
Such agreement hints at that the error mainly lies in the shared tidal dephasing model \textsc{NRTv3} of the waveform, and that the difference between the underlying BBH models -- \textsc{SEOBNRv5} and \textsc{IMRPhenomXAS} -- is smaller.
However, small difference between BBH baseline does not necessarily imply that they contribute negligibly to the NR-analytic waveforms discrepancies. 
Mapping the phase difference in the frequency domain (\cref{fig:dphase_ome}), shows a trend of $\sim2.5$ PN order at late time and a higher-order trend in earlier phase.
We propose that the former behavior arises from inconsistencies in the construction of BNS waveforms, which are often built by augmenting a BBH waveform with a tidal contribution.
The undesired horizon absorption effects that are irrelevant for BNS mergers cannot be removed in this approach, thus introducing systematic biases at the corresponding PN order.

In addition, we observed that the dephasing prior to two NSs contact is more significant for stiffer EOS as expected.
After contact, however, the phase differences increase to a commensurate level at late times.
We interpret this as the system transitioning into the essential-one-body regime, where current EOB and phenomenological models lose accuracy.
From the results, we saw that \textsc{TEOBResumS} exhibits the smallest phase shift from the numerical waveforms among the other analytic waveform models (cf.~\cref{fig:vs_eob_sly,fig:vs_eob_h4,fig:dphase_ome}).
We have also performed a phase acceleration analysis (\cref{sec:Qw}).
In contrast to the above, the $Q_w$-analysis suggests that the \textsc{SEOBNRv2/4T} models provide a more accurate approximation to the NR data compared to \textsc{TEOBResumS} (\cref{secIV}).
Therefore, the current dataset is not sufficient to validate one specific model from the others.
That said, the mismatch between NR results and current waveform models unambiguously lies outside the numerical uncertainties in both analyses even if the most conservative estimates (cf.~the lower boundary of the shaded area in \cref{fig:summary}) were assumed.
For developing next-generation waveform models to enhance signal-to-noise ratio in the search pipelines and to reduce systematic biases in parameter estimation in subsequent Bayesian inferences, our simulations of BNS inspirals and mergers in the high-spin parameter space provide an important addition to the cross-code NR waveform database.

Our analysis has presented a benchmark for numerical waveforms of BNS mergers and an assessment of the latest waveform models for the high-aligned spin cases.
However, the situation for the retrograde spin remains to be explored (see \cite{Kuan:2024jnw} for the recent attempt to numerically study BNS with high anti-aligned spins) since the $f$--mode resonance could become enhanced in that case.

%%%%%%%%%%%%%%%%%%%%%%%%%%%%%%%%%%%%%%%%
\section*{Acknowledgements}
We thank Kyohei Kawaguchi for fruitful discussions on the data analysis of gravitational waves, Harald Pfeiffer for instruction on using waveforms from the SXS catalog for this work, Adrian Abac for helping generate the NRTidalv3 models used here, Jan Steinhoff for valuable discussion on parts of the manuscript and for initiating the discussion about horizon absorption effects that lead to Sec.~\ref{Sec:astrayhorizon}. We also thank Marcus Haberland for discussions on the GW modeling.

Numerical computations were performed on the clusters Sakura, Cobra, Raven, and Viper at the Max Planck Computing and Data Facility, the national supercomputer HPE Apollo Hawk at the High-Performance Computing Center Stuttgart (HLRS) under grant number GWanalysis/44189, and the GCS Supercomputer SuperMUC-NG at the Leibniz Supercomputing Centre (LRZ) under project pn29ba. I.\ M.\ gratefully acknowledges the support of Deutsche Forschungsgemeinschaft (DFG) through Project No. 504148597. Furthermore, T.D.\ acknowledges funding from the EU Horizon under ERC Starting Grant, no.\ SMArt-101076369. Views and opinions expressed are however those of the authors only and do not necessarily reflect those of the European Union or the European Research Council. Neither the European Union nor the granting authority can be held responsible for them. W.T.\ was supported by the National Science Foundation under grants PHY-2136036 and PHY-2408903. MS and KK were in part supported by Grant-in-Aid for Scientific Research (grant No.~23H04900 and No.~23K25869) of Japanese MEXT/JSPS, respectively.

\appendix

\section{Setup differences}\label{appendixA}

\begin{table}%[b]
    \centering    
    \caption{Relevant parameters of the numerical setups for each evolution code. %\kk{I prefer to write the dimension of $\kappa_1$ as $M_\odot^{-1}$.}
    }
    \begin{tabular}{l|cc}
        \hline
         Parameter & BAM & SACRA \\
         \hline
         Z4c $\kappa_1$ ($M_\odot^{-1}$) & 0.02 & 0.005 \\
         $\eta_{\rm B}$ ($M_\odot^{-1}$) & 0.3 & 0.15 \\
         Puncture tracker & Minimum lapse & Maximum density \\
         Riemann solver & HO-LLF & HLLE \\
         Reconstruction scheme & WENO-Z & PPM \\
         CFL factor & 0.25 & 0.5 \\
         C2P threshold & $10^{-11}$ & none \\
         C2P iterations & up to 250 & always 5 \\
         Atmosphere level & $10^{-12} \rho_{\mathrm{max}}$ & $10^{-12} \rho_{\mathrm{max}}$ \\ 
         $\Gamma_{\mathrm{th}}$ & 1.75 & 1.67 \\
         GW angle discretization ($N_{\theta}$, $N_{\phi}$) & (47, 46) & (200, 400) \\
         \hline
    \end{tabular}
    \label{tab:setup_parameters}
\end{table}
In addition to the details of the evolution codes adopted (viz.~\textsc{SACRA} and \textsc{BAM}) provided in the main text, \cref{tab:setup_parameters} lists more details about the parameter ($\kappa_1$) set for the Z4c constraint propagation, the parameter ($\eta_{\rm B}$) for the moving puncture gauge, how the puncture point is tracked, the used Riemann solver and reconstruction method, the choice of the Courant–Friedrichs–Lewy (CFL) factor, threshold and iteration for the primitive recovery procedure (C2P), the lower bound on the rest-mass density for an artificial atmosphere, the adiabatic index of the gamma-law approximated heated matter ($\Gamma_{\rm th}$), and the grid used for surface integration to extract GWs.

\begin{figure}
    \centering
    \includegraphics[width=1\columnwidth]{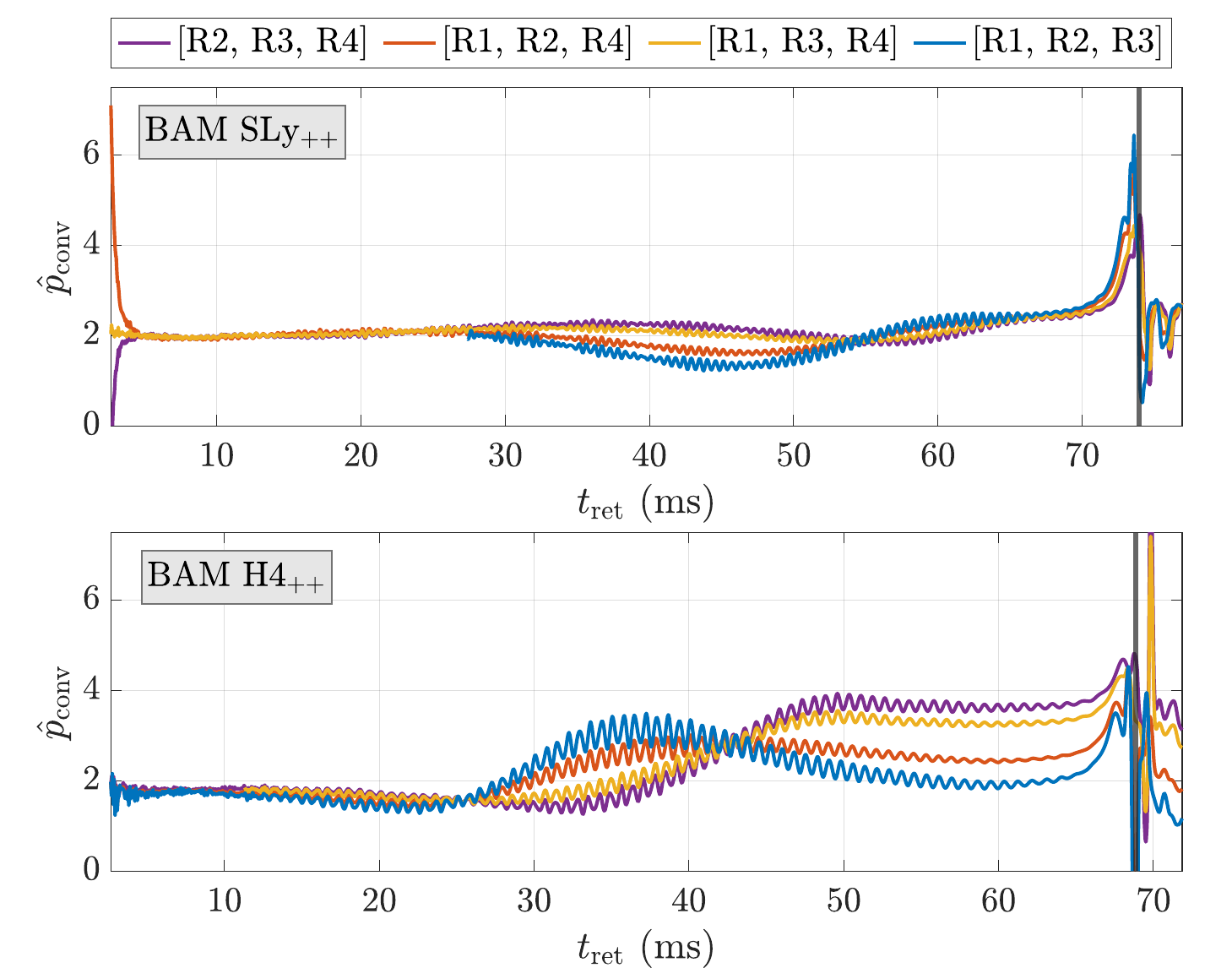}
    \caption{Convergence order estimated by \cref{eq:pconv_0}, where three out of four resolutions are used at once (see legend). Results are shown for the SLy$_{++}$ and H4$_{++}$ models in the top and bottom panels, respectively, using numerical data from BAM runs. Merger times are indicated by vertical lines for each model.
    }
    \label{fig:bam_3res}
\end{figure}

\section{Three-resolution estimate of convergence}\label{appendixB}

Here we provide the convergence power estimated via \cref{eq:pconv_0}, which can be calculated from any three resolutions in the dataset.
The results for BAM waveforms are plotted in \cref{fig:bam_3res}.
Initially, the waveforms exhibit clean 2nd-order convergence, but the convergence order starts to vary between resolutions in both configurations from approximately 30 ms onward.
The origin of this behavior is unknown, despite our efforts to identify any anomalies in the metrics of the running simulations, such as Hamiltonian constraint violation, maximum density, and baryonic mass.
This method is also sensitive to initial relaxation effects and is therefore not well-suited to SACRA’s numerical features, hence, we restrict us to the BAM data in Fig.~\ref{fig:bam_3res}

\section{Error budget of EOB point-particle baselines}\label{appendixC}

\begin{figure}[htp]
    \centering
    \includegraphics[width=1\columnwidth]{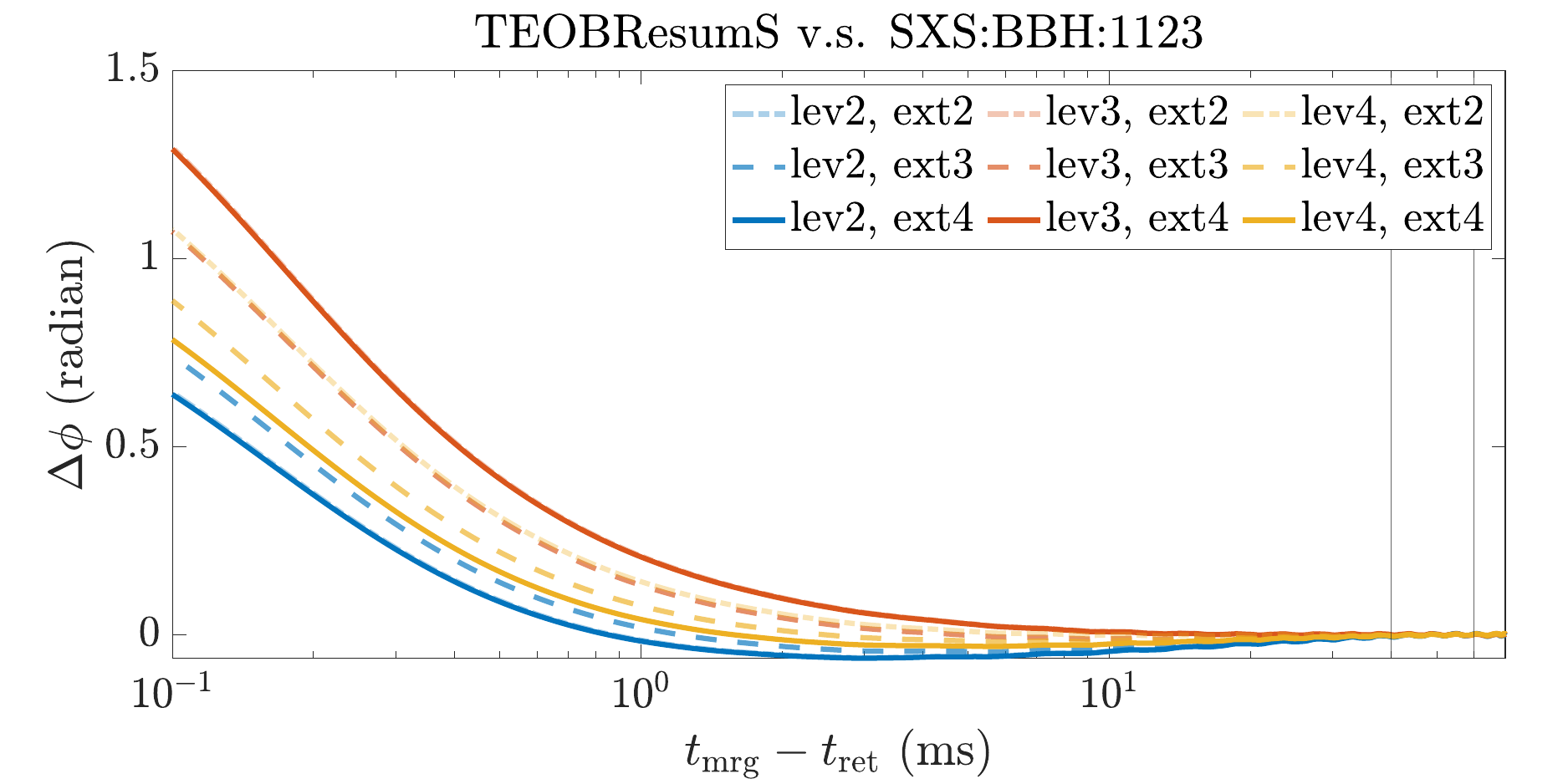}
    \caption{Comparison between the TEOBResumS approximants to the SXS NR waveforms with the openly accessible resolutions (from lower lev2 to higher lev4) and extrapolation orders 2--4 (dumed as ext2--4). Vertical lines indicate the time window where the matching process is carried out, which ranges from 60 to 40 ms prior to the merger. Two pairs of curves overlap and thus difficult to see on the plot: (lev2, ext2) and (lev2, ext4); (lev3, ext2) and (lev3, ext4).
    }
    \label{fig:sxs1123}
\end{figure}

In the main text, we quantified the dephasing between our NR waveforms and the selected waveform models.
However, the dephasing consists of two sources: discrepancies in modeling the inspiral of skeletonized objects and inaccuracy in describing tidal (finite size) effects.
In this Appendix, we focus on evaluating how well the considered EOB models agree with numerical BBH waveforms. These latter waveforms are expected to approximate signals emitted by skeletonized BNS systems, effectively neglecting finite size effects in the sense of effective field theory. 
For this purpose, we utilize waveforms from the SXS collaboration's open catalog~\cite{Mroue:2013xna,Boyle:2019kee} with a setup identical to the BNS systems studied in the main text. 
Specifically, we consider equal-mass binary with both components having a dimensionless spin of $\chi_1=\chi_2=0.5$. 
For this case, simulations with three resolutions are available and are labeled as lev2--4 in \cref{fig:sxs1123}.
The numerical waveforms are then extrapolated to infinity based on an assumption of the polynomial behavior \cref{eq:polyfit} at large distances.
In the catalog, extrapolated waveforms of orders 2 to 4, denoted as ext2 to ext4 in the plot, are provided.
We analyze the dephasing between the numerical and EOB waveforms after alignment through the minimization of \cref{eq:I_phase}. 
Particular attention is given to how the dephasing at the moment of merger depends on variations in the matching windows, resolutions, and extraction orders.

Taking \textsc{TEOBResumS} as an example while noting that the results are qualitatively the same for \textsc{SEOBNRv2/4T}, we plot in \cref{fig:sxs1123} the dephasing, $\Delta\phi=\phi_{\rm EOB}-\phi_{\rm NR}$, for the obtainable resolutions and extrapolation orders; notably, there are lev2--4 for the lower to higher resolutions, and ext2--4 for 2nd to 4th order extrapolation to future null infinity.
The deviation between different resolutions is comparable to $\Delta\phi_{\rm mrg}$, while the deviation due to the extraction order is a factor of a few less than $\Delta\phi_{\rm mrg}$.
The dephasing is overall much less than the deviation seen in the BNS cases shown in the main text.

%%%%%%%%%%%%%%%%%%%%%%%%%%%%%%%%%%%%%%%%
\bibliography{references}

\begin{thebibliography}{196}%
\makeatletter
\providecommand \@ifxundefined [1]{%
 \@ifx{#1\undefined}
}%
\providecommand \@ifnum [1]{%
 \ifnum #1\expandafter \@firstoftwo
 \else \expandafter \@secondoftwo
 \fi
}%
\providecommand \@ifx [1]{%
 \ifx #1\expandafter \@firstoftwo
 \else \expandafter \@secondoftwo
 \fi
}%
\providecommand \natexlab [1]{#1}%
\providecommand \enquote  [1]{``#1''}%
\providecommand \bibnamefont  [1]{#1}%
\providecommand \bibfnamefont [1]{#1}%
\providecommand \citenamefont [1]{#1}%
\providecommand \href@noop [0]{\@secondoftwo}%
\providecommand \href [0]{\begingroup \@sanitize@url \@href}%
\providecommand \@href[1]{\@@startlink{#1}\@@href}%
\providecommand \@@href[1]{\endgroup#1\@@endlink}%
\providecommand \@sanitize@url [0]{\catcode `\\12\catcode `\$12\catcode `\&12\catcode `\#12\catcode `\^12\catcode `\_12\catcode `\%12\relax}%
\providecommand \@@startlink[1]{}%
\providecommand \@@endlink[0]{}%
\providecommand \url  [0]{\begingroup\@sanitize@url \@url }%
\providecommand \@url [1]{\endgroup\@href {#1}{\urlprefix }}%
\providecommand \urlprefix  [0]{URL }%
\providecommand \Eprint [0]{\href }%
\providecommand \doibase [0]{https://doi.org/}%
\providecommand \selectlanguage [0]{\@gobble}%
\providecommand \bibinfo  [0]{\@secondoftwo}%
\providecommand \bibfield  [0]{\@secondoftwo}%
\providecommand \translation [1]{[#1]}%
\providecommand \BibitemOpen [0]{}%
\providecommand \bibitemStop [0]{}%
\providecommand \bibitemNoStop [0]{.\EOS\space}%
\providecommand \EOS [0]{\spacefactor3000\relax}%
\providecommand \BibitemShut  [1]{\csname bibitem#1\endcsname}%
\let\auto@bib@innerbib\@empty
%</preamble>
\bibitem [{\citenamefont {Dietrich}\ \emph {et~al.}(2021)\citenamefont {Dietrich}, \citenamefont {Hinderer},\ and\ \citenamefont {Samajdar}}]{Dietrich:2020eud}%
  \BibitemOpen
  \bibfield  {author} {\bibinfo {author} {\bibfnamefont {T.}~\bibnamefont {Dietrich}}, \bibinfo {author} {\bibfnamefont {T.}~\bibnamefont {Hinderer}},\ and\ \bibinfo {author} {\bibfnamefont {A.}~\bibnamefont {Samajdar}},\ }\bibfield  {title} {\bibinfo {title} {{Interpreting Binary Neutron Star Mergers: Describing the Binary Neutron Star Dynamics, Modelling Gravitational Waveforms, and Analyzing Detections}},\ }\href {https://doi.org/10.1007/s10714-020-02751-6} {\bibfield  {journal} {\bibinfo  {journal} {Gen. Rel. Grav.}\ }\textbf {\bibinfo {volume} {53}},\ \bibinfo {pages} {27} (\bibinfo {year} {2021})},\ \Eprint {https://arxiv.org/abs/2004.02527} {arXiv:2004.02527 [gr-qc]} \BibitemShut {NoStop}%
\bibitem [{\citenamefont {Chatziioannou}(2020)}]{Chatziioannou:2020pqz}%
  \BibitemOpen
  \bibfield  {author} {\bibinfo {author} {\bibfnamefont {K.}~\bibnamefont {Chatziioannou}},\ }\bibfield  {title} {\bibinfo {title} {{Neutron star tidal deformability and equation of state constraints}},\ }\href {https://doi.org/10.1007/s10714-020-02754-3} {\bibfield  {journal} {\bibinfo  {journal} {Gen. Rel. Grav.}\ }\textbf {\bibinfo {volume} {52}},\ \bibinfo {pages} {109} (\bibinfo {year} {2020})},\ \Eprint {https://arxiv.org/abs/2006.03168} {arXiv:2006.03168 [gr-qc]} \BibitemShut {NoStop}%
\bibitem [{\citenamefont {Radice}\ \emph {et~al.}(2020)\citenamefont {Radice}, \citenamefont {Bernuzzi},\ and\ \citenamefont {Perego}}]{Radice:2020ddv}%
  \BibitemOpen
  \bibfield  {author} {\bibinfo {author} {\bibfnamefont {D.}~\bibnamefont {Radice}}, \bibinfo {author} {\bibfnamefont {S.}~\bibnamefont {Bernuzzi}},\ and\ \bibinfo {author} {\bibfnamefont {A.}~\bibnamefont {Perego}},\ }\bibfield  {title} {\bibinfo {title} {{The Dynamics of Binary Neutron Star Mergers and GW170817}},\ }\href {https://doi.org/10.1146/annurev-nucl-013120-114541} {\bibfield  {journal} {\bibinfo  {journal} {Ann. Rev. Nucl. Part. Sci.}\ }\textbf {\bibinfo {volume} {70}},\ \bibinfo {pages} {95} (\bibinfo {year} {2020})},\ \Eprint {https://arxiv.org/abs/2002.03863} {arXiv:2002.03863 [astro-ph.HE]} \BibitemShut {NoStop}%
\bibitem [{\citenamefont {Suvorov}\ \emph {et~al.}(2024)\citenamefont {Suvorov}, \citenamefont {Kuan},\ and\ \citenamefont {Kokkotas}}]{Suvorov:2024cff}%
  \BibitemOpen
  \bibfield  {author} {\bibinfo {author} {\bibfnamefont {A.~G.}\ \bibnamefont {Suvorov}}, \bibinfo {author} {\bibfnamefont {H.-J.}\ \bibnamefont {Kuan}},\ and\ \bibinfo {author} {\bibfnamefont {K.~D.}\ \bibnamefont {Kokkotas}},\ }\bibfield  {title} {\bibinfo {title} {{Premerger phenomena in neutron-star binary coalescences}},\ }\href@noop {} {\bibfield  {journal} {\bibinfo  {journal} {arXiv:2408.16283}\ } (\bibinfo {year} {2024})},\ \Eprint {https://arxiv.org/abs/2408.16283} {arXiv:2408.16283 [astro-ph.HE]} \BibitemShut {NoStop}%
\bibitem [{\citenamefont {Figura}\ \emph {et~al.}(2020)\citenamefont {Figura}, \citenamefont {Lu}, \citenamefont {Burgio}, \citenamefont {Li},\ and\ \citenamefont {Schulze}}]{Figura:2020fkj}%
  \BibitemOpen
  \bibfield  {author} {\bibinfo {author} {\bibfnamefont {A.}~\bibnamefont {Figura}}, \bibinfo {author} {\bibfnamefont {J.~J.}\ \bibnamefont {Lu}}, \bibinfo {author} {\bibfnamefont {G.~F.}\ \bibnamefont {Burgio}}, \bibinfo {author} {\bibfnamefont {Z.~H.}\ \bibnamefont {Li}},\ and\ \bibinfo {author} {\bibfnamefont {H.~J.}\ \bibnamefont {Schulze}},\ }\bibfield  {title} {\bibinfo {title} {{Hybrid equation of state approach in binary neutron-star merger simulations}},\ }\href {https://doi.org/10.1103/PhysRevD.102.043006} {\bibfield  {journal} {\bibinfo  {journal} {Phys. Rev. D}\ }\textbf {\bibinfo {volume} {102}},\ \bibinfo {pages} {043006} (\bibinfo {year} {2020})},\ \Eprint {https://arxiv.org/abs/2005.08691} {arXiv:2005.08691 [gr-qc]} \BibitemShut {NoStop}%
\bibitem [{\citenamefont {Raithel}\ and\ \citenamefont {Paschalidis}(2024)}]{Raithel:2023gct}%
  \BibitemOpen
  \bibfield  {author} {\bibinfo {author} {\bibfnamefont {C.~A.}\ \bibnamefont {Raithel}}\ and\ \bibinfo {author} {\bibfnamefont {V.}~\bibnamefont {Paschalidis}},\ }\bibfield  {title} {\bibinfo {title} {{Detectability of finite-temperature effects from neutron star mergers with next-generation gravitational wave detectors}},\ }\href {https://doi.org/10.1103/PhysRevD.110.043002} {\bibfield  {journal} {\bibinfo  {journal} {Phys. Rev. D}\ }\textbf {\bibinfo {volume} {110}},\ \bibinfo {pages} {043002} (\bibinfo {year} {2024})},\ \Eprint {https://arxiv.org/abs/2312.14046} {arXiv:2312.14046 [astro-ph.HE]} \BibitemShut {NoStop}%
\bibitem [{\citenamefont {Fields}\ \emph {et~al.}(2023)\citenamefont {Fields}, \citenamefont {Prakash}, \citenamefont {Breschi}, \citenamefont {Radice}, \citenamefont {Bernuzzi},\ and\ \citenamefont {da~Silva~Schneider}}]{Fields:2023bhs}%
  \BibitemOpen
  \bibfield  {author} {\bibinfo {author} {\bibfnamefont {J.}~\bibnamefont {Fields}}, \bibinfo {author} {\bibfnamefont {A.}~\bibnamefont {Prakash}}, \bibinfo {author} {\bibfnamefont {M.}~\bibnamefont {Breschi}}, \bibinfo {author} {\bibfnamefont {D.}~\bibnamefont {Radice}}, \bibinfo {author} {\bibfnamefont {S.}~\bibnamefont {Bernuzzi}},\ and\ \bibinfo {author} {\bibfnamefont {A.}~\bibnamefont {da~Silva~Schneider}},\ }\bibfield  {title} {\bibinfo {title} {{Thermal Effects in Binary Neutron Star Mergers}},\ }\href {https://doi.org/10.3847/2041-8213/ace5b2} {\bibfield  {journal} {\bibinfo  {journal} {Astrophys. J. Lett.}\ }\textbf {\bibinfo {volume} {952}},\ \bibinfo {pages} {L36} (\bibinfo {year} {2023})},\ \Eprint {https://arxiv.org/abs/2302.11359} {arXiv:2302.11359 [astro-ph.HE]} \BibitemShut {NoStop}%
\bibitem [{\citenamefont {Miravet-Ten\'es}\ \emph {et~al.}(2024)\citenamefont {Miravet-Ten\'es}, \citenamefont {Guerra}, \citenamefont {Ruiz}, \citenamefont {Cerd\'a-Dur\'an},\ and\ \citenamefont {Font}}]{Miravet-Tenes:2024vba}%
  \BibitemOpen
  \bibfield  {author} {\bibinfo {author} {\bibfnamefont {M.}~\bibnamefont {Miravet-Ten\'es}}, \bibinfo {author} {\bibfnamefont {D.}~\bibnamefont {Guerra}}, \bibinfo {author} {\bibfnamefont {M.}~\bibnamefont {Ruiz}}, \bibinfo {author} {\bibfnamefont {P.}~\bibnamefont {Cerd\'a-Dur\'an}},\ and\ \bibinfo {author} {\bibfnamefont {J.~A.}\ \bibnamefont {Font}},\ }\bibfield  {title} {\bibinfo {title} {{Identifying thermal effects in neutron star merger remnants with model-agnostic waveform reconstructions and third-generation detectors}},\ }\href@noop {} {\bibfield  {journal} {\bibinfo  {journal} {arXiv:2401.02493}\ } (\bibinfo {year} {2024})},\ \Eprint {https://arxiv.org/abs/2401.02493} {arXiv:2401.02493 [gr-qc]} \BibitemShut {NoStop}%
\bibitem [{\citenamefont {Aasi}\ \emph {et~al.}(2015)\citenamefont {Aasi} \emph {et~al.}}]{LIGOScientific:2014pky}%
  \BibitemOpen
  \bibfield  {author} {\bibinfo {author} {\bibfnamefont {J.}~\bibnamefont {Aasi}} \emph {et~al.} (\bibinfo {collaboration} {LIGO Scientific}),\ }\bibfield  {title} {\bibinfo {title} {{Advanced LIGO}},\ }\href {https://doi.org/10.1088/0264-9381/32/7/074001} {\bibfield  {journal} {\bibinfo  {journal} {Class. Quant. Grav.}\ }\textbf {\bibinfo {volume} {32}},\ \bibinfo {pages} {074001} (\bibinfo {year} {2015})},\ \Eprint {https://arxiv.org/abs/1411.4547} {arXiv:1411.4547 [gr-qc]} \BibitemShut {NoStop}%
\bibitem [{\citenamefont {Acernese}\ \emph {et~al.}(2015)\citenamefont {Acernese} \emph {et~al.}}]{VIRGO:2014yos}%
  \BibitemOpen
  \bibfield  {author} {\bibinfo {author} {\bibfnamefont {F.}~\bibnamefont {Acernese}} \emph {et~al.} (\bibinfo {collaboration} {VIRGO}),\ }\bibfield  {title} {\bibinfo {title} {{Advanced Virgo: a second-generation interferometric gravitational wave detector}},\ }\href {https://doi.org/10.1088/0264-9381/32/2/024001} {\bibfield  {journal} {\bibinfo  {journal} {Class. Quant. Grav.}\ }\textbf {\bibinfo {volume} {32}},\ \bibinfo {pages} {024001} (\bibinfo {year} {2015})},\ \Eprint {https://arxiv.org/abs/1408.3978} {arXiv:1408.3978 [gr-qc]} \BibitemShut {NoStop}%
\bibitem [{\citenamefont {Hild}\ \emph {et~al.}(2011)\citenamefont {Hild} \emph {et~al.}}]{Hild:2010id}%
  \BibitemOpen
  \bibfield  {author} {\bibinfo {author} {\bibfnamefont {S.}~\bibnamefont {Hild}} \emph {et~al.},\ }\bibfield  {title} {\bibinfo {title} {{Sensitivity Studies for Third-Generation Gravitational Wave Observatories}},\ }\href {https://doi.org/10.1088/0264-9381/28/9/094013} {\bibfield  {journal} {\bibinfo  {journal} {Class. Quant. Grav.}\ }\textbf {\bibinfo {volume} {28}},\ \bibinfo {pages} {094013} (\bibinfo {year} {2011})}\BibitemShut {NoStop}%
\bibitem [{\citenamefont {Punturo}\ \emph {et~al.}(2010)\citenamefont {Punturo} \emph {et~al.}}]{Punturo:2010zz}%
  \BibitemOpen
  \bibfield  {author} {\bibinfo {author} {\bibfnamefont {M.}~\bibnamefont {Punturo}} \emph {et~al.},\ }\bibfield  {title} {\bibinfo {title} {{The Einstein Telescope: A third-generation gravitational wave observatory}},\ }\href {https://doi.org/10.1088/0264-9381/27/19/194002} {\bibfield  {journal} {\bibinfo  {journal} {Class. Quant. Grav.}\ }\textbf {\bibinfo {volume} {27}},\ \bibinfo {pages} {194002} (\bibinfo {year} {2010})}\BibitemShut {NoStop}%
\bibitem [{\citenamefont {Branchesi}\ \emph {et~al.}(2023)\citenamefont {Branchesi} \emph {et~al.}}]{Branchesi:2023mws}%
  \BibitemOpen
  \bibfield  {author} {\bibinfo {author} {\bibfnamefont {M.}~\bibnamefont {Branchesi}} \emph {et~al.},\ }\bibfield  {title} {\bibinfo {title} {{Science with the Einstein Telescope: a comparison of different designs}},\ }\href {https://doi.org/10.1088/1475-7516/2023/07/068} {\bibfield  {journal} {\bibinfo  {journal} {JCAP}\ }\textbf {\bibinfo {volume} {07}},\ \bibinfo {pages} {068}},\ \Eprint {https://arxiv.org/abs/2303.15923} {arXiv:2303.15923 [gr-qc]} \BibitemShut {NoStop}%
\bibitem [{\citenamefont {Abac}\ \emph {et~al.}(2025)\citenamefont {Abac} \emph {et~al.}}]{Abac:2025saz}%
  \BibitemOpen
  \bibfield  {author} {\bibinfo {author} {\bibfnamefont {A.}~\bibnamefont {Abac}} \emph {et~al.},\ }\bibfield  {title} {\bibinfo {title} {{The Science of the Einstein Telescope}},\ }\href@noop {} {\bibfield  {journal} {\bibinfo  {journal} {arXiv:2503.12263}\ } (\bibinfo {year} {2025})},\ \Eprint {https://arxiv.org/abs/2503.12263} {arXiv:2503.12263 [gr-qc]} \BibitemShut {NoStop}%
\bibitem [{\citenamefont {Abbott}\ \emph {et~al.}(2017{\natexlab{a}})\citenamefont {Abbott} \emph {et~al.}}]{LIGOScientific:2016wof}%
  \BibitemOpen
  \bibfield  {author} {\bibinfo {author} {\bibfnamefont {B.~P.}\ \bibnamefont {Abbott}} \emph {et~al.} (\bibinfo {collaboration} {LIGO Scientific}),\ }\bibfield  {title} {\bibinfo {title} {{Exploring the Sensitivity of Next Generation Gravitational Wave Detectors}},\ }\href {https://doi.org/10.1088/1361-6382/aa51f4} {\bibfield  {journal} {\bibinfo  {journal} {Class. Quant. Grav.}\ }\textbf {\bibinfo {volume} {34}},\ \bibinfo {pages} {044001} (\bibinfo {year} {2017}{\natexlab{a}})},\ \Eprint {https://arxiv.org/abs/1607.08697} {arXiv:1607.08697 [astro-ph.IM]} \BibitemShut {NoStop}%
\bibitem [{\citenamefont {Reitze}\ \emph {et~al.}(2019)\citenamefont {Reitze} \emph {et~al.}}]{Reitze:2019iox}%
  \BibitemOpen
  \bibfield  {author} {\bibinfo {author} {\bibfnamefont {D.}~\bibnamefont {Reitze}} \emph {et~al.},\ }\bibfield  {title} {\bibinfo {title} {{Cosmic Explorer: The U.S. Contribution to Gravitational-Wave Astronomy beyond LIGO}},\ }\href@noop {} {\bibfield  {journal} {\bibinfo  {journal} {Bull. Am. Astron. Soc.}\ }\textbf {\bibinfo {volume} {51}},\ \bibinfo {pages} {035} (\bibinfo {year} {2019})},\ \Eprint {https://arxiv.org/abs/1907.04833} {arXiv:1907.04833 [astro-ph.IM]} \BibitemShut {NoStop}%
\bibitem [{\citenamefont {Evans}\ \emph {et~al.}(2021)\citenamefont {Evans} \emph {et~al.}}]{Evans:2021gyd}%
  \BibitemOpen
  \bibfield  {author} {\bibinfo {author} {\bibfnamefont {M.}~\bibnamefont {Evans}} \emph {et~al.},\ }\bibfield  {title} {\bibinfo {title} {{A Horizon Study for Cosmic Explorer: Science, Observatories, and Community}},\ }\href@noop {} {\bibfield  {journal} {\bibinfo  {journal} {{arXiv:2109.09882}}\ } (\bibinfo {year} {2021})},\ \Eprint {https://arxiv.org/abs/2109.09882} {arXiv:2109.09882 [astro-ph.IM]} \BibitemShut {NoStop}%
\bibitem [{\citenamefont {Flanagan}\ and\ \citenamefont {Hinderer}(2008)}]{Flanagan:2007ix}%
  \BibitemOpen
  \bibfield  {author} {\bibinfo {author} {\bibfnamefont {E.~E.}\ \bibnamefont {Flanagan}}\ and\ \bibinfo {author} {\bibfnamefont {T.}~\bibnamefont {Hinderer}},\ }\bibfield  {title} {\bibinfo {title} {{Constraining neutron star tidal Love numbers with gravitational wave detectors}},\ }\href {https://doi.org/10.1103/PhysRevD.77.021502} {\bibfield  {journal} {\bibinfo  {journal} {Phys. Rev. D}\ }\textbf {\bibinfo {volume} {77}},\ \bibinfo {pages} {021502} (\bibinfo {year} {2008})},\ \Eprint {https://arxiv.org/abs/0709.1915} {arXiv:0709.1915 [astro-ph]} \BibitemShut {NoStop}%
\bibitem [{\citenamefont {Hinderer}(2008)}]{Hinderer:2007mb}%
  \BibitemOpen
  \bibfield  {author} {\bibinfo {author} {\bibfnamefont {T.}~\bibnamefont {Hinderer}},\ }\bibfield  {title} {\bibinfo {title} {{Tidal Love numbers of neutron stars}},\ }\href {https://doi.org/10.1086/533487} {\bibfield  {journal} {\bibinfo  {journal} {Astrophys. J.}\ }\textbf {\bibinfo {volume} {677}},\ \bibinfo {pages} {1216} (\bibinfo {year} {2008})},\ \bibinfo {note} {[Erratum: Astrophys.J. 697, 964 (2009)]},\ \Eprint {https://arxiv.org/abs/0711.2420} {arXiv:0711.2420 [astro-ph]} \BibitemShut {NoStop}%
\bibitem [{\citenamefont {Hinderer}\ \emph {et~al.}(2010)\citenamefont {Hinderer}, \citenamefont {Lackey}, \citenamefont {Lang},\ and\ \citenamefont {Read}}]{Hinderer:2009ca}%
  \BibitemOpen
  \bibfield  {author} {\bibinfo {author} {\bibfnamefont {T.}~\bibnamefont {Hinderer}}, \bibinfo {author} {\bibfnamefont {B.~D.}\ \bibnamefont {Lackey}}, \bibinfo {author} {\bibfnamefont {R.~N.}\ \bibnamefont {Lang}},\ and\ \bibinfo {author} {\bibfnamefont {J.~S.}\ \bibnamefont {Read}},\ }\bibfield  {title} {\bibinfo {title} {{Tidal deformability of neutron stars with realistic equations of state and their gravitational wave signatures in binary inspiral}},\ }\href {https://doi.org/10.1103/PhysRevD.81.123016} {\bibfield  {journal} {\bibinfo  {journal} {Phys. Rev. D}\ }\textbf {\bibinfo {volume} {81}},\ \bibinfo {pages} {123016} (\bibinfo {year} {2010})},\ \Eprint {https://arxiv.org/abs/0911.3535} {arXiv:0911.3535 [astro-ph.HE]} \BibitemShut {NoStop}%
\bibitem [{\citenamefont {Damour}\ \emph {et~al.}(2012)\citenamefont {Damour}, \citenamefont {Nagar},\ and\ \citenamefont {Villain}}]{Damour:2012yf}%
  \BibitemOpen
  \bibfield  {author} {\bibinfo {author} {\bibfnamefont {T.}~\bibnamefont {Damour}}, \bibinfo {author} {\bibfnamefont {A.}~\bibnamefont {Nagar}},\ and\ \bibinfo {author} {\bibfnamefont {L.}~\bibnamefont {Villain}},\ }\bibfield  {title} {\bibinfo {title} {{Measurability of the tidal polarizability of neutron stars in late-inspiral gravitational-wave signals}},\ }\href {https://doi.org/10.1103/PhysRevD.85.123007} {\bibfield  {journal} {\bibinfo  {journal} {Phys. Rev. D}\ }\textbf {\bibinfo {volume} {85}},\ \bibinfo {pages} {123007} (\bibinfo {year} {2012})},\ \Eprint {https://arxiv.org/abs/1203.4352} {arXiv:1203.4352 [gr-qc]} \BibitemShut {NoStop}%
\bibitem [{\citenamefont {Read}\ \emph {et~al.}(2013)\citenamefont {Read}, \citenamefont {Baiotti}, \citenamefont {Creighton}, \citenamefont {Friedman}, \citenamefont {Giacomazzo}, \citenamefont {Kyutoku}, \citenamefont {Markakis}, \citenamefont {Rezzolla}, \citenamefont {Shibata},\ and\ \citenamefont {Taniguchi}}]{Read:2013zra}%
  \BibitemOpen
  \bibfield  {author} {\bibinfo {author} {\bibfnamefont {J.~S.}\ \bibnamefont {Read}}, \bibinfo {author} {\bibfnamefont {L.}~\bibnamefont {Baiotti}}, \bibinfo {author} {\bibfnamefont {J.~D.~E.}\ \bibnamefont {Creighton}}, \bibinfo {author} {\bibfnamefont {J.~L.}\ \bibnamefont {Friedman}}, \bibinfo {author} {\bibfnamefont {B.}~\bibnamefont {Giacomazzo}}, \bibinfo {author} {\bibfnamefont {K.}~\bibnamefont {Kyutoku}}, \bibinfo {author} {\bibfnamefont {C.}~\bibnamefont {Markakis}}, \bibinfo {author} {\bibfnamefont {L.}~\bibnamefont {Rezzolla}}, \bibinfo {author} {\bibfnamefont {M.}~\bibnamefont {Shibata}},\ and\ \bibinfo {author} {\bibfnamefont {K.}~\bibnamefont {Taniguchi}},\ }\bibfield  {title} {\bibinfo {title} {{Matter effects on binary neutron star waveforms}},\ }\href {https://doi.org/10.1103/PhysRevD.88.044042} {\bibfield  {journal} {\bibinfo  {journal} {Phys. Rev. D}\ }\textbf {\bibinfo {volume} {88}},\ \bibinfo {pages} {044042} (\bibinfo {year} {2013})},\ \Eprint {https://arxiv.org/abs/1306.4065}
  {arXiv:1306.4065 [gr-qc]} \BibitemShut {NoStop}%
\bibitem [{\citenamefont {Abbott}\ \emph {et~al.}(2017{\natexlab{b}})\citenamefont {Abbott} \emph {et~al.}}]{LIGOScientific:2017vwq}%
  \BibitemOpen
  \bibfield  {author} {\bibinfo {author} {\bibfnamefont {B.~P.}\ \bibnamefont {Abbott}} \emph {et~al.} (\bibinfo {collaboration} {LIGO Scientific, Virgo}),\ }\bibfield  {title} {\bibinfo {title} {{GW170817: Observation of Gravitational Waves from a Binary Neutron Star Inspiral}},\ }\href {https://doi.org/10.1103/PhysRevLett.119.161101} {\bibfield  {journal} {\bibinfo  {journal} {Phys. Rev. Lett.}\ }\textbf {\bibinfo {volume} {119}},\ \bibinfo {pages} {161101} (\bibinfo {year} {2017}{\natexlab{b}})},\ \Eprint {https://arxiv.org/abs/1710.05832} {arXiv:1710.05832 [gr-qc]} \BibitemShut {NoStop}%
\bibitem [{\citenamefont {Abbott}\ \emph {et~al.}(2018)\citenamefont {Abbott} \emph {et~al.}}]{LIGOScientific:2018cki}%
  \BibitemOpen
  \bibfield  {author} {\bibinfo {author} {\bibfnamefont {B.~P.}\ \bibnamefont {Abbott}} \emph {et~al.} (\bibinfo {collaboration} {LIGO Scientific, Virgo}),\ }\bibfield  {title} {\bibinfo {title} {{GW170817: Measurements of neutron star radii and equation of state}},\ }\href {https://doi.org/10.1103/PhysRevLett.121.161101} {\bibfield  {journal} {\bibinfo  {journal} {Phys. Rev. Lett.}\ }\textbf {\bibinfo {volume} {121}},\ \bibinfo {pages} {161101} (\bibinfo {year} {2018})},\ \Eprint {https://arxiv.org/abs/1805.11581} {arXiv:1805.11581 [gr-qc]} \BibitemShut {NoStop}%
\bibitem [{\citenamefont {Abbott}\ \emph {et~al.}(2019)\citenamefont {Abbott} \emph {et~al.}}]{LIGOScientific:2018hze}%
  \BibitemOpen
  \bibfield  {author} {\bibinfo {author} {\bibfnamefont {B.~P.}\ \bibnamefont {Abbott}} \emph {et~al.} (\bibinfo {collaboration} {LIGO Scientific, Virgo}),\ }\bibfield  {title} {\bibinfo {title} {{Properties of the binary neutron star merger GW170817}},\ }\href {https://doi.org/10.1103/PhysRevX.9.011001} {\bibfield  {journal} {\bibinfo  {journal} {Phys. Rev. X}\ }\textbf {\bibinfo {volume} {9}},\ \bibinfo {pages} {011001} (\bibinfo {year} {2019})},\ \Eprint {https://arxiv.org/abs/1805.11579} {arXiv:1805.11579 [gr-qc]} \BibitemShut {NoStop}%
\bibitem [{\citenamefont {Annala}\ \emph {et~al.}(2018)\citenamefont {Annala}, \citenamefont {Gorda}, \citenamefont {Kurkela},\ and\ \citenamefont {Vuorinen}}]{Annala:2017llu}%
  \BibitemOpen
  \bibfield  {author} {\bibinfo {author} {\bibfnamefont {E.}~\bibnamefont {Annala}}, \bibinfo {author} {\bibfnamefont {T.}~\bibnamefont {Gorda}}, \bibinfo {author} {\bibfnamefont {A.}~\bibnamefont {Kurkela}},\ and\ \bibinfo {author} {\bibfnamefont {A.}~\bibnamefont {Vuorinen}},\ }\bibfield  {title} {\bibinfo {title} {{Gravitational-wave constraints on the neutron-star-matter Equation of State}},\ }\href {https://doi.org/10.1103/PhysRevLett.120.172703} {\bibfield  {journal} {\bibinfo  {journal} {Phys. Rev. Lett.}\ }\textbf {\bibinfo {volume} {120}},\ \bibinfo {pages} {172703} (\bibinfo {year} {2018})},\ \Eprint {https://arxiv.org/abs/1711.02644} {arXiv:1711.02644 [astro-ph.HE]} \BibitemShut {NoStop}%
\bibitem [{\citenamefont {Dietrich}\ \emph {et~al.}(2020)\citenamefont {Dietrich}, \citenamefont {Coughlin}, \citenamefont {Pang}, \citenamefont {Bulla}, \citenamefont {Heinzel}, \citenamefont {Issa}, \citenamefont {Tews},\ and\ \citenamefont {Antier}}]{Dietrich:2020efo}%
  \BibitemOpen
  \bibfield  {author} {\bibinfo {author} {\bibfnamefont {T.}~\bibnamefont {Dietrich}}, \bibinfo {author} {\bibfnamefont {M.~W.}\ \bibnamefont {Coughlin}}, \bibinfo {author} {\bibfnamefont {P.~T.~H.}\ \bibnamefont {Pang}}, \bibinfo {author} {\bibfnamefont {M.}~\bibnamefont {Bulla}}, \bibinfo {author} {\bibfnamefont {J.}~\bibnamefont {Heinzel}}, \bibinfo {author} {\bibfnamefont {L.}~\bibnamefont {Issa}}, \bibinfo {author} {\bibfnamefont {I.}~\bibnamefont {Tews}},\ and\ \bibinfo {author} {\bibfnamefont {S.}~\bibnamefont {Antier}},\ }\bibfield  {title} {\bibinfo {title} {{Multimessenger constraints on the neutron-star equation of state and the Hubble constant}},\ }\href {https://doi.org/10.1126/science.abb4317} {\bibfield  {journal} {\bibinfo  {journal} {Science}\ }\textbf {\bibinfo {volume} {370}},\ \bibinfo {pages} {1450} (\bibinfo {year} {2020})},\ \Eprint {https://arxiv.org/abs/2002.11355} {arXiv:2002.11355 [astro-ph.HE]} \BibitemShut {NoStop}%
\bibitem [{\citenamefont {Lattimer}(2021)}]{Lattimer:2021emm}%
  \BibitemOpen
  \bibfield  {author} {\bibinfo {author} {\bibfnamefont {J.~M.}\ \bibnamefont {Lattimer}},\ }\bibfield  {title} {\bibinfo {title} {{Neutron Stars and the Nuclear Matter Equation of State}},\ }\href {https://doi.org/10.1146/annurev-nucl-102419-124827} {\bibfield  {journal} {\bibinfo  {journal} {Ann. Rev. Nucl. Part. Sci.}\ }\textbf {\bibinfo {volume} {71}},\ \bibinfo {pages} {433} (\bibinfo {year} {2021})}\BibitemShut {NoStop}%
\bibitem [{\citenamefont {Pang}\ \emph {et~al.}(2021)\citenamefont {Pang}, \citenamefont {Tews}, \citenamefont {Coughlin}, \citenamefont {Bulla}, \citenamefont {Van Den~Broeck},\ and\ \citenamefont {Dietrich}}]{Pang:2021jta}%
  \BibitemOpen
  \bibfield  {author} {\bibinfo {author} {\bibfnamefont {P.~T.~H.}\ \bibnamefont {Pang}}, \bibinfo {author} {\bibfnamefont {I.}~\bibnamefont {Tews}}, \bibinfo {author} {\bibfnamefont {M.~W.}\ \bibnamefont {Coughlin}}, \bibinfo {author} {\bibfnamefont {M.}~\bibnamefont {Bulla}}, \bibinfo {author} {\bibfnamefont {C.}~\bibnamefont {Van Den~Broeck}},\ and\ \bibinfo {author} {\bibfnamefont {T.}~\bibnamefont {Dietrich}},\ }\bibfield  {title} {\bibinfo {title} {{Nuclear Physics Multimessenger Astrophysics Constraints on the Neutron Star Equation of State: Adding NICER\textquoteright{}s PSR J0740+6620 Measurement}},\ }\href {https://doi.org/10.3847/1538-4357/ac19ab} {\bibfield  {journal} {\bibinfo  {journal} {Astrophys. J.}\ }\textbf {\bibinfo {volume} {922}},\ \bibinfo {pages} {14} (\bibinfo {year} {2021})},\ \Eprint {https://arxiv.org/abs/2105.08688} {arXiv:2105.08688 [astro-ph.HE]} \BibitemShut {NoStop}%
\bibitem [{\citenamefont {Ascenzi}\ \emph {et~al.}(2024)\citenamefont {Ascenzi}, \citenamefont {Graber},\ and\ \citenamefont {Rea}}]{Ascenzi:2024wws}%
  \BibitemOpen
  \bibfield  {author} {\bibinfo {author} {\bibfnamefont {S.}~\bibnamefont {Ascenzi}}, \bibinfo {author} {\bibfnamefont {V.}~\bibnamefont {Graber}},\ and\ \bibinfo {author} {\bibfnamefont {N.}~\bibnamefont {Rea}},\ }\bibfield  {title} {\bibinfo {title} {{Neutron-star measurements in the multi-messenger Era}},\ }\href {https://doi.org/10.1016/j.astropartphys.2024.102935} {\bibfield  {journal} {\bibinfo  {journal} {Astropart. Phys.}\ }\textbf {\bibinfo {volume} {158}},\ \bibinfo {pages} {102935} (\bibinfo {year} {2024})},\ \Eprint {https://arxiv.org/abs/2401.14930} {arXiv:2401.14930 [astro-ph.HE]} \BibitemShut {NoStop}%
\bibitem [{\citenamefont {Abbott}\ \emph {et~al.}(2017{\natexlab{c}})\citenamefont {Abbott} \emph {et~al.}}]{LIGOScientific:2017ync}%
  \BibitemOpen
  \bibfield  {author} {\bibinfo {author} {\bibfnamefont {B.~P.}\ \bibnamefont {Abbott}} \emph {et~al.} (\bibinfo {collaboration} {LIGO Scientific, Virgo, Fermi GBM, INTEGRAL, IceCube, AstroSat Cadmium Zinc Telluride Imager Team, IPN, Insight-Hxmt, ANTARES, Swift, AGILE Team, 1M2H Team, Dark Energy Camera GW-EM, DES, DLT40, GRAWITA, Fermi-LAT, ATCA, ASKAP, Las Cumbres Observatory Group, OzGrav, DWF (Deeper Wider Faster Program), AST3, CAASTRO, VINROUGE, MASTER, J-GEM, GROWTH, JAGWAR, CaltechNRAO, TTU-NRAO, NuSTAR, Pan-STARRS, MAXI Team, TZAC Consortium, KU, Nordic Optical Telescope, ePESSTO, GROND, Texas Tech University, SALT Group, TOROS, BOOTES, MWA, CALET, IKI-GW Follow-up, H.E.S.S., LOFAR, LWA, HAWC, Pierre Auger, ALMA, Euro VLBI Team, Pi of Sky, Chandra Team at McGill University, DFN, ATLAS Telescopes, High Time Resolution Universe Survey, RIMAS, RATIR, SKA South Africa/MeerKAT}),\ }\bibfield  {title} {\bibinfo {title} {{Multi-messenger Observations of a Binary Neutron Star Merger}},\ }\href
  {https://doi.org/10.3847/2041-8213/aa91c9} {\bibfield  {journal} {\bibinfo  {journal} {Astrophys. J. Lett.}\ }\textbf {\bibinfo {volume} {848}},\ \bibinfo {pages} {L12} (\bibinfo {year} {2017}{\natexlab{c}})},\ \Eprint {https://arxiv.org/abs/1710.05833} {arXiv:1710.05833 [astro-ph.HE]} \BibitemShut {NoStop}%
\bibitem [{\citenamefont {Radice}\ \emph {et~al.}(2018)\citenamefont {Radice}, \citenamefont {Perego}, \citenamefont {Zappa},\ and\ \citenamefont {Bernuzzi}}]{Radice:2017lry}%
  \BibitemOpen
  \bibfield  {author} {\bibinfo {author} {\bibfnamefont {D.}~\bibnamefont {Radice}}, \bibinfo {author} {\bibfnamefont {A.}~\bibnamefont {Perego}}, \bibinfo {author} {\bibfnamefont {F.}~\bibnamefont {Zappa}},\ and\ \bibinfo {author} {\bibfnamefont {S.}~\bibnamefont {Bernuzzi}},\ }\bibfield  {title} {\bibinfo {title} {{GW170817: Joint Constraint on the Neutron Star Equation of State from Multimessenger Observations}},\ }\href {https://doi.org/10.3847/2041-8213/aaa402} {\bibfield  {journal} {\bibinfo  {journal} {Astrophys. J. Lett.}\ }\textbf {\bibinfo {volume} {852}},\ \bibinfo {pages} {L29} (\bibinfo {year} {2018})},\ \Eprint {https://arxiv.org/abs/1711.03647} {arXiv:1711.03647 [astro-ph.HE]} \BibitemShut {NoStop}%
\bibitem [{\citenamefont {Coughlin}\ \emph {et~al.}(2019)\citenamefont {Coughlin}, \citenamefont {Dietrich}, \citenamefont {Margalit},\ and\ \citenamefont {Metzger}}]{Coughlin:2018fis}%
  \BibitemOpen
  \bibfield  {author} {\bibinfo {author} {\bibfnamefont {M.~W.}\ \bibnamefont {Coughlin}}, \bibinfo {author} {\bibfnamefont {T.}~\bibnamefont {Dietrich}}, \bibinfo {author} {\bibfnamefont {B.}~\bibnamefont {Margalit}},\ and\ \bibinfo {author} {\bibfnamefont {B.~D.}\ \bibnamefont {Metzger}},\ }\bibfield  {title} {\bibinfo {title} {{Multimessenger Bayesian parameter inference of a binary neutron star merger}},\ }\href {https://doi.org/10.1093/mnrasl/slz133} {\bibfield  {journal} {\bibinfo  {journal} {Mon. Not. Roy. Astron. Soc.}\ }\textbf {\bibinfo {volume} {489}},\ \bibinfo {pages} {L91} (\bibinfo {year} {2019})},\ \Eprint {https://arxiv.org/abs/1812.04803} {arXiv:1812.04803 [astro-ph.HE]} \BibitemShut {NoStop}%
\bibitem [{\citenamefont {Kiuchi}\ \emph {et~al.}(2019)\citenamefont {Kiuchi}, \citenamefont {Kyutoku}, \citenamefont {Shibata},\ and\ \citenamefont {Taniguchi}}]{Kiuchi:2019lls}%
  \BibitemOpen
  \bibfield  {author} {\bibinfo {author} {\bibfnamefont {K.}~\bibnamefont {Kiuchi}}, \bibinfo {author} {\bibfnamefont {K.}~\bibnamefont {Kyutoku}}, \bibinfo {author} {\bibfnamefont {M.}~\bibnamefont {Shibata}},\ and\ \bibinfo {author} {\bibfnamefont {K.}~\bibnamefont {Taniguchi}},\ }\bibfield  {title} {\bibinfo {title} {{Revisiting the lower bound on tidal deformability derived by AT 2017gfo}},\ }\href {https://doi.org/10.3847/2041-8213/ab1e45} {\bibfield  {journal} {\bibinfo  {journal} {Astrophys. J. Lett.}\ }\textbf {\bibinfo {volume} {876}},\ \bibinfo {pages} {L31} (\bibinfo {year} {2019})},\ \Eprint {https://arxiv.org/abs/1903.01466} {arXiv:1903.01466 [astro-ph.HE]} \BibitemShut {NoStop}%
\bibitem [{\citenamefont {Binnington}\ and\ \citenamefont {Poisson}(2009)}]{Binnington:2009bb}%
  \BibitemOpen
  \bibfield  {author} {\bibinfo {author} {\bibfnamefont {T.}~\bibnamefont {Binnington}}\ and\ \bibinfo {author} {\bibfnamefont {E.}~\bibnamefont {Poisson}},\ }\bibfield  {title} {\bibinfo {title} {{Relativistic theory of tidal Love numbers}},\ }\href {https://doi.org/10.1103/PhysRevD.80.084018} {\bibfield  {journal} {\bibinfo  {journal} {Phys. Rev. D}\ }\textbf {\bibinfo {volume} {80}},\ \bibinfo {pages} {084018} (\bibinfo {year} {2009})},\ \Eprint {https://arxiv.org/abs/0906.1366} {arXiv:0906.1366 [gr-qc]} \BibitemShut {NoStop}%
\bibitem [{\citenamefont {Damour}\ and\ \citenamefont {Nagar}(2009)}]{Damour:2009vw}%
  \BibitemOpen
  \bibfield  {author} {\bibinfo {author} {\bibfnamefont {T.}~\bibnamefont {Damour}}\ and\ \bibinfo {author} {\bibfnamefont {A.}~\bibnamefont {Nagar}},\ }\bibfield  {title} {\bibinfo {title} {{Relativistic tidal properties of neutron stars}},\ }\href {https://doi.org/10.1103/PhysRevD.80.084035} {\bibfield  {journal} {\bibinfo  {journal} {Phys. Rev. D}\ }\textbf {\bibinfo {volume} {80}},\ \bibinfo {pages} {084035} (\bibinfo {year} {2009})},\ \Eprint {https://arxiv.org/abs/0906.0096} {arXiv:0906.0096 [gr-qc]} \BibitemShut {NoStop}%
\bibitem [{\citenamefont {Yagi}(2014)}]{Yagi:2013sva}%
  \BibitemOpen
  \bibfield  {author} {\bibinfo {author} {\bibfnamefont {K.}~\bibnamefont {Yagi}},\ }\bibfield  {title} {\bibinfo {title} {{Multipole Love Relations}},\ }\href {https://doi.org/10.1103/PhysRevD.89.043011} {\bibfield  {journal} {\bibinfo  {journal} {Phys. Rev. D}\ }\textbf {\bibinfo {volume} {89}},\ \bibinfo {pages} {043011} (\bibinfo {year} {2014})},\ \bibinfo {note} {[Erratum: Phys.Rev.D 96, 129904 (2017), Erratum: Phys.Rev.D 97, 129901 (2018)]},\ \Eprint {https://arxiv.org/abs/1311.0872} {arXiv:1311.0872 [gr-qc]} \BibitemShut {NoStop}%
\bibitem [{\citenamefont {Banihashemi}\ and\ \citenamefont {Vines}(2020)}]{Banihashemi:2018xfb}%
  \BibitemOpen
  \bibfield  {author} {\bibinfo {author} {\bibfnamefont {B.}~\bibnamefont {Banihashemi}}\ and\ \bibinfo {author} {\bibfnamefont {J.}~\bibnamefont {Vines}},\ }\bibfield  {title} {\bibinfo {title} {{Gravitomagnetic tidal effects in gravitational waves from neutron star binaries}},\ }\href {https://doi.org/10.1103/PhysRevD.101.064003} {\bibfield  {journal} {\bibinfo  {journal} {Phys. Rev. D}\ }\textbf {\bibinfo {volume} {101}},\ \bibinfo {pages} {064003} (\bibinfo {year} {2020})},\ \Eprint {https://arxiv.org/abs/1805.07266} {arXiv:1805.07266 [gr-qc]} \BibitemShut {NoStop}%
\bibitem [{\citenamefont {Henry}\ \emph {et~al.}(2020)\citenamefont {Henry}, \citenamefont {Faye},\ and\ \citenamefont {Blanchet}}]{Henry:2020ski}%
  \BibitemOpen
  \bibfield  {author} {\bibinfo {author} {\bibfnamefont {Q.}~\bibnamefont {Henry}}, \bibinfo {author} {\bibfnamefont {G.}~\bibnamefont {Faye}},\ and\ \bibinfo {author} {\bibfnamefont {L.}~\bibnamefont {Blanchet}},\ }\bibfield  {title} {\bibinfo {title} {{Tidal effects in the gravitational-wave phase evolution of compact binary systems to next-to-next-to-leading post-Newtonian order}},\ }\href {https://doi.org/10.1103/PhysRevD.102.044033} {\bibfield  {journal} {\bibinfo  {journal} {Phys. Rev. D}\ }\textbf {\bibinfo {volume} {102}},\ \bibinfo {pages} {044033} (\bibinfo {year} {2020})},\ \bibinfo {note} {[Erratum: Phys.Rev.D 108, 089901 (2023)]},\ \Eprint {https://arxiv.org/abs/2005.13367} {arXiv:2005.13367 [gr-qc]} \BibitemShut {NoStop}%
\bibitem [{\citenamefont {Narikawa}(2023)}]{Narikawa:2023deu}%
  \BibitemOpen
  \bibfield  {author} {\bibinfo {author} {\bibfnamefont {T.}~\bibnamefont {Narikawa}},\ }\bibfield  {title} {\bibinfo {title} {{Multipole tidal effects in the post-Newtonian gravitational-wave phase of compact binary coalescences}},\ }\href {https://doi.org/10.1103/PhysRevD.108.063029} {\bibfield  {journal} {\bibinfo  {journal} {Phys. Rev. D}\ }\textbf {\bibinfo {volume} {108}},\ \bibinfo {pages} {063029} (\bibinfo {year} {2023})},\ \Eprint {https://arxiv.org/abs/2307.02033} {arXiv:2307.02033 [gr-qc]} \BibitemShut {NoStop}%
\bibitem [{\citenamefont {Ripley}\ \emph {et~al.}(2023{\natexlab{a}})\citenamefont {Ripley}, \citenamefont {Hegade K.~R.},\ and\ \citenamefont {Yunes}}]{Ripley:2023qxo}%
  \BibitemOpen
  \bibfield  {author} {\bibinfo {author} {\bibfnamefont {J.~L.}\ \bibnamefont {Ripley}}, \bibinfo {author} {\bibfnamefont {A.}~\bibnamefont {Hegade K.~R.}},\ and\ \bibinfo {author} {\bibfnamefont {N.}~\bibnamefont {Yunes}},\ }\bibfield  {title} {\bibinfo {title} {{Probing internal dissipative processes of neutron stars with gravitational waves during the inspiral of neutron star binaries}},\ }\href {https://doi.org/10.1103/PhysRevD.108.103037} {\bibfield  {journal} {\bibinfo  {journal} {Phys. Rev. D}\ }\textbf {\bibinfo {volume} {108}},\ \bibinfo {pages} {103037} (\bibinfo {year} {2023}{\natexlab{a}})},\ \Eprint {https://arxiv.org/abs/2306.15633} {arXiv:2306.15633 [gr-qc]} \BibitemShut {NoStop}%
\bibitem [{\citenamefont {Hegade K.~R.}\ \emph {et~al.}(2024)\citenamefont {Hegade K.~R.}, \citenamefont {Ripley},\ and\ \citenamefont {Yunes}}]{HegadeKR:2024agt}%
  \BibitemOpen
  \bibfield  {author} {\bibinfo {author} {\bibfnamefont {A.}~\bibnamefont {Hegade K.~R.}}, \bibinfo {author} {\bibfnamefont {J.~L.}\ \bibnamefont {Ripley}},\ and\ \bibinfo {author} {\bibfnamefont {N.}~\bibnamefont {Yunes}},\ }\bibfield  {title} {\bibinfo {title} {{Dynamical tidal response of nonrotating relativistic stars}},\ }\href {https://doi.org/10.1103/PhysRevD.109.104064} {\bibfield  {journal} {\bibinfo  {journal} {Phys. Rev. D}\ }\textbf {\bibinfo {volume} {109}},\ \bibinfo {pages} {104064} (\bibinfo {year} {2024})},\ \Eprint {https://arxiv.org/abs/2403.03254} {arXiv:2403.03254 [gr-qc]} \BibitemShut {NoStop}%
\bibitem [{\citenamefont {Saketh}\ \emph {et~al.}(2024)\citenamefont {Saketh}, \citenamefont {Zhou}, \citenamefont {Ghosh}, \citenamefont {Steinhoff},\ and\ \citenamefont {Chatterjee}}]{Saketh:2024juq}%
  \BibitemOpen
  \bibfield  {author} {\bibinfo {author} {\bibfnamefont {M.~V.~S.}\ \bibnamefont {Saketh}}, \bibinfo {author} {\bibfnamefont {Z.}~\bibnamefont {Zhou}}, \bibinfo {author} {\bibfnamefont {S.}~\bibnamefont {Ghosh}}, \bibinfo {author} {\bibfnamefont {J.}~\bibnamefont {Steinhoff}},\ and\ \bibinfo {author} {\bibfnamefont {D.}~\bibnamefont {Chatterjee}},\ }\bibfield  {title} {\bibinfo {title} {{Investigating tidal heating in neutron stars via gravitational Raman scattering}},\ }\href {https://doi.org/10.1103/PhysRevD.110.103001} {\bibfield  {journal} {\bibinfo  {journal} {Phys. Rev. D}\ }\textbf {\bibinfo {volume} {110}},\ \bibinfo {pages} {103001} (\bibinfo {year} {2024})},\ \Eprint {https://arxiv.org/abs/2407.08327} {arXiv:2407.08327 [gr-qc]} \BibitemShut {NoStop}%
\bibitem [{\citenamefont {Ripley}\ \emph {et~al.}(2023{\natexlab{b}})\citenamefont {Ripley}, \citenamefont {Hegade K.~R.}, \citenamefont {Chandramouli},\ and\ \citenamefont {Yunes}}]{Ripley:2023lsq}%
  \BibitemOpen
  \bibfield  {author} {\bibinfo {author} {\bibfnamefont {J.~L.}\ \bibnamefont {Ripley}}, \bibinfo {author} {\bibfnamefont {A.}~\bibnamefont {Hegade K.~R.}}, \bibinfo {author} {\bibfnamefont {R.~S.}\ \bibnamefont {Chandramouli}},\ and\ \bibinfo {author} {\bibfnamefont {N.}~\bibnamefont {Yunes}},\ }\bibfield  {title} {\bibinfo {title} {{First constraint on the dissipative tidal deformability of neutron stars}},\ }\href@noop {} {\bibfield  {journal} {\bibinfo  {journal} {arXiv}\ } (\bibinfo {year} {2023}{\natexlab{b}})},\ \Eprint {https://arxiv.org/abs/2312.11659} {2312.11659 [gr-qc]} \BibitemShut {NoStop}%
\bibitem [{\citenamefont {Andersson}\ and\ \citenamefont {Pnigouras}(2020)}]{Andersson:2019ahb}%
  \BibitemOpen
  \bibfield  {author} {\bibinfo {author} {\bibfnamefont {N.}~\bibnamefont {Andersson}}\ and\ \bibinfo {author} {\bibfnamefont {P.}~\bibnamefont {Pnigouras}},\ }\bibfield  {title} {\bibinfo {title} {{Exploring the effective tidal deformability of neutron stars}},\ }\href {https://doi.org/10.1103/PhysRevD.101.083001} {\bibfield  {journal} {\bibinfo  {journal} {Phys. Rev. D}\ }\textbf {\bibinfo {volume} {101}},\ \bibinfo {pages} {083001} (\bibinfo {year} {2020})},\ \Eprint {https://arxiv.org/abs/1906.08982} {arXiv:1906.08982 [astro-ph.HE]} \BibitemShut {NoStop}%
\bibitem [{\citenamefont {Gupta}\ \emph {et~al.}(2021)\citenamefont {Gupta}, \citenamefont {Steinhoff},\ and\ \citenamefont {Hinderer}}]{Gupta:2020lnv}%
  \BibitemOpen
  \bibfield  {author} {\bibinfo {author} {\bibfnamefont {P.~K.}\ \bibnamefont {Gupta}}, \bibinfo {author} {\bibfnamefont {J.}~\bibnamefont {Steinhoff}},\ and\ \bibinfo {author} {\bibfnamefont {T.}~\bibnamefont {Hinderer}},\ }\bibfield  {title} {\bibinfo {title} {{Relativistic effective action of dynamical gravitomagnetic tides for slowly rotating neutron stars}},\ }\href {https://doi.org/10.1103/PhysRevResearch.3.013147} {\bibfield  {journal} {\bibinfo  {journal} {Phys. Rev. Res.}\ }\textbf {\bibinfo {volume} {3}},\ \bibinfo {pages} {013147} (\bibinfo {year} {2021})},\ \Eprint {https://arxiv.org/abs/2011.03508} {arXiv:2011.03508 [gr-qc]} \BibitemShut {NoStop}%
\bibitem [{\citenamefont {Pitre}\ and\ \citenamefont {Poisson}(2024)}]{Pitre:2023xsr}%
  \BibitemOpen
  \bibfield  {author} {\bibinfo {author} {\bibfnamefont {T.}~\bibnamefont {Pitre}}\ and\ \bibinfo {author} {\bibfnamefont {E.}~\bibnamefont {Poisson}},\ }\bibfield  {title} {\bibinfo {title} {{General relativistic dynamical tides in binary inspirals without modes}},\ }\href {https://doi.org/10.1103/PhysRevD.109.064004} {\bibfield  {journal} {\bibinfo  {journal} {Phys. Rev. D}\ }\textbf {\bibinfo {volume} {109}},\ \bibinfo {pages} {064004} (\bibinfo {year} {2024})},\ \Eprint {https://arxiv.org/abs/2311.04075} {arXiv:2311.04075 [gr-qc]} \BibitemShut {NoStop}%
\bibitem [{\citenamefont {Gupta}\ \emph {et~al.}(2023)\citenamefont {Gupta}, \citenamefont {Steinhoff},\ and\ \citenamefont {Hinderer}}]{Gupta:2023oyy}%
  \BibitemOpen
  \bibfield  {author} {\bibinfo {author} {\bibfnamefont {P.~K.}\ \bibnamefont {Gupta}}, \bibinfo {author} {\bibfnamefont {J.}~\bibnamefont {Steinhoff}},\ and\ \bibinfo {author} {\bibfnamefont {T.}~\bibnamefont {Hinderer}},\ }\bibfield  {title} {\bibinfo {title} {{Effect of dynamical gravitomagnetic tides on measurability of tidal parameters for binary neutron stars using gravitational waves}},\ }\href {https://doi.org/10.1103/PhysRevD.108.124040} {\bibfield  {journal} {\bibinfo  {journal} {Phys. Rev. D}\ }\textbf {\bibinfo {volume} {108}},\ \bibinfo {pages} {124040} (\bibinfo {year} {2023})},\ \Eprint {https://arxiv.org/abs/2302.11274} {arXiv:2302.11274 [gr-qc]} \BibitemShut {NoStop}%
\bibitem [{\citenamefont {Mandal}\ \emph {et~al.}(2023{\natexlab{a}})\citenamefont {Mandal}, \citenamefont {Mastrolia}, \citenamefont {Silva}, \citenamefont {Patil},\ and\ \citenamefont {Steinhoff}}]{Mandal:2023lgy}%
  \BibitemOpen
  \bibfield  {author} {\bibinfo {author} {\bibfnamefont {M.~K.}\ \bibnamefont {Mandal}}, \bibinfo {author} {\bibfnamefont {P.}~\bibnamefont {Mastrolia}}, \bibinfo {author} {\bibfnamefont {H.~O.}\ \bibnamefont {Silva}}, \bibinfo {author} {\bibfnamefont {R.}~\bibnamefont {Patil}},\ and\ \bibinfo {author} {\bibfnamefont {J.}~\bibnamefont {Steinhoff}},\ }\bibfield  {title} {\bibinfo {title} {{Gravitoelectric dynamical tides at second post-Newtonian order}},\ }\href {https://doi.org/10.1007/JHEP11(2023)067} {\bibfield  {journal} {\bibinfo  {journal} {JHEP}\ }\textbf {\bibinfo {volume} {11}},\ \bibinfo {pages} {067}},\ \Eprint {https://arxiv.org/abs/2304.02030} {arXiv:2304.02030 [hep-th]} \BibitemShut {NoStop}%
\bibitem [{\citenamefont {Flanagan}\ and\ \citenamefont {Racine}(2007)}]{Flanagan:2006sb}%
  \BibitemOpen
  \bibfield  {author} {\bibinfo {author} {\bibfnamefont {E.~E.}\ \bibnamefont {Flanagan}}\ and\ \bibinfo {author} {\bibfnamefont {E.}~\bibnamefont {Racine}},\ }\bibfield  {title} {\bibinfo {title} {{Gravitomagnetic resonant excitation of Rossby modes in coalescing neutron star binaries}},\ }\href {https://doi.org/10.1103/PhysRevD.75.044001} {\bibfield  {journal} {\bibinfo  {journal} {Phys. Rev. D}\ }\textbf {\bibinfo {volume} {75}},\ \bibinfo {pages} {044001} (\bibinfo {year} {2007})},\ \Eprint {https://arxiv.org/abs/gr-qc/0601029} {arXiv:gr-qc/0601029} \BibitemShut {NoStop}%
\bibitem [{\citenamefont {Lai}\ and\ \citenamefont {Wu}(2006)}]{Lai:2006pr}%
  \BibitemOpen
  \bibfield  {author} {\bibinfo {author} {\bibfnamefont {D.}~\bibnamefont {Lai}}\ and\ \bibinfo {author} {\bibfnamefont {Y.}~\bibnamefont {Wu}},\ }\bibfield  {title} {\bibinfo {title} {{Resonant Tidal Excitations of Inertial Modes in Coalescing Neutron Star Binaries}},\ }\href {https://doi.org/10.1103/PhysRevD.74.024007} {\bibfield  {journal} {\bibinfo  {journal} {Phys. Rev. D}\ }\textbf {\bibinfo {volume} {74}},\ \bibinfo {pages} {024007} (\bibinfo {year} {2006})},\ \Eprint {https://arxiv.org/abs/astro-ph/0604163} {arXiv:astro-ph/0604163} \BibitemShut {NoStop}%
\bibitem [{\citenamefont {Ma}\ \emph {et~al.}(2021)\citenamefont {Ma}, \citenamefont {Yu},\ and\ \citenamefont {Chen}}]{Ma:2020oni}%
  \BibitemOpen
  \bibfield  {author} {\bibinfo {author} {\bibfnamefont {S.}~\bibnamefont {Ma}}, \bibinfo {author} {\bibfnamefont {H.}~\bibnamefont {Yu}},\ and\ \bibinfo {author} {\bibfnamefont {Y.}~\bibnamefont {Chen}},\ }\bibfield  {title} {\bibinfo {title} {{Detecting resonant tidal excitations of Rossby modes in coalescing neutron-star binaries with third-generation gravitational-wave detectors}},\ }\href {https://doi.org/10.1103/PhysRevD.103.063020} {\bibfield  {journal} {\bibinfo  {journal} {Phys. Rev. D}\ }\textbf {\bibinfo {volume} {103}},\ \bibinfo {pages} {063020} (\bibinfo {year} {2021})},\ \Eprint {https://arxiv.org/abs/2010.03066} {arXiv:2010.03066 [gr-qc]} \BibitemShut {NoStop}%
\bibitem [{\citenamefont {Shibata}(1994)}]{Shibata:1993qc}%
  \BibitemOpen
  \bibfield  {author} {\bibinfo {author} {\bibfnamefont {M.}~\bibnamefont {Shibata}},\ }\bibfield  {title} {\bibinfo {title} {{Effects of tidal resonances in coalescing compact binary systems}},\ }\href {https://doi.org/10.1143/PTP.91.871} {\bibfield  {journal} {\bibinfo  {journal} {Prog. Theor. Phys.}\ }\textbf {\bibinfo {volume} {91}},\ \bibinfo {pages} {871} (\bibinfo {year} {1994})}\BibitemShut {NoStop}%
\bibitem [{\citenamefont {Kokkotas}\ and\ \citenamefont {Schaefer}(1995)}]{Kokkotas:1995xe}%
  \BibitemOpen
  \bibfield  {author} {\bibinfo {author} {\bibfnamefont {K.~D.}\ \bibnamefont {Kokkotas}}\ and\ \bibinfo {author} {\bibfnamefont {G.}~\bibnamefont {Schaefer}},\ }\bibfield  {title} {\bibinfo {title} {{Tidal and tidal resonant effects in coalescing binaries}},\ }\href {https://doi.org/10.1093/mnras/275.2.301} {\bibfield  {journal} {\bibinfo  {journal} {Mon. Not. Roy. Astron. Soc.}\ }\textbf {\bibinfo {volume} {275}},\ \bibinfo {pages} {301} (\bibinfo {year} {1995})},\ \Eprint {https://arxiv.org/abs/gr-qc/9502034} {arXiv:gr-qc/9502034} \BibitemShut {NoStop}%
\bibitem [{\citenamefont {Lai}(1994)}]{Lai:1993di}%
  \BibitemOpen
  \bibfield  {author} {\bibinfo {author} {\bibfnamefont {D.}~\bibnamefont {Lai}},\ }\bibfield  {title} {\bibinfo {title} {{Resonant oscillations and tidal heating in coalescing binary neutron stars}},\ }\href {https://doi.org/10.1093/mnras/270.3.611} {\bibfield  {journal} {\bibinfo  {journal} {Mon. Not. Roy. Astron. Soc.}\ }\textbf {\bibinfo {volume} {270}},\ \bibinfo {pages} {611} (\bibinfo {year} {1994})},\ \Eprint {https://arxiv.org/abs/astro-ph/9404062} {arXiv:astro-ph/9404062} \BibitemShut {NoStop}%
\bibitem [{\citenamefont {Ho}\ and\ \citenamefont {Lai}(1999)}]{Ho:1998hq}%
  \BibitemOpen
  \bibfield  {author} {\bibinfo {author} {\bibfnamefont {W.~C.~G.}\ \bibnamefont {Ho}}\ and\ \bibinfo {author} {\bibfnamefont {D.}~\bibnamefont {Lai}},\ }\bibfield  {title} {\bibinfo {title} {{Resonant tidal excitations of rotating neutron stars in coalescing binaries}},\ }\href {https://doi.org/10.1046/j.1365-8711.1999.02703.x} {\bibfield  {journal} {\bibinfo  {journal} {Mon. Not. Roy. Astron. Soc.}\ }\textbf {\bibinfo {volume} {308}},\ \bibinfo {pages} {153} (\bibinfo {year} {1999})},\ \Eprint {https://arxiv.org/abs/astro-ph/9812116} {arXiv:astro-ph/9812116} \BibitemShut {NoStop}%
\bibitem [{\citenamefont {Ma}\ \emph {et~al.}(2020)\citenamefont {Ma}, \citenamefont {Yu},\ and\ \citenamefont {Chen}}]{Ma:2020rak}%
  \BibitemOpen
  \bibfield  {author} {\bibinfo {author} {\bibfnamefont {S.}~\bibnamefont {Ma}}, \bibinfo {author} {\bibfnamefont {H.}~\bibnamefont {Yu}},\ and\ \bibinfo {author} {\bibfnamefont {Y.}~\bibnamefont {Chen}},\ }\bibfield  {title} {\bibinfo {title} {{Excitation of f-modes during mergers of spinning binary neutron star}},\ }\href {https://doi.org/10.1103/PhysRevD.101.123020} {\bibfield  {journal} {\bibinfo  {journal} {Phys. Rev. D}\ }\textbf {\bibinfo {volume} {101}},\ \bibinfo {pages} {123020} (\bibinfo {year} {2020})},\ \Eprint {https://arxiv.org/abs/2003.02373} {arXiv:2003.02373 [gr-qc]} \BibitemShut {NoStop}%
\bibitem [{\citenamefont {Kuan}\ and\ \citenamefont {Kokkotas}(2022)}]{Kuan:2022etu}%
  \BibitemOpen
  \bibfield  {author} {\bibinfo {author} {\bibfnamefont {H.-J.}\ \bibnamefont {Kuan}}\ and\ \bibinfo {author} {\bibfnamefont {K.~D.}\ \bibnamefont {Kokkotas}},\ }\bibfield  {title} {\bibinfo {title} {{f-mode imprints on gravitational waves from coalescing binaries involving aligned spinning neutron stars}},\ }\href {https://doi.org/10.1103/PhysRevD.106.064052} {\bibfield  {journal} {\bibinfo  {journal} {Phys. Rev. D}\ }\textbf {\bibinfo {volume} {106}},\ \bibinfo {pages} {064052} (\bibinfo {year} {2022})},\ \Eprint {https://arxiv.org/abs/2205.01705} {arXiv:2205.01705 [gr-qc]} \BibitemShut {NoStop}%
\bibitem [{\citenamefont {Gamba}\ and\ \citenamefont {Bernuzzi}(2023)}]{Gamba:2022mgx}%
  \BibitemOpen
  \bibfield  {author} {\bibinfo {author} {\bibfnamefont {R.}~\bibnamefont {Gamba}}\ and\ \bibinfo {author} {\bibfnamefont {S.}~\bibnamefont {Bernuzzi}},\ }\bibfield  {title} {\bibinfo {title} {{Resonant tides in binary neutron star mergers: Analytical-numerical relativity study}},\ }\href {https://doi.org/10.1103/PhysRevD.107.044014} {\bibfield  {journal} {\bibinfo  {journal} {Phys. Rev. D}\ }\textbf {\bibinfo {volume} {107}},\ \bibinfo {pages} {044014} (\bibinfo {year} {2023})},\ \Eprint {https://arxiv.org/abs/2207.13106} {arXiv:2207.13106 [gr-qc]} \BibitemShut {NoStop}%
\bibitem [{\citenamefont {Kuan}\ and\ \citenamefont {Kokkotas}(2023)}]{Kuan:2023qxo}%
  \BibitemOpen
  \bibfield  {author} {\bibinfo {author} {\bibfnamefont {H.-J.}\ \bibnamefont {Kuan}}\ and\ \bibinfo {author} {\bibfnamefont {K.~D.}\ \bibnamefont {Kokkotas}},\ }\bibfield  {title} {\bibinfo {title} {{Last three seconds: Packed message delivered by tides in binary neutron star mergers}},\ }\href {https://doi.org/10.1103/PhysRevD.108.063026} {\bibfield  {journal} {\bibinfo  {journal} {Phys. Rev. D}\ }\textbf {\bibinfo {volume} {108}},\ \bibinfo {pages} {063026} (\bibinfo {year} {2023})},\ \Eprint {https://arxiv.org/abs/2309.04622} {arXiv:2309.04622 [gr-qc]} \BibitemShut {NoStop}%
\bibitem [{\citenamefont {Steinhoff}\ \emph {et~al.}(2016)\citenamefont {Steinhoff}, \citenamefont {Hinderer}, \citenamefont {Buonanno},\ and\ \citenamefont {Taracchini}}]{Steinhoff:2016rfi}%
  \BibitemOpen
  \bibfield  {author} {\bibinfo {author} {\bibfnamefont {J.}~\bibnamefont {Steinhoff}}, \bibinfo {author} {\bibfnamefont {T.}~\bibnamefont {Hinderer}}, \bibinfo {author} {\bibfnamefont {A.}~\bibnamefont {Buonanno}},\ and\ \bibinfo {author} {\bibfnamefont {A.}~\bibnamefont {Taracchini}},\ }\bibfield  {title} {\bibinfo {title} {{Dynamical Tides in General Relativity: Effective Action and Effective-One-Body Hamiltonian}},\ }\href {https://doi.org/10.1103/PhysRevD.94.104028} {\bibfield  {journal} {\bibinfo  {journal} {Phys. Rev. D}\ }\textbf {\bibinfo {volume} {94}},\ \bibinfo {pages} {104028} (\bibinfo {year} {2016})},\ \Eprint {https://arxiv.org/abs/1608.01907} {arXiv:1608.01907 [gr-qc]} \BibitemShut {NoStop}%
\bibitem [{\citenamefont {Hinderer}\ \emph {et~al.}(2016)\citenamefont {Hinderer} \emph {et~al.}}]{Hinderer:2016eia}%
  \BibitemOpen
  \bibfield  {author} {\bibinfo {author} {\bibfnamefont {T.}~\bibnamefont {Hinderer}} \emph {et~al.},\ }\bibfield  {title} {\bibinfo {title} {{Effects of neutron-star dynamic tides on gravitational waveforms within the effective-one-body approach}},\ }\href {https://doi.org/10.1103/PhysRevLett.116.181101} {\bibfield  {journal} {\bibinfo  {journal} {Phys. Rev. Lett.}\ }\textbf {\bibinfo {volume} {116}},\ \bibinfo {pages} {181101} (\bibinfo {year} {2016})},\ \Eprint {https://arxiv.org/abs/1602.00599} {arXiv:1602.00599 [gr-qc]} \BibitemShut {NoStop}%
\bibitem [{\citenamefont {Andersson}\ and\ \citenamefont {Pnigouras}(2021)}]{Andersson:2019dwg}%
  \BibitemOpen
  \bibfield  {author} {\bibinfo {author} {\bibfnamefont {N.}~\bibnamefont {Andersson}}\ and\ \bibinfo {author} {\bibfnamefont {P.}~\bibnamefont {Pnigouras}},\ }\bibfield  {title} {\bibinfo {title} {{The phenomenology of dynamical neutron star tides}},\ }\href {https://doi.org/10.1093/mnras/stab371} {\bibfield  {journal} {\bibinfo  {journal} {Mon. Not. Roy. Astron. Soc.}\ }\textbf {\bibinfo {volume} {503}},\ \bibinfo {pages} {533} (\bibinfo {year} {2021})},\ \Eprint {https://arxiv.org/abs/1905.00012} {arXiv:1905.00012 [gr-qc]} \BibitemShut {NoStop}%
\bibitem [{\citenamefont {Dietrich}\ \emph {et~al.}(2017{\natexlab{a}})\citenamefont {Dietrich}, \citenamefont {Bernuzzi},\ and\ \citenamefont {Tichy}}]{Dietrich:2017aum}%
  \BibitemOpen
  \bibfield  {author} {\bibinfo {author} {\bibfnamefont {T.}~\bibnamefont {Dietrich}}, \bibinfo {author} {\bibfnamefont {S.}~\bibnamefont {Bernuzzi}},\ and\ \bibinfo {author} {\bibfnamefont {W.}~\bibnamefont {Tichy}},\ }\bibfield  {title} {\bibinfo {title} {{Closed-form tidal approximants for binary neutron star gravitational waveforms constructed from high-resolution numerical relativity simulations}},\ }\href {https://doi.org/10.1103/PhysRevD.96.121501} {\bibfield  {journal} {\bibinfo  {journal} {Phys. Rev. D}\ }\textbf {\bibinfo {volume} {96}},\ \bibinfo {pages} {121501} (\bibinfo {year} {2017}{\natexlab{a}})},\ \Eprint {https://arxiv.org/abs/1706.02969} {arXiv:1706.02969 [gr-qc]} \BibitemShut {NoStop}%
\bibitem [{\citenamefont {Kawaguchi}\ \emph {et~al.}(2018)\citenamefont {Kawaguchi}, \citenamefont {Kiuchi}, \citenamefont {Kyutoku}, \citenamefont {Sekiguchi}, \citenamefont {Shibata},\ and\ \citenamefont {Taniguchi}}]{Kawaguchi:2018gvj}%
  \BibitemOpen
  \bibfield  {author} {\bibinfo {author} {\bibfnamefont {K.}~\bibnamefont {Kawaguchi}}, \bibinfo {author} {\bibfnamefont {K.}~\bibnamefont {Kiuchi}}, \bibinfo {author} {\bibfnamefont {K.}~\bibnamefont {Kyutoku}}, \bibinfo {author} {\bibfnamefont {Y.}~\bibnamefont {Sekiguchi}}, \bibinfo {author} {\bibfnamefont {M.}~\bibnamefont {Shibata}},\ and\ \bibinfo {author} {\bibfnamefont {K.}~\bibnamefont {Taniguchi}},\ }\bibfield  {title} {\bibinfo {title} {{Frequency-domain gravitational waveform models for inspiraling binary neutron stars}},\ }\href {https://doi.org/10.1103/PhysRevD.97.044044} {\bibfield  {journal} {\bibinfo  {journal} {Phys. Rev. D}\ }\textbf {\bibinfo {volume} {97}},\ \bibinfo {pages} {044044} (\bibinfo {year} {2018})},\ \Eprint {https://arxiv.org/abs/1802.06518} {arXiv:1802.06518 [gr-qc]} \BibitemShut {NoStop}%
\bibitem [{\citenamefont {Dietrich}\ \emph {et~al.}(2019{\natexlab{a}})\citenamefont {Dietrich} \emph {et~al.}}]{Dietrich:2018uni}%
  \BibitemOpen
  \bibfield  {author} {\bibinfo {author} {\bibfnamefont {T.}~\bibnamefont {Dietrich}} \emph {et~al.},\ }\bibfield  {title} {\bibinfo {title} {{Matter imprints in waveform models for neutron star binaries: Tidal and self-spin effects}},\ }\href {https://doi.org/10.1103/PhysRevD.99.024029} {\bibfield  {journal} {\bibinfo  {journal} {Phys. Rev. D}\ }\textbf {\bibinfo {volume} {99}},\ \bibinfo {pages} {024029} (\bibinfo {year} {2019}{\natexlab{a}})},\ \Eprint {https://arxiv.org/abs/1804.02235} {arXiv:1804.02235 [gr-qc]} \BibitemShut {NoStop}%
\bibitem [{\citenamefont {Dietrich}\ \emph {et~al.}(2019{\natexlab{b}})\citenamefont {Dietrich}, \citenamefont {Samajdar}, \citenamefont {Khan}, \citenamefont {Johnson-McDaniel}, \citenamefont {Dudi},\ and\ \citenamefont {Tichy}}]{Dietrich:2019kaq}%
  \BibitemOpen
  \bibfield  {author} {\bibinfo {author} {\bibfnamefont {T.}~\bibnamefont {Dietrich}}, \bibinfo {author} {\bibfnamefont {A.}~\bibnamefont {Samajdar}}, \bibinfo {author} {\bibfnamefont {S.}~\bibnamefont {Khan}}, \bibinfo {author} {\bibfnamefont {N.~K.}\ \bibnamefont {Johnson-McDaniel}}, \bibinfo {author} {\bibfnamefont {R.}~\bibnamefont {Dudi}},\ and\ \bibinfo {author} {\bibfnamefont {W.}~\bibnamefont {Tichy}},\ }\bibfield  {title} {\bibinfo {title} {{Improving the NRTidal model for binary neutron star systems}},\ }\href {https://doi.org/10.1103/PhysRevD.100.044003} {\bibfield  {journal} {\bibinfo  {journal} {Phys. Rev. D}\ }\textbf {\bibinfo {volume} {100}},\ \bibinfo {pages} {044003} (\bibinfo {year} {2019}{\natexlab{b}})},\ \Eprint {https://arxiv.org/abs/1905.06011} {arXiv:1905.06011 [gr-qc]} \BibitemShut {NoStop}%
\bibitem [{\citenamefont {Abac}\ \emph {et~al.}(2024)\citenamefont {Abac}, \citenamefont {Dietrich}, \citenamefont {Buonanno}, \citenamefont {Steinhoff},\ and\ \citenamefont {Ujevic}}]{Abac:2023ujg}%
  \BibitemOpen
  \bibfield  {author} {\bibinfo {author} {\bibfnamefont {A.}~\bibnamefont {Abac}}, \bibinfo {author} {\bibfnamefont {T.}~\bibnamefont {Dietrich}}, \bibinfo {author} {\bibfnamefont {A.}~\bibnamefont {Buonanno}}, \bibinfo {author} {\bibfnamefont {J.}~\bibnamefont {Steinhoff}},\ and\ \bibinfo {author} {\bibfnamefont {M.}~\bibnamefont {Ujevic}},\ }\bibfield  {title} {\bibinfo {title} {{New and robust gravitational-waveform model for high-mass-ratio binary neutron star systems with dynamical tidal effects}},\ }\href {https://doi.org/10.1103/PhysRevD.109.024062} {\bibfield  {journal} {\bibinfo  {journal} {Phys. Rev. D}\ }\textbf {\bibinfo {volume} {109}},\ \bibinfo {pages} {024062} (\bibinfo {year} {2024})},\ \Eprint {https://arxiv.org/abs/2311.07456} {arXiv:2311.07456 [gr-qc]} \BibitemShut {NoStop}%
\bibitem [{\citenamefont {Dietrich}\ and\ \citenamefont {Hinderer}(2017)}]{Dietrich:2017feu}%
  \BibitemOpen
  \bibfield  {author} {\bibinfo {author} {\bibfnamefont {T.}~\bibnamefont {Dietrich}}\ and\ \bibinfo {author} {\bibfnamefont {T.}~\bibnamefont {Hinderer}},\ }\bibfield  {title} {\bibinfo {title} {{Comprehensive comparison of numerical relativity and effective-one-body results to inform improvements in waveform models for binary neutron star systems}},\ }\href {https://doi.org/10.1103/PhysRevD.95.124006} {\bibfield  {journal} {\bibinfo  {journal} {Phys. Rev. D}\ }\textbf {\bibinfo {volume} {95}},\ \bibinfo {pages} {124006} (\bibinfo {year} {2017})},\ \Eprint {https://arxiv.org/abs/1702.02053} {arXiv:1702.02053 [gr-qc]} \BibitemShut {NoStop}%
\bibitem [{\citenamefont {Rettegno}\ \emph {et~al.}(2020)\citenamefont {Rettegno}, \citenamefont {Martinetti}, \citenamefont {Nagar}, \citenamefont {Bini}, \citenamefont {Riemenschneider},\ and\ \citenamefont {Damour}}]{Rettegno:2019tzh}%
  \BibitemOpen
  \bibfield  {author} {\bibinfo {author} {\bibfnamefont {P.}~\bibnamefont {Rettegno}}, \bibinfo {author} {\bibfnamefont {F.}~\bibnamefont {Martinetti}}, \bibinfo {author} {\bibfnamefont {A.}~\bibnamefont {Nagar}}, \bibinfo {author} {\bibfnamefont {D.}~\bibnamefont {Bini}}, \bibinfo {author} {\bibfnamefont {G.}~\bibnamefont {Riemenschneider}},\ and\ \bibinfo {author} {\bibfnamefont {T.}~\bibnamefont {Damour}},\ }\bibfield  {title} {\bibinfo {title} {{Comparing Effective One Body Hamiltonians for spin-aligned coalescing binaries}},\ }\href {https://doi.org/10.1103/PhysRevD.101.104027} {\bibfield  {journal} {\bibinfo  {journal} {Phys. Rev. D}\ }\textbf {\bibinfo {volume} {101}},\ \bibinfo {pages} {104027} (\bibinfo {year} {2020})},\ \Eprint {https://arxiv.org/abs/1911.10818} {arXiv:1911.10818 [gr-qc]} \BibitemShut {NoStop}%
\bibitem [{\citenamefont {Damour}\ and\ \citenamefont {Nagar}(2010)}]{Damour:2009wj}%
  \BibitemOpen
  \bibfield  {author} {\bibinfo {author} {\bibfnamefont {T.}~\bibnamefont {Damour}}\ and\ \bibinfo {author} {\bibfnamefont {A.}~\bibnamefont {Nagar}},\ }\bibfield  {title} {\bibinfo {title} {{Effective One Body description of tidal effects in inspiralling compact binaries}},\ }\href {https://doi.org/10.1103/PhysRevD.81.084016} {\bibfield  {journal} {\bibinfo  {journal} {Phys. Rev. D}\ }\textbf {\bibinfo {volume} {81}},\ \bibinfo {pages} {084016} (\bibinfo {year} {2010})},\ \Eprint {https://arxiv.org/abs/0911.5041} {arXiv:0911.5041 [gr-qc]} \BibitemShut {NoStop}%
\bibitem [{\citenamefont {Bernuzzi}\ \emph {et~al.}(2015)\citenamefont {Bernuzzi}, \citenamefont {Nagar}, \citenamefont {Dietrich},\ and\ \citenamefont {Damour}}]{Bernuzzi:2014owa}%
  \BibitemOpen
  \bibfield  {author} {\bibinfo {author} {\bibfnamefont {S.}~\bibnamefont {Bernuzzi}}, \bibinfo {author} {\bibfnamefont {A.}~\bibnamefont {Nagar}}, \bibinfo {author} {\bibfnamefont {T.}~\bibnamefont {Dietrich}},\ and\ \bibinfo {author} {\bibfnamefont {T.}~\bibnamefont {Damour}},\ }\bibfield  {title} {\bibinfo {title} {{Modeling the Dynamics of Tidally Interacting Binary Neutron Stars up to the Merger}},\ }\href {https://doi.org/10.1103/PhysRevLett.114.161103} {\bibfield  {journal} {\bibinfo  {journal} {Phys. Rev. Lett.}\ }\textbf {\bibinfo {volume} {114}},\ \bibinfo {pages} {161103} (\bibinfo {year} {2015})},\ \Eprint {https://arxiv.org/abs/1412.4553} {arXiv:1412.4553 [gr-qc]} \BibitemShut {NoStop}%
\bibitem [{\citenamefont {Nagar}\ \emph {et~al.}(2018)\citenamefont {Nagar} \emph {et~al.}}]{Nagar:2018zoe}%
  \BibitemOpen
  \bibfield  {author} {\bibinfo {author} {\bibfnamefont {A.}~\bibnamefont {Nagar}} \emph {et~al.},\ }\bibfield  {title} {\bibinfo {title} {{Time-domain effective-one-body gravitational waveforms for coalescing compact binaries with nonprecessing spins, tides and self-spin effects}},\ }\href {https://doi.org/10.1103/PhysRevD.98.104052} {\bibfield  {journal} {\bibinfo  {journal} {Phys. Rev. D}\ }\textbf {\bibinfo {volume} {98}},\ \bibinfo {pages} {104052} (\bibinfo {year} {2018})},\ \Eprint {https://arxiv.org/abs/1806.01772} {arXiv:1806.01772 [gr-qc]} \BibitemShut {NoStop}%
\bibitem [{\citenamefont {Akcay}\ \emph {et~al.}(2019)\citenamefont {Akcay}, \citenamefont {Bernuzzi}, \citenamefont {Messina}, \citenamefont {Nagar}, \citenamefont {Ortiz},\ and\ \citenamefont {Rettegno}}]{Akcay:2018yyh}%
  \BibitemOpen
  \bibfield  {author} {\bibinfo {author} {\bibfnamefont {S.}~\bibnamefont {Akcay}}, \bibinfo {author} {\bibfnamefont {S.}~\bibnamefont {Bernuzzi}}, \bibinfo {author} {\bibfnamefont {F.}~\bibnamefont {Messina}}, \bibinfo {author} {\bibfnamefont {A.}~\bibnamefont {Nagar}}, \bibinfo {author} {\bibfnamefont {N.}~\bibnamefont {Ortiz}},\ and\ \bibinfo {author} {\bibfnamefont {P.}~\bibnamefont {Rettegno}},\ }\bibfield  {title} {\bibinfo {title} {{Effective-one-body multipolar waveform for tidally interacting binary neutron stars up to merger}},\ }\href {https://doi.org/10.1103/PhysRevD.99.044051} {\bibfield  {journal} {\bibinfo  {journal} {Phys. Rev. D}\ }\textbf {\bibinfo {volume} {99}},\ \bibinfo {pages} {044051} (\bibinfo {year} {2019})},\ \Eprint {https://arxiv.org/abs/1812.02744} {arXiv:1812.02744 [gr-qc]} \BibitemShut {NoStop}%
\bibitem [{\citenamefont {Nagar}\ \emph {et~al.}(2019)\citenamefont {Nagar}, \citenamefont {Messina}, \citenamefont {Rettegno}, \citenamefont {Bini}, \citenamefont {Damour}, \citenamefont {Geralico}, \citenamefont {Akcay},\ and\ \citenamefont {Bernuzzi}}]{Nagar:2018plt}%
  \BibitemOpen
  \bibfield  {author} {\bibinfo {author} {\bibfnamefont {A.}~\bibnamefont {Nagar}}, \bibinfo {author} {\bibfnamefont {F.}~\bibnamefont {Messina}}, \bibinfo {author} {\bibfnamefont {P.}~\bibnamefont {Rettegno}}, \bibinfo {author} {\bibfnamefont {D.}~\bibnamefont {Bini}}, \bibinfo {author} {\bibfnamefont {T.}~\bibnamefont {Damour}}, \bibinfo {author} {\bibfnamefont {A.}~\bibnamefont {Geralico}}, \bibinfo {author} {\bibfnamefont {S.}~\bibnamefont {Akcay}},\ and\ \bibinfo {author} {\bibfnamefont {S.}~\bibnamefont {Bernuzzi}},\ }\bibfield  {title} {\bibinfo {title} {{Nonlinear-in-spin effects in effective-one-body waveform models of spin-aligned, inspiralling, neutron star binaries}},\ }\href {https://doi.org/10.1103/PhysRevD.99.044007} {\bibfield  {journal} {\bibinfo  {journal} {Phys. Rev. D}\ }\textbf {\bibinfo {volume} {99}},\ \bibinfo {pages} {044007} (\bibinfo {year} {2019})},\ \Eprint {https://arxiv.org/abs/1812.07923} {arXiv:1812.07923 [gr-qc]} \BibitemShut {NoStop}%
\bibitem [{\citenamefont {Gamba}\ \emph {et~al.}(2023)\citenamefont {Gamba} \emph {et~al.}}]{Gamba:2023mww}%
  \BibitemOpen
  \bibfield  {author} {\bibinfo {author} {\bibfnamefont {R.}~\bibnamefont {Gamba}} \emph {et~al.},\ }\bibfield  {title} {\bibinfo {title} {{Analytically improved and numerical-relativity informed effective-one-body model for coalescing binary neutron stars}},\ }\href@noop {} {\bibfield  {journal} {\bibinfo  {journal} {arXiv 2307.15125}\ } (\bibinfo {year} {2023})},\ \Eprint {https://arxiv.org/abs/2307.15125} {arXiv:2307.15125 [gr-qc]} \BibitemShut {NoStop}%
\bibitem [{\citenamefont {Steinhoff}\ \emph {et~al.}(2021)\citenamefont {Steinhoff}, \citenamefont {Hinderer}, \citenamefont {Dietrich},\ and\ \citenamefont {Foucart}}]{Steinhoff:2021dsn}%
  \BibitemOpen
  \bibfield  {author} {\bibinfo {author} {\bibfnamefont {J.}~\bibnamefont {Steinhoff}}, \bibinfo {author} {\bibfnamefont {T.}~\bibnamefont {Hinderer}}, \bibinfo {author} {\bibfnamefont {T.}~\bibnamefont {Dietrich}},\ and\ \bibinfo {author} {\bibfnamefont {F.}~\bibnamefont {Foucart}},\ }\bibfield  {title} {\bibinfo {title} {{Spin effects on neutron star fundamental-mode dynamical tides: Phenomenology and comparison to numerical simulations}},\ }\href {https://doi.org/10.1103/PhysRevResearch.3.033129} {\bibfield  {journal} {\bibinfo  {journal} {Phys. Rev. Res.}\ }\textbf {\bibinfo {volume} {3}},\ \bibinfo {pages} {033129} (\bibinfo {year} {2021})},\ \Eprint {https://arxiv.org/abs/2103.06100} {arXiv:2103.06100 [gr-qc]} \BibitemShut {NoStop}%
\bibitem [{\citenamefont {Haberland}\ \emph {et~al.}(2025)\citenamefont {Haberland}, \citenamefont {Buonanno},\ and\ \citenamefont {Steinhoff}}]{Haberland:2025luz}%
  \BibitemOpen
  \bibfield  {author} {\bibinfo {author} {\bibfnamefont {M.}~\bibnamefont {Haberland}}, \bibinfo {author} {\bibfnamefont {A.}~\bibnamefont {Buonanno}},\ and\ \bibinfo {author} {\bibfnamefont {J.}~\bibnamefont {Steinhoff}},\ }\bibfield  {title} {\bibinfo {title} {{Modeling matter(s) in SEOBNRv5THM: Generating fast and accurate effective-one-body waveforms for spin-aligned binary neutron stars}},\ }\href@noop {} {\bibfield  {journal} {\bibinfo  {journal} {arXiv:2503.18934}\ } (\bibinfo {year} {2025})},\ \Eprint {https://arxiv.org/abs/2503.18934} {arXiv:2503.18934 [gr-qc]} \BibitemShut {NoStop}%
\bibitem [{\citenamefont {Abdelsalhin}\ \emph {et~al.}(2018)\citenamefont {Abdelsalhin}, \citenamefont {Gualtieri},\ and\ \citenamefont {Pani}}]{Abdelsalhin:2018reg}%
  \BibitemOpen
  \bibfield  {author} {\bibinfo {author} {\bibfnamefont {T.}~\bibnamefont {Abdelsalhin}}, \bibinfo {author} {\bibfnamefont {L.}~\bibnamefont {Gualtieri}},\ and\ \bibinfo {author} {\bibfnamefont {P.}~\bibnamefont {Pani}},\ }\bibfield  {title} {\bibinfo {title} {{Post-Newtonian spin-tidal couplings for compact binaries}},\ }\href {https://doi.org/10.1103/PhysRevD.98.104046} {\bibfield  {journal} {\bibinfo  {journal} {Phys. Rev. D}\ }\textbf {\bibinfo {volume} {98}},\ \bibinfo {pages} {104046} (\bibinfo {year} {2018})},\ \Eprint {https://arxiv.org/abs/1805.01487} {arXiv:1805.01487 [gr-qc]} \BibitemShut {NoStop}%
\bibitem [{\citenamefont {Castro}\ \emph {et~al.}(2022)\citenamefont {Castro}, \citenamefont {Gualtieri}, \citenamefont {Maselli},\ and\ \citenamefont {Pani}}]{Castro:2022mpw}%
  \BibitemOpen
  \bibfield  {author} {\bibinfo {author} {\bibfnamefont {G.}~\bibnamefont {Castro}}, \bibinfo {author} {\bibfnamefont {L.}~\bibnamefont {Gualtieri}}, \bibinfo {author} {\bibfnamefont {A.}~\bibnamefont {Maselli}},\ and\ \bibinfo {author} {\bibfnamefont {P.}~\bibnamefont {Pani}},\ }\bibfield  {title} {\bibinfo {title} {{Impact and detectability of spin-tidal couplings in neutron star inspirals}},\ }\href {https://doi.org/10.1103/PhysRevD.106.024011} {\bibfield  {journal} {\bibinfo  {journal} {Phys. Rev. D}\ }\textbf {\bibinfo {volume} {106}},\ \bibinfo {pages} {024011} (\bibinfo {year} {2022})},\ \Eprint {https://arxiv.org/abs/2204.12510} {arXiv:2204.12510 [gr-qc]} \BibitemShut {NoStop}%
\bibitem [{\citenamefont {Lai}(1997)}]{Lai:1997wh}%
  \BibitemOpen
  \bibfield  {author} {\bibinfo {author} {\bibfnamefont {D.}~\bibnamefont {Lai}},\ }\bibfield  {title} {\bibinfo {title} {{Dynamical tides in rotating binary stars}},\ }\href {https://doi.org/10.1086/304899} {\bibfield  {journal} {\bibinfo  {journal} {Astrophys. J.}\ }\textbf {\bibinfo {volume} {490}},\ \bibinfo {pages} {847} (\bibinfo {year} {1997})},\ \Eprint {https://arxiv.org/abs/astro-ph/9704132} {arXiv:astro-ph/9704132} \BibitemShut {NoStop}%
\bibitem [{\citenamefont {Kr\"uger}\ and\ \citenamefont {Kokkotas}(2020)}]{Kruger:2019zuz}%
  \BibitemOpen
  \bibfield  {author} {\bibinfo {author} {\bibfnamefont {C.~J.}\ \bibnamefont {Kr\"uger}}\ and\ \bibinfo {author} {\bibfnamefont {K.~D.}\ \bibnamefont {Kokkotas}},\ }\bibfield  {title} {\bibinfo {title} {{Fast Rotating Relativistic Stars: Spectra and Stability without Approximation}},\ }\href {https://doi.org/10.1103/PhysRevLett.125.111106} {\bibfield  {journal} {\bibinfo  {journal} {Phys. Rev. Lett.}\ }\textbf {\bibinfo {volume} {125}},\ \bibinfo {pages} {111106} (\bibinfo {year} {2020})},\ \Eprint {https://arxiv.org/abs/1910.08370} {arXiv:1910.08370 [gr-qc]} \BibitemShut {NoStop}%
\bibitem [{\citenamefont {Kr\"uger}\ \emph {et~al.}(2021)\citenamefont {Kr\"uger}, \citenamefont {Kokkotas}, \citenamefont {Manoharan},\ and\ \citenamefont {V\"olkel}}]{Kruger:2021zta}%
  \BibitemOpen
  \bibfield  {author} {\bibinfo {author} {\bibfnamefont {C.~J.}\ \bibnamefont {Kr\"uger}}, \bibinfo {author} {\bibfnamefont {K.~D.}\ \bibnamefont {Kokkotas}}, \bibinfo {author} {\bibfnamefont {P.}~\bibnamefont {Manoharan}},\ and\ \bibinfo {author} {\bibfnamefont {S.~H.}\ \bibnamefont {V\"olkel}},\ }\bibfield  {title} {\bibinfo {title} {{Fast Rotating Neutron Stars: Oscillations and Instabilities}},\ }\href {https://doi.org/10.3389/fspas.2021.736918} {\bibfield  {journal} {\bibinfo  {journal} {Front. Astron. Space Sci.}\ }\textbf {\bibinfo {volume} {8}},\ \bibinfo {pages} {736918} (\bibinfo {year} {2021})},\ \Eprint {https://arxiv.org/abs/2110.00393} {arXiv:2110.00393 [gr-qc]} \BibitemShut {NoStop}%
\bibitem [{\citenamefont {{Dewberry}}\ and\ \citenamefont {{Lai}}(2022)}]{Dewberry:2022}%
  \BibitemOpen
  \bibfield  {author} {\bibinfo {author} {\bibfnamefont {J.~W.}\ \bibnamefont {{Dewberry}}}\ and\ \bibinfo {author} {\bibfnamefont {D.}~\bibnamefont {{Lai}}},\ }\bibfield  {title} {\bibinfo {title} {{Dynamical Tidal Love Numbers of Rapidly Rotating Planets and Stars}},\ }\href {https://doi.org/10.3847/1538-4357/ac3ede} {\bibfield  {journal} {\bibinfo  {journal} {\apj}\ }\textbf {\bibinfo {volume} {925}},\ \bibinfo {eid} {124} (\bibinfo {year} {2022})},\ \Eprint {https://arxiv.org/abs/2110.12129} {arXiv:2110.12129 [astro-ph.EP]} \BibitemShut {NoStop}%
\bibitem [{\citenamefont {Kuan}\ \emph {et~al.}(2024)\citenamefont {Kuan}, \citenamefont {Kiuchi},\ and\ \citenamefont {Shibata}}]{Kuan:2024jnw}%
  \BibitemOpen
  \bibfield  {author} {\bibinfo {author} {\bibfnamefont {H.-J.}\ \bibnamefont {Kuan}}, \bibinfo {author} {\bibfnamefont {K.}~\bibnamefont {Kiuchi}},\ and\ \bibinfo {author} {\bibfnamefont {M.}~\bibnamefont {Shibata}},\ }\bibfield  {title} {\bibinfo {title} {{Tidal Resonance in Binary Neutron Star Inspirals: A High-Precision Study in Numerical Relativity}},\ }\href@noop {} {\bibfield  {journal} {\bibinfo  {journal} {arXiv:2411.16850}\ } (\bibinfo {year} {2024})},\ \Eprint {https://arxiv.org/abs/2411.16850} {arXiv:2411.16850 [hep-ph]} \BibitemShut {NoStop}%
\bibitem [{\citenamefont {Yu}\ \emph {et~al.}(2023)\citenamefont {Yu}, \citenamefont {Weinberg}, \citenamefont {Arras}, \citenamefont {Kwon},\ and\ \citenamefont {Venumadhav}}]{Yu:2022fzw}%
  \BibitemOpen
  \bibfield  {author} {\bibinfo {author} {\bibfnamefont {H.}~\bibnamefont {Yu}}, \bibinfo {author} {\bibfnamefont {N.~N.}\ \bibnamefont {Weinberg}}, \bibinfo {author} {\bibfnamefont {P.}~\bibnamefont {Arras}}, \bibinfo {author} {\bibfnamefont {J.}~\bibnamefont {Kwon}},\ and\ \bibinfo {author} {\bibfnamefont {T.}~\bibnamefont {Venumadhav}},\ }\bibfield  {title} {\bibinfo {title} {{Beyond the linear tide: impact of the non-linear tidal response of neutron stars on gravitational waveforms from binary inspirals}},\ }\href {https://doi.org/10.1093/mnras/stac3614} {\bibfield  {journal} {\bibinfo  {journal} {Mon. Not. Roy. Astron. Soc.}\ }\textbf {\bibinfo {volume} {519}},\ \bibinfo {pages} {4325} (\bibinfo {year} {2023})},\ \Eprint {https://arxiv.org/abs/2211.07002} {arXiv:2211.07002 [gr-qc]} \BibitemShut {NoStop}%
\bibitem [{\citenamefont {Yu}\ and\ \citenamefont {Lau}(2025)}]{Yu:2025ptm}%
  \BibitemOpen
  \bibfield  {author} {\bibinfo {author} {\bibfnamefont {H.}~\bibnamefont {Yu}}\ and\ \bibinfo {author} {\bibfnamefont {S.~Y.}\ \bibnamefont {Lau}},\ }\bibfield  {title} {\bibinfo {title} {{Effective-one-body model for coalescing binary neutron stars: Incorporating tidal spin and enhanced radiation from dynamical tides}},\ }\href@noop {} {\bibfield  {journal} {\bibinfo  {journal} {Arxiv}\ } (\bibinfo {year} {2025})},\ \Eprint {https://arxiv.org/abs/2501.13064} {arXiv:2501.13064 [gr-qc]} \BibitemShut {NoStop}%
\bibitem [{\citenamefont {Ryan}(1995)}]{Ryan:1995wh}%
  \BibitemOpen
  \bibfield  {author} {\bibinfo {author} {\bibfnamefont {F.~D.}\ \bibnamefont {Ryan}},\ }\bibfield  {title} {\bibinfo {title} {{Gravitational waves from the inspiral of a compact object into a massive, axisymmetric body with arbitrary multipole moments}},\ }\href {https://doi.org/10.1103/PhysRevD.52.5707} {\bibfield  {journal} {\bibinfo  {journal} {Phys. Rev. D}\ }\textbf {\bibinfo {volume} {52}},\ \bibinfo {pages} {5707} (\bibinfo {year} {1995})}\BibitemShut {NoStop}%
\bibitem [{\citenamefont {Mandal}\ \emph {et~al.}(2023{\natexlab{b}})\citenamefont {Mandal}, \citenamefont {Mastrolia}, \citenamefont {Patil},\ and\ \citenamefont {Steinhoff}}]{Mandal:2022nty}%
  \BibitemOpen
  \bibfield  {author} {\bibinfo {author} {\bibfnamefont {M.~K.}\ \bibnamefont {Mandal}}, \bibinfo {author} {\bibfnamefont {P.}~\bibnamefont {Mastrolia}}, \bibinfo {author} {\bibfnamefont {R.}~\bibnamefont {Patil}},\ and\ \bibinfo {author} {\bibfnamefont {J.}~\bibnamefont {Steinhoff}},\ }\bibfield  {title} {\bibinfo {title} {{Gravitational spin-orbit Hamiltonian at NNNLO in the post-Newtonian framework}},\ }\href {https://doi.org/10.1007/JHEP03(2023)130} {\bibfield  {journal} {\bibinfo  {journal} {JHEP}\ }\textbf {\bibinfo {volume} {03}},\ \bibinfo {pages} {130}},\ \Eprint {https://arxiv.org/abs/2209.00611} {arXiv:2209.00611 [hep-th]} \BibitemShut {NoStop}%
\bibitem [{\citenamefont {Mandal}\ \emph {et~al.}(2023{\natexlab{c}})\citenamefont {Mandal}, \citenamefont {Mastrolia}, \citenamefont {Patil},\ and\ \citenamefont {Steinhoff}}]{Mandal:2022ufb}%
  \BibitemOpen
  \bibfield  {author} {\bibinfo {author} {\bibfnamefont {M.~K.}\ \bibnamefont {Mandal}}, \bibinfo {author} {\bibfnamefont {P.}~\bibnamefont {Mastrolia}}, \bibinfo {author} {\bibfnamefont {R.}~\bibnamefont {Patil}},\ and\ \bibinfo {author} {\bibfnamefont {J.}~\bibnamefont {Steinhoff}},\ }\bibfield  {title} {\bibinfo {title} {{Gravitational quadratic-in-spin Hamiltonian at NNNLO in the post-Newtonian framework}},\ }\href {https://doi.org/10.1007/JHEP07(2023)128} {\bibfield  {journal} {\bibinfo  {journal} {JHEP}\ }\textbf {\bibinfo {volume} {07}},\ \bibinfo {pages} {128}},\ \Eprint {https://arxiv.org/abs/2210.09176} {arXiv:2210.09176 [hep-th]} \BibitemShut {NoStop}%
\bibitem [{\citenamefont {Yu}\ \emph {et~al.}(2024)\citenamefont {Yu}, \citenamefont {Arras},\ and\ \citenamefont {Weinberg}}]{Yu:2024uxt}%
  \BibitemOpen
  \bibfield  {author} {\bibinfo {author} {\bibfnamefont {H.}~\bibnamefont {Yu}}, \bibinfo {author} {\bibfnamefont {P.}~\bibnamefont {Arras}},\ and\ \bibinfo {author} {\bibfnamefont {N.~N.}\ \bibnamefont {Weinberg}},\ }\bibfield  {title} {\bibinfo {title} {{Dynamical tides during the inspiral of rapidly spinning neutron stars: Solutions beyond mode resonance}},\ }\href@noop {} {\bibfield  {journal} {\bibinfo  {journal} {Arxiv}\ } (\bibinfo {year} {2024})},\ \Eprint {https://arxiv.org/abs/2404.00147} {arXiv:2404.00147 [gr-qc]} \BibitemShut {NoStop}%
\bibitem [{\citenamefont {Schmidt}\ and\ \citenamefont {Hinderer}(2019)}]{Schmidt:2019wrl}%
  \BibitemOpen
  \bibfield  {author} {\bibinfo {author} {\bibfnamefont {P.}~\bibnamefont {Schmidt}}\ and\ \bibinfo {author} {\bibfnamefont {T.}~\bibnamefont {Hinderer}},\ }\bibfield  {title} {\bibinfo {title} {{Frequency domain model of $f$-mode dynamic tides in gravitational waveforms from compact binary inspirals}},\ }\href {https://doi.org/10.1103/PhysRevD.100.021501} {\bibfield  {journal} {\bibinfo  {journal} {Phys. Rev. D}\ }\textbf {\bibinfo {volume} {100}},\ \bibinfo {pages} {021501} (\bibinfo {year} {2019})},\ \Eprint {https://arxiv.org/abs/1905.00818} {arXiv:1905.00818 [gr-qc]} \BibitemShut {NoStop}%
\bibitem [{\citenamefont {Baiotti}\ \emph {et~al.}(2011)\citenamefont {Baiotti}, \citenamefont {Damour}, \citenamefont {Giacomazzo}, \citenamefont {Nagar},\ and\ \citenamefont {Rezzolla}}]{Baiotti:2011am}%
  \BibitemOpen
  \bibfield  {author} {\bibinfo {author} {\bibfnamefont {L.}~\bibnamefont {Baiotti}}, \bibinfo {author} {\bibfnamefont {T.}~\bibnamefont {Damour}}, \bibinfo {author} {\bibfnamefont {B.}~\bibnamefont {Giacomazzo}}, \bibinfo {author} {\bibfnamefont {A.}~\bibnamefont {Nagar}},\ and\ \bibinfo {author} {\bibfnamefont {L.}~\bibnamefont {Rezzolla}},\ }\bibfield  {title} {\bibinfo {title} {{Accurate numerical simulations of inspiralling binary neutron stars and their comparison with effective-one-body analytical models}},\ }\href {https://doi.org/10.1103/PhysRevD.84.024017} {\bibfield  {journal} {\bibinfo  {journal} {Phys. Rev. D}\ }\textbf {\bibinfo {volume} {84}},\ \bibinfo {pages} {024017} (\bibinfo {year} {2011})},\ \Eprint {https://arxiv.org/abs/1103.3874} {arXiv:1103.3874 [gr-qc]} \BibitemShut {NoStop}%
\bibitem [{\citenamefont {Bernuzzi}\ \emph {et~al.}(2012{\natexlab{a}})\citenamefont {Bernuzzi}, \citenamefont {Nagar}, \citenamefont {Thierfelder},\ and\ \citenamefont {Brugmann}}]{Bernuzzi:2012ci}%
  \BibitemOpen
  \bibfield  {author} {\bibinfo {author} {\bibfnamefont {S.}~\bibnamefont {Bernuzzi}}, \bibinfo {author} {\bibfnamefont {A.}~\bibnamefont {Nagar}}, \bibinfo {author} {\bibfnamefont {M.}~\bibnamefont {Thierfelder}},\ and\ \bibinfo {author} {\bibfnamefont {B.}~\bibnamefont {Brugmann}},\ }\bibfield  {title} {\bibinfo {title} {{Tidal effects in binary neutron star coalescence}},\ }\href {https://doi.org/10.1103/PhysRevD.86.044030} {\bibfield  {journal} {\bibinfo  {journal} {Phys. Rev. D}\ }\textbf {\bibinfo {volume} {86}},\ \bibinfo {pages} {044030} (\bibinfo {year} {2012}{\natexlab{a}})},\ \Eprint {https://arxiv.org/abs/1205.3403} {arXiv:1205.3403 [gr-qc]} \BibitemShut {NoStop}%
\bibitem [{\citenamefont {Hotokezaka}\ \emph {et~al.}(2013)\citenamefont {Hotokezaka}, \citenamefont {Kyutoku},\ and\ \citenamefont {Shibata}}]{Hotokezaka:2013mm}%
  \BibitemOpen
  \bibfield  {author} {\bibinfo {author} {\bibfnamefont {K.}~\bibnamefont {Hotokezaka}}, \bibinfo {author} {\bibfnamefont {K.}~\bibnamefont {Kyutoku}},\ and\ \bibinfo {author} {\bibfnamefont {M.}~\bibnamefont {Shibata}},\ }\bibfield  {title} {\bibinfo {title} {{Exploring tidal effects of coalescing binary neutron stars in numerical relativity}},\ }\href {https://doi.org/10.1103/PhysRevD.87.044001} {\bibfield  {journal} {\bibinfo  {journal} {Phys. Rev. D}\ }\textbf {\bibinfo {volume} {87}},\ \bibinfo {pages} {044001} (\bibinfo {year} {2013})},\ \Eprint {https://arxiv.org/abs/1301.3555} {arXiv:1301.3555 [gr-qc]} \BibitemShut {NoStop}%
\bibitem [{\citenamefont {Mroue}\ \emph {et~al.}(2013)\citenamefont {Mroue} \emph {et~al.}}]{Mroue:2013xna}%
  \BibitemOpen
  \bibfield  {author} {\bibinfo {author} {\bibfnamefont {A.~H.}\ \bibnamefont {Mroue}} \emph {et~al.},\ }\bibfield  {title} {\bibinfo {title} {{Catalog of 174 Binary Black Hole Simulations for Gravitational Wave Astronomy}},\ }\href {https://doi.org/10.1103/PhysRevLett.111.241104} {\bibfield  {journal} {\bibinfo  {journal} {Phys. Rev. Lett.}\ }\textbf {\bibinfo {volume} {111}},\ \bibinfo {pages} {241104} (\bibinfo {year} {2013})},\ \Eprint {https://arxiv.org/abs/1304.6077} {arXiv:1304.6077 [gr-qc]} \BibitemShut {NoStop}%
\bibitem [{\citenamefont {Boyle}\ \emph {et~al.}(2019)\citenamefont {Boyle} \emph {et~al.}}]{Boyle:2019kee}%
  \BibitemOpen
  \bibfield  {author} {\bibinfo {author} {\bibfnamefont {M.}~\bibnamefont {Boyle}} \emph {et~al.},\ }\bibfield  {title} {\bibinfo {title} {{The SXS Collaboration catalog of binary black hole simulations}},\ }\href {https://doi.org/10.1088/1361-6382/ab34e2} {\bibfield  {journal} {\bibinfo  {journal} {Class. Quant. Grav.}\ }\textbf {\bibinfo {volume} {36}},\ \bibinfo {pages} {195006} (\bibinfo {year} {2019})},\ \Eprint {https://arxiv.org/abs/1904.04831} {arXiv:1904.04831 [gr-qc]} \BibitemShut {NoStop}%
\bibitem [{\citenamefont {Dietrich}\ \emph {et~al.}(2018{\natexlab{a}})\citenamefont {Dietrich}, \citenamefont {Radice}, \citenamefont {Bernuzzi}, \citenamefont {Zappa}, \citenamefont {Perego}, \citenamefont {Br\"ugmann}, \citenamefont {Chaurasia}, \citenamefont {Dudi}, \citenamefont {Tichy},\ and\ \citenamefont {Ujevic}}]{Dietrich:2018phi}%
  \BibitemOpen
  \bibfield  {author} {\bibinfo {author} {\bibfnamefont {T.}~\bibnamefont {Dietrich}}, \bibinfo {author} {\bibfnamefont {D.}~\bibnamefont {Radice}}, \bibinfo {author} {\bibfnamefont {S.}~\bibnamefont {Bernuzzi}}, \bibinfo {author} {\bibfnamefont {F.}~\bibnamefont {Zappa}}, \bibinfo {author} {\bibfnamefont {A.}~\bibnamefont {Perego}}, \bibinfo {author} {\bibfnamefont {B.}~\bibnamefont {Br\"ugmann}}, \bibinfo {author} {\bibfnamefont {S.~V.}\ \bibnamefont {Chaurasia}}, \bibinfo {author} {\bibfnamefont {R.}~\bibnamefont {Dudi}}, \bibinfo {author} {\bibfnamefont {W.}~\bibnamefont {Tichy}},\ and\ \bibinfo {author} {\bibfnamefont {M.}~\bibnamefont {Ujevic}},\ }\bibfield  {title} {\bibinfo {title} {{CoRe database of binary neutron star merger waveforms}},\ }\href {https://doi.org/10.1088/1361-6382/aaebc0} {\bibfield  {journal} {\bibinfo  {journal} {Class. Quant. Grav.}\ }\textbf {\bibinfo {volume} {35}},\ \bibinfo {pages} {24LT01} (\bibinfo {year} {2018}{\natexlab{a}})},\ \Eprint {https://arxiv.org/abs/1806.01625}
  {arXiv:1806.01625 [gr-qc]} \BibitemShut {NoStop}%
\bibitem [{\citenamefont {Gonzalez}\ \emph {et~al.}(2023)\citenamefont {Gonzalez} \emph {et~al.}}]{Gonzalez:2022mgo}%
  \BibitemOpen
  \bibfield  {author} {\bibinfo {author} {\bibfnamefont {A.}~\bibnamefont {Gonzalez}} \emph {et~al.},\ }\bibfield  {title} {\bibinfo {title} {{Second release of the CoRe database of binary neutron star merger waveforms}},\ }\href {https://doi.org/10.1088/1361-6382/acc231} {\bibfield  {journal} {\bibinfo  {journal} {Class. Quant. Grav.}\ }\textbf {\bibinfo {volume} {40}},\ \bibinfo {pages} {085011} (\bibinfo {year} {2023})},\ \Eprint {https://arxiv.org/abs/2210.16366} {arXiv:2210.16366 [gr-qc]} \BibitemShut {NoStop}%
\bibitem [{\citenamefont {Kiuchi}\ \emph {et~al.}(2017)\citenamefont {Kiuchi}, \citenamefont {Kawaguchi}, \citenamefont {Kyutoku}, \citenamefont {Sekiguchi}, \citenamefont {Shibata},\ and\ \citenamefont {Taniguchi}}]{Kiuchi:2017pte}%
  \BibitemOpen
  \bibfield  {author} {\bibinfo {author} {\bibfnamefont {K.}~\bibnamefont {Kiuchi}}, \bibinfo {author} {\bibfnamefont {K.}~\bibnamefont {Kawaguchi}}, \bibinfo {author} {\bibfnamefont {K.}~\bibnamefont {Kyutoku}}, \bibinfo {author} {\bibfnamefont {Y.}~\bibnamefont {Sekiguchi}}, \bibinfo {author} {\bibfnamefont {M.}~\bibnamefont {Shibata}},\ and\ \bibinfo {author} {\bibfnamefont {K.}~\bibnamefont {Taniguchi}},\ }\bibfield  {title} {\bibinfo {title} {{Sub-radian-accuracy gravitational waveforms of coalescing binary neutron stars in numerical relativity}},\ }\href {https://doi.org/10.1103/PhysRevD.96.084060} {\bibfield  {journal} {\bibinfo  {journal} {Phys. Rev. D}\ }\textbf {\bibinfo {volume} {96}},\ \bibinfo {pages} {084060} (\bibinfo {year} {2017})},\ \Eprint {https://arxiv.org/abs/1708.08926} {arXiv:1708.08926 [astro-ph.HE]} \BibitemShut {NoStop}%
\bibitem [{\citenamefont {Kiuchi}\ \emph {et~al.}(2020)\citenamefont {Kiuchi}, \citenamefont {Kawaguchi}, \citenamefont {Kyutoku}, \citenamefont {Sekiguchi},\ and\ \citenamefont {Shibata}}]{Kiuchi:2019kzt}%
  \BibitemOpen
  \bibfield  {author} {\bibinfo {author} {\bibfnamefont {K.}~\bibnamefont {Kiuchi}}, \bibinfo {author} {\bibfnamefont {K.}~\bibnamefont {Kawaguchi}}, \bibinfo {author} {\bibfnamefont {K.}~\bibnamefont {Kyutoku}}, \bibinfo {author} {\bibfnamefont {Y.}~\bibnamefont {Sekiguchi}},\ and\ \bibinfo {author} {\bibfnamefont {M.}~\bibnamefont {Shibata}},\ }\bibfield  {title} {\bibinfo {title} {{Sub-radian-accuracy gravitational waves from coalescing binary neutron stars in numerical relativity. II. Systematic study on the equation of state, binary mass, and mass ratio}},\ }\href {https://doi.org/10.1103/PhysRevD.101.084006} {\bibfield  {journal} {\bibinfo  {journal} {Phys. Rev. D}\ }\textbf {\bibinfo {volume} {101}},\ \bibinfo {pages} {084006} (\bibinfo {year} {2020})},\ \Eprint {https://arxiv.org/abs/1907.03790} {arXiv:1907.03790 [astro-ph.HE]} \BibitemShut {NoStop}%
\bibitem [{\citenamefont {Radice}\ \emph {et~al.}(2014)\citenamefont {Radice}, \citenamefont {Rezzolla},\ and\ \citenamefont {Galeazzi}}]{Radice:2013hxh}%
  \BibitemOpen
  \bibfield  {author} {\bibinfo {author} {\bibfnamefont {D.}~\bibnamefont {Radice}}, \bibinfo {author} {\bibfnamefont {L.}~\bibnamefont {Rezzolla}},\ and\ \bibinfo {author} {\bibfnamefont {F.}~\bibnamefont {Galeazzi}},\ }\bibfield  {title} {\bibinfo {title} {{Beyond second-order convergence in simulations of binary neutron stars in full general-relativity}},\ }\href {https://doi.org/10.1093/mnrasl/slt137} {\bibfield  {journal} {\bibinfo  {journal} {Mon. Not. Roy. Astron. Soc.}\ }\textbf {\bibinfo {volume} {437}},\ \bibinfo {pages} {L46} (\bibinfo {year} {2014})},\ \Eprint {https://arxiv.org/abs/1306.6052} {arXiv:1306.6052 [gr-qc]} \BibitemShut {NoStop}%
\bibitem [{\citenamefont {Foucart}\ \emph {et~al.}(2019)\citenamefont {Foucart} \emph {et~al.}}]{Foucart:2018lhe}%
  \BibitemOpen
  \bibfield  {author} {\bibinfo {author} {\bibfnamefont {F.}~\bibnamefont {Foucart}} \emph {et~al.},\ }\bibfield  {title} {\bibinfo {title} {{Gravitational waveforms from spectral Einstein code simulations: Neutron star-neutron star and low-mass black hole-neutron star binaries}},\ }\href {https://doi.org/10.1103/PhysRevD.99.044008} {\bibfield  {journal} {\bibinfo  {journal} {Phys. Rev. D}\ }\textbf {\bibinfo {volume} {99}},\ \bibinfo {pages} {044008} (\bibinfo {year} {2019})},\ \Eprint {https://arxiv.org/abs/1812.06988} {arXiv:1812.06988 [gr-qc]} \BibitemShut {NoStop}%
\bibitem [{\citenamefont {Baiotti}\ \emph {et~al.}(2010{\natexlab{a}})\citenamefont {Baiotti}, \citenamefont {Shibata},\ and\ \citenamefont {Yamamoto}}]{Baiotti:2010ka}%
  \BibitemOpen
  \bibfield  {author} {\bibinfo {author} {\bibfnamefont {L.}~\bibnamefont {Baiotti}}, \bibinfo {author} {\bibfnamefont {M.}~\bibnamefont {Shibata}},\ and\ \bibinfo {author} {\bibfnamefont {T.}~\bibnamefont {Yamamoto}},\ }\bibfield  {title} {\bibinfo {title} {{Binary neutron-star mergers with Whisky and SACRA: First quantitative comparison of results from independent general-relativistic hydrodynamics codes}},\ }\href {https://doi.org/10.1103/PhysRevD.82.064015} {\bibfield  {journal} {\bibinfo  {journal} {Phys. Rev. D}\ }\textbf {\bibinfo {volume} {82}},\ \bibinfo {pages} {064015} (\bibinfo {year} {2010}{\natexlab{a}})},\ \Eprint {https://arxiv.org/abs/1007.1754} {arXiv:1007.1754 [gr-qc]} \BibitemShut {NoStop}%
\bibitem [{\citenamefont {Gourgoulhon}\ \emph {et~al.}(2001)\citenamefont {Gourgoulhon}, \citenamefont {Grandclement}, \citenamefont {Taniguchi}, \citenamefont {Marck},\ and\ \citenamefont {Bonazzola}}]{Gourgoulhon:2000nn}%
  \BibitemOpen
  \bibfield  {author} {\bibinfo {author} {\bibfnamefont {E.}~\bibnamefont {Gourgoulhon}}, \bibinfo {author} {\bibfnamefont {P.}~\bibnamefont {Grandclement}}, \bibinfo {author} {\bibfnamefont {K.}~\bibnamefont {Taniguchi}}, \bibinfo {author} {\bibfnamefont {J.-A.}\ \bibnamefont {Marck}},\ and\ \bibinfo {author} {\bibfnamefont {S.}~\bibnamefont {Bonazzola}},\ }\bibfield  {title} {\bibinfo {title} {{Quasiequilibrium sequences of synchronized and irrotational binary neutron stars in general relativity: 1. Method and tests}},\ }\href {https://doi.org/10.1103/PhysRevD.63.064029} {\bibfield  {journal} {\bibinfo  {journal} {Phys. Rev. D}\ }\textbf {\bibinfo {volume} {63}},\ \bibinfo {pages} {064029} (\bibinfo {year} {2001})},\ \Eprint {https://arxiv.org/abs/gr-qc/0007028} {arXiv:gr-qc/0007028} \BibitemShut {NoStop}%
\bibitem [{\citenamefont {Taniguchi}\ \emph {et~al.}(2001)\citenamefont {Taniguchi}, \citenamefont {Gourgoulhon},\ and\ \citenamefont {Bonazzola}}]{Taniguchi:2001qv}%
  \BibitemOpen
  \bibfield  {author} {\bibinfo {author} {\bibfnamefont {K.}~\bibnamefont {Taniguchi}}, \bibinfo {author} {\bibfnamefont {E.}~\bibnamefont {Gourgoulhon}},\ and\ \bibinfo {author} {\bibfnamefont {S.}~\bibnamefont {Bonazzola}},\ }\bibfield  {title} {\bibinfo {title} {{Quasiequilibrium sequences of synchronized and irrotational binary neutron stars in general relativity. 2. Newtonian limits}},\ }\href {https://doi.org/10.1103/PhysRevD.64.064012} {\bibfield  {journal} {\bibinfo  {journal} {Phys. Rev. D}\ }\textbf {\bibinfo {volume} {64}},\ \bibinfo {pages} {064012} (\bibinfo {year} {2001})},\ \Eprint {https://arxiv.org/abs/gr-qc/0103041} {arXiv:gr-qc/0103041} \BibitemShut {NoStop}%
\bibitem [{\citenamefont {Hamilton}\ and\ \citenamefont {Messman}(2024)}]{Hamilton:2024ziw}%
  \BibitemOpen
  \bibfield  {author} {\bibinfo {author} {\bibfnamefont {M.~C.~B.}\ \bibnamefont {Hamilton}}\ and\ \bibinfo {author} {\bibfnamefont {W.~A.}\ \bibnamefont {Messman}},\ }\bibfield  {title} {\bibinfo {title} {{Insights into Binary Neutron Star Merger Simulations: A Multi-Code Comparison}},\ }\href@noop {} {\bibfield  {journal} {\bibinfo  {journal} {arXiv:2411.10552}\ } (\bibinfo {year} {2024})},\ \Eprint {https://arxiv.org/abs/2411.10552} {arXiv:2411.10552 [gr-qc]} \BibitemShut {NoStop}%
\bibitem [{\citenamefont {Dietrich}\ \emph {et~al.}(2015{\natexlab{a}})\citenamefont {Dietrich}, \citenamefont {Moldenhauer}, \citenamefont {Johnson-McDaniel}, \citenamefont {Bernuzzi}, \citenamefont {Markakis}, \citenamefont {Br\"ugmann},\ and\ \citenamefont {Tichy}}]{Dietrich:2015pxa}%
  \BibitemOpen
  \bibfield  {author} {\bibinfo {author} {\bibfnamefont {T.}~\bibnamefont {Dietrich}}, \bibinfo {author} {\bibfnamefont {N.}~\bibnamefont {Moldenhauer}}, \bibinfo {author} {\bibfnamefont {N.~K.}\ \bibnamefont {Johnson-McDaniel}}, \bibinfo {author} {\bibfnamefont {S.}~\bibnamefont {Bernuzzi}}, \bibinfo {author} {\bibfnamefont {C.~M.}\ \bibnamefont {Markakis}}, \bibinfo {author} {\bibfnamefont {B.}~\bibnamefont {Br\"ugmann}},\ and\ \bibinfo {author} {\bibfnamefont {W.}~\bibnamefont {Tichy}},\ }\bibfield  {title} {\bibinfo {title} {{Binary Neutron Stars with Generic Spin, Eccentricity, Mass ratio, and Compactness - Quasi-equilibrium Sequences and First Evolutions}},\ }\href {https://doi.org/10.1103/PhysRevD.92.124007} {\bibfield  {journal} {\bibinfo  {journal} {Phys. Rev. D}\ }\textbf {\bibinfo {volume} {92}},\ \bibinfo {pages} {124007} (\bibinfo {year} {2015}{\natexlab{a}})},\ \Eprint {https://arxiv.org/abs/1507.07100} {arXiv:1507.07100 [gr-qc]} \BibitemShut {NoStop}%
\bibitem [{\citenamefont {Dietrich}\ \emph {et~al.}(2017{\natexlab{b}})\citenamefont {Dietrich}, \citenamefont {Bernuzzi}, \citenamefont {Ujevic},\ and\ \citenamefont {Tichy}}]{Dietrich:2016lyp}%
  \BibitemOpen
  \bibfield  {author} {\bibinfo {author} {\bibfnamefont {T.}~\bibnamefont {Dietrich}}, \bibinfo {author} {\bibfnamefont {S.}~\bibnamefont {Bernuzzi}}, \bibinfo {author} {\bibfnamefont {M.}~\bibnamefont {Ujevic}},\ and\ \bibinfo {author} {\bibfnamefont {W.}~\bibnamefont {Tichy}},\ }\bibfield  {title} {\bibinfo {title} {{Gravitational waves and mass ejecta from binary neutron star mergers: Effect of the stars' rotation}},\ }\href {https://doi.org/10.1103/PhysRevD.95.044045} {\bibfield  {journal} {\bibinfo  {journal} {Phys. Rev. D}\ }\textbf {\bibinfo {volume} {95}},\ \bibinfo {pages} {044045} (\bibinfo {year} {2017}{\natexlab{b}})},\ \Eprint {https://arxiv.org/abs/1611.07367} {arXiv:1611.07367 [gr-qc]} \BibitemShut {NoStop}%
\bibitem [{\citenamefont {Dietrich}\ \emph {et~al.}(2018{\natexlab{b}})\citenamefont {Dietrich}, \citenamefont {Bernuzzi}, \citenamefont {Bruegmann},\ and\ \citenamefont {Tichy}}]{Dietrich:2018upm}%
  \BibitemOpen
  \bibfield  {author} {\bibinfo {author} {\bibfnamefont {T.}~\bibnamefont {Dietrich}}, \bibinfo {author} {\bibfnamefont {S.}~\bibnamefont {Bernuzzi}}, \bibinfo {author} {\bibfnamefont {B.}~\bibnamefont {Bruegmann}},\ and\ \bibinfo {author} {\bibfnamefont {W.}~\bibnamefont {Tichy}},\ }\bibfield  {title} {\bibinfo {title} {{High-resolution numerical relativity simulations of spinning binary neutron star mergers}},\ }in\ \href {https://doi.org/10.1109/PDP2018.2018.00113} {\emph {\bibinfo {booktitle} {{26th Euromicro International Conference on Parallel, Distributed and Network-based Processing}}}}\ (\bibinfo {year} {2018})\ pp.\ \bibinfo {pages} {682--689},\ \Eprint {https://arxiv.org/abs/1803.07965} {arXiv:1803.07965 [gr-qc]} \BibitemShut {NoStop}%
\bibitem [{\citenamefont {Tacik}\ \emph {et~al.}(2015)\citenamefont {Tacik} \emph {et~al.}}]{Tacik:2015tja}%
  \BibitemOpen
  \bibfield  {author} {\bibinfo {author} {\bibfnamefont {N.}~\bibnamefont {Tacik}} \emph {et~al.},\ }\bibfield  {title} {\bibinfo {title} {{Binary Neutron Stars with Arbitrary Spins in Numerical Relativity}},\ }\href {https://doi.org/10.1103/PhysRevD.92.124012} {\bibfield  {journal} {\bibinfo  {journal} {Phys. Rev. D}\ }\textbf {\bibinfo {volume} {92}},\ \bibinfo {pages} {124012} (\bibinfo {year} {2015})},\ \bibinfo {note} {[Erratum: Phys.Rev.D 94, 049903 (2016)]},\ \Eprint {https://arxiv.org/abs/1508.06986} {arXiv:1508.06986 [gr-qc]} \BibitemShut {NoStop}%
\bibitem [{\citenamefont {Most}\ \emph {et~al.}(2019)\citenamefont {Most}, \citenamefont {Papenfort}, \citenamefont {Tsokaros},\ and\ \citenamefont {Rezzolla}}]{Most:2019pac}%
  \BibitemOpen
  \bibfield  {author} {\bibinfo {author} {\bibfnamefont {E.~R.}\ \bibnamefont {Most}}, \bibinfo {author} {\bibfnamefont {L.~J.}\ \bibnamefont {Papenfort}}, \bibinfo {author} {\bibfnamefont {A.}~\bibnamefont {Tsokaros}},\ and\ \bibinfo {author} {\bibfnamefont {L.}~\bibnamefont {Rezzolla}},\ }\bibfield  {title} {\bibinfo {title} {{Impact of high spins on the ejection of mass in GW170817}},\ }\href {https://doi.org/10.3847/1538-4357/ab3ebb} {\bibfield  {journal} {\bibinfo  {journal} {Astrophys. J.}\ }\textbf {\bibinfo {volume} {884}},\ \bibinfo {pages} {40} (\bibinfo {year} {2019})},\ \Eprint {https://arxiv.org/abs/1904.04220} {arXiv:1904.04220 [astro-ph.HE]} \BibitemShut {NoStop}%
\bibitem [{\citenamefont {Tichy}\ \emph {et~al.}(2019)\citenamefont {Tichy}, \citenamefont {Rashti}, \citenamefont {Dietrich}, \citenamefont {Dudi},\ and\ \citenamefont {Br\"ugmann}}]{Tichy:2019ouu}%
  \BibitemOpen
  \bibfield  {author} {\bibinfo {author} {\bibfnamefont {W.}~\bibnamefont {Tichy}}, \bibinfo {author} {\bibfnamefont {A.}~\bibnamefont {Rashti}}, \bibinfo {author} {\bibfnamefont {T.}~\bibnamefont {Dietrich}}, \bibinfo {author} {\bibfnamefont {R.}~\bibnamefont {Dudi}},\ and\ \bibinfo {author} {\bibfnamefont {B.}~\bibnamefont {Br\"ugmann}},\ }\bibfield  {title} {\bibinfo {title} {{Constructing binary neutron star initial data with high spins, high compactnesses, and high mass ratios}},\ }\href {https://doi.org/10.1103/PhysRevD.100.124046} {\bibfield  {journal} {\bibinfo  {journal} {Phys. Rev. D}\ }\textbf {\bibinfo {volume} {100}},\ \bibinfo {pages} {124046} (\bibinfo {year} {2019})},\ \Eprint {https://arxiv.org/abs/1910.09690} {arXiv:1910.09690 [gr-qc]} \BibitemShut {NoStop}%
\bibitem [{\citenamefont {Dudi}\ \emph {et~al.}(2022)\citenamefont {Dudi}, \citenamefont {Dietrich}, \citenamefont {Rashti}, \citenamefont {Bruegmann}, \citenamefont {Steinhoff},\ and\ \citenamefont {Tichy}}]{Dudi:2021wcf}%
  \BibitemOpen
  \bibfield  {author} {\bibinfo {author} {\bibfnamefont {R.}~\bibnamefont {Dudi}}, \bibinfo {author} {\bibfnamefont {T.}~\bibnamefont {Dietrich}}, \bibinfo {author} {\bibfnamefont {A.}~\bibnamefont {Rashti}}, \bibinfo {author} {\bibfnamefont {B.}~\bibnamefont {Bruegmann}}, \bibinfo {author} {\bibfnamefont {J.}~\bibnamefont {Steinhoff}},\ and\ \bibinfo {author} {\bibfnamefont {W.}~\bibnamefont {Tichy}},\ }\bibfield  {title} {\bibinfo {title} {{High-accuracy simulations of highly spinning binary neutron star systems}},\ }\href {https://doi.org/10.1103/PhysRevD.105.064050} {\bibfield  {journal} {\bibinfo  {journal} {Phys. Rev. D}\ }\textbf {\bibinfo {volume} {105}},\ \bibinfo {pages} {064050} (\bibinfo {year} {2022})},\ \Eprint {https://arxiv.org/abs/2108.10429} {arXiv:2108.10429 [gr-qc]} \BibitemShut {NoStop}%
\bibitem [{\citenamefont {Rosswog}\ \emph {et~al.}(2024)\citenamefont {Rosswog}, \citenamefont {Diener}, \citenamefont {Torsello}, \citenamefont {Tauris},\ and\ \citenamefont {Sarin}}]{Rosswog:2023rqa}%
  \BibitemOpen
  \bibfield  {author} {\bibinfo {author} {\bibfnamefont {S.}~\bibnamefont {Rosswog}}, \bibinfo {author} {\bibfnamefont {P.}~\bibnamefont {Diener}}, \bibinfo {author} {\bibfnamefont {F.}~\bibnamefont {Torsello}}, \bibinfo {author} {\bibfnamefont {T.~M.}\ \bibnamefont {Tauris}},\ and\ \bibinfo {author} {\bibfnamefont {N.}~\bibnamefont {Sarin}},\ }\bibfield  {title} {\bibinfo {title} {{Mergers of double NSs with one high-spin component: brighter kilonovae and fallback accretion, weaker gravitational waves}},\ }\href {https://doi.org/10.1093/mnras/stae454} {\bibfield  {journal} {\bibinfo  {journal} {Mon. Not. Roy. Astron. Soc.}\ }\textbf {\bibinfo {volume} {530}},\ \bibinfo {pages} {2336} (\bibinfo {year} {2024})},\ \Eprint {https://arxiv.org/abs/2310.15920} {arXiv:2310.15920 [astro-ph.HE]} \BibitemShut {NoStop}%
\bibitem [{\citenamefont {Bernuzzi}\ \emph {et~al.}(2014)\citenamefont {Bernuzzi}, \citenamefont {Dietrich}, \citenamefont {Tichy},\ and\ \citenamefont {Br\"ugmann}}]{Bernuzzi:2013rza}%
  \BibitemOpen
  \bibfield  {author} {\bibinfo {author} {\bibfnamefont {S.}~\bibnamefont {Bernuzzi}}, \bibinfo {author} {\bibfnamefont {T.}~\bibnamefont {Dietrich}}, \bibinfo {author} {\bibfnamefont {W.}~\bibnamefont {Tichy}},\ and\ \bibinfo {author} {\bibfnamefont {B.}~\bibnamefont {Br\"ugmann}},\ }\bibfield  {title} {\bibinfo {title} {{Mergers of binary neutron stars with realistic spin}},\ }\href {https://doi.org/10.1103/PhysRevD.89.104021} {\bibfield  {journal} {\bibinfo  {journal} {Phys. Rev. D}\ }\textbf {\bibinfo {volume} {89}},\ \bibinfo {pages} {104021} (\bibinfo {year} {2014})},\ \Eprint {https://arxiv.org/abs/1311.4443} {arXiv:1311.4443 [gr-qc]} \BibitemShut {NoStop}%
\bibitem [{\citenamefont {Tsokaros}\ \emph {et~al.}(2016)\citenamefont {Tsokaros}, \citenamefont {Mundim}, \citenamefont {Galeazzi}, \citenamefont {Rezzolla},\ and\ \citenamefont {Ury\={u}}}]{Tsokaros:2016eik}%
  \BibitemOpen
  \bibfield  {author} {\bibinfo {author} {\bibfnamefont {A.}~\bibnamefont {Tsokaros}}, \bibinfo {author} {\bibfnamefont {B.~C.}\ \bibnamefont {Mundim}}, \bibinfo {author} {\bibfnamefont {F.}~\bibnamefont {Galeazzi}}, \bibinfo {author} {\bibfnamefont {L.}~\bibnamefont {Rezzolla}},\ and\ \bibinfo {author} {\bibfnamefont {K.}~\bibnamefont {Ury\={u}}},\ }\bibfield  {title} {\bibinfo {title} {{Initial-data contribution to the error budget of gravitational waves from neutron-star binaries}},\ }\href {https://doi.org/10.1103/PhysRevD.94.044049} {\bibfield  {journal} {\bibinfo  {journal} {Phys. Rev. D}\ }\textbf {\bibinfo {volume} {94}},\ \bibinfo {pages} {044049} (\bibinfo {year} {2016})},\ \Eprint {https://arxiv.org/abs/1605.07205} {arXiv:1605.07205 [gr-qc]} \BibitemShut {NoStop}%
\bibitem [{\citenamefont {Read}\ \emph {et~al.}(2009)\citenamefont {Read}, \citenamefont {Lackey}, \citenamefont {Owen},\ and\ \citenamefont {Friedman}}]{Read:2008iy}%
  \BibitemOpen
  \bibfield  {author} {\bibinfo {author} {\bibfnamefont {J.~S.}\ \bibnamefont {Read}}, \bibinfo {author} {\bibfnamefont {B.~D.}\ \bibnamefont {Lackey}}, \bibinfo {author} {\bibfnamefont {B.~J.}\ \bibnamefont {Owen}},\ and\ \bibinfo {author} {\bibfnamefont {J.~L.}\ \bibnamefont {Friedman}},\ }\bibfield  {title} {\bibinfo {title} {{Constraints on a phenomenologically parameterized neutron-star equation of state}},\ }\href {https://doi.org/10.1103/PhysRevD.79.124032} {\bibfield  {journal} {\bibinfo  {journal} {Phys. Rev. D}\ }\textbf {\bibinfo {volume} {79}},\ \bibinfo {pages} {124032} (\bibinfo {year} {2009})},\ \Eprint {https://arxiv.org/abs/0812.2163} {arXiv:0812.2163 [astro-ph]} \BibitemShut {NoStop}%
\bibitem [{\citenamefont {Lackey}\ \emph {et~al.}(2006)\citenamefont {Lackey}, \citenamefont {Nayyar},\ and\ \citenamefont {Owen}}]{Lackey:2005tk}%
  \BibitemOpen
  \bibfield  {author} {\bibinfo {author} {\bibfnamefont {B.~D.}\ \bibnamefont {Lackey}}, \bibinfo {author} {\bibfnamefont {M.}~\bibnamefont {Nayyar}},\ and\ \bibinfo {author} {\bibfnamefont {B.~J.}\ \bibnamefont {Owen}},\ }\bibfield  {title} {\bibinfo {title} {{Observational constraints on hyperons in neutron stars}},\ }\href {https://doi.org/10.1103/PhysRevD.73.024021} {\bibfield  {journal} {\bibinfo  {journal} {Phys. Rev. D}\ }\textbf {\bibinfo {volume} {73}},\ \bibinfo {pages} {024021} (\bibinfo {year} {2006})},\ \Eprint {https://arxiv.org/abs/astro-ph/0507312} {arXiv:astro-ph/0507312} \BibitemShut {NoStop}%
\bibitem [{\citenamefont {Douchin}\ and\ \citenamefont {Haensel}(2001)}]{Douchin:2001sv}%
  \BibitemOpen
  \bibfield  {author} {\bibinfo {author} {\bibfnamefont {F.}~\bibnamefont {Douchin}}\ and\ \bibinfo {author} {\bibfnamefont {P.}~\bibnamefont {Haensel}},\ }\bibfield  {title} {\bibinfo {title} {{A unified equation of state of dense matter and neutron star structure}},\ }\href {https://doi.org/10.1051/0004-6361:20011402} {\bibfield  {journal} {\bibinfo  {journal} {Astron. Astrophys.}\ }\textbf {\bibinfo {volume} {380}},\ \bibinfo {pages} {151} (\bibinfo {year} {2001})},\ \Eprint {https://arxiv.org/abs/astro-ph/0111092} {arXiv:astro-ph/0111092} \BibitemShut {NoStop}%
\bibitem [{\citenamefont {Shibata}\ \emph {et~al.}(2017)\citenamefont {Shibata}, \citenamefont {Fujibayashi}, \citenamefont {Hotokezaka}, \citenamefont {Kiuchi}, \citenamefont {Kyutoku}, \citenamefont {Sekiguchi},\ and\ \citenamefont {Tanaka}}]{Shibata:2017xdx}%
  \BibitemOpen
  \bibfield  {author} {\bibinfo {author} {\bibfnamefont {M.}~\bibnamefont {Shibata}}, \bibinfo {author} {\bibfnamefont {S.}~\bibnamefont {Fujibayashi}}, \bibinfo {author} {\bibfnamefont {K.}~\bibnamefont {Hotokezaka}}, \bibinfo {author} {\bibfnamefont {K.}~\bibnamefont {Kiuchi}}, \bibinfo {author} {\bibfnamefont {K.}~\bibnamefont {Kyutoku}}, \bibinfo {author} {\bibfnamefont {Y.}~\bibnamefont {Sekiguchi}},\ and\ \bibinfo {author} {\bibfnamefont {M.}~\bibnamefont {Tanaka}},\ }\bibfield  {title} {\bibinfo {title} {{Modeling GW170817 based on numerical relativity and its implications}},\ }\href {https://doi.org/10.1103/PhysRevD.96.123012} {\bibfield  {journal} {\bibinfo  {journal} {Phys. Rev. D}\ }\textbf {\bibinfo {volume} {96}},\ \bibinfo {pages} {123012} (\bibinfo {year} {2017})},\ \Eprint {https://arxiv.org/abs/1710.07579} {arXiv:1710.07579 [astro-ph.HE]} \BibitemShut {NoStop}%
\bibitem [{\citenamefont {Rezzolla}\ \emph {et~al.}(2018)\citenamefont {Rezzolla}, \citenamefont {Most},\ and\ \citenamefont {Weih}}]{Rezzolla:2017aly}%
  \BibitemOpen
  \bibfield  {author} {\bibinfo {author} {\bibfnamefont {L.}~\bibnamefont {Rezzolla}}, \bibinfo {author} {\bibfnamefont {E.~R.}\ \bibnamefont {Most}},\ and\ \bibinfo {author} {\bibfnamefont {L.~R.}\ \bibnamefont {Weih}},\ }\bibfield  {title} {\bibinfo {title} {{Using gravitational-wave observations and quasi-universal relations to constrain the maximum mass of neutron stars}},\ }\href {https://doi.org/10.3847/2041-8213/aaa401} {\bibfield  {journal} {\bibinfo  {journal} {Astrophys. J. Lett.}\ }\textbf {\bibinfo {volume} {852}},\ \bibinfo {pages} {L25} (\bibinfo {year} {2018})},\ \Eprint {https://arxiv.org/abs/1711.00314} {arXiv:1711.00314 [astro-ph.HE]} \BibitemShut {NoStop}%
\bibitem [{\citenamefont {Margalit}\ and\ \citenamefont {Metzger}(2017)}]{Margalit:2017dij}%
  \BibitemOpen
  \bibfield  {author} {\bibinfo {author} {\bibfnamefont {B.}~\bibnamefont {Margalit}}\ and\ \bibinfo {author} {\bibfnamefont {B.~D.}\ \bibnamefont {Metzger}},\ }\bibfield  {title} {\bibinfo {title} {{Constraining the Maximum Mass of Neutron Stars From Multi-Messenger Observations of GW170817}},\ }\href {https://doi.org/10.3847/2041-8213/aa991c} {\bibfield  {journal} {\bibinfo  {journal} {Astrophys. J. Lett.}\ }\textbf {\bibinfo {volume} {850}},\ \bibinfo {pages} {L19} (\bibinfo {year} {2017})},\ \Eprint {https://arxiv.org/abs/1710.05938} {arXiv:1710.05938 [astro-ph.HE]} \BibitemShut {NoStop}%
\bibitem [{\citenamefont {De}\ \emph {et~al.}(2018)\citenamefont {De}, \citenamefont {Finstad}, \citenamefont {Lattimer}, \citenamefont {Brown}, \citenamefont {Berger},\ and\ \citenamefont {Biwer}}]{De:2018uhw}%
  \BibitemOpen
  \bibfield  {author} {\bibinfo {author} {\bibfnamefont {S.}~\bibnamefont {De}}, \bibinfo {author} {\bibfnamefont {D.}~\bibnamefont {Finstad}}, \bibinfo {author} {\bibfnamefont {J.~M.}\ \bibnamefont {Lattimer}}, \bibinfo {author} {\bibfnamefont {D.~A.}\ \bibnamefont {Brown}}, \bibinfo {author} {\bibfnamefont {E.}~\bibnamefont {Berger}},\ and\ \bibinfo {author} {\bibfnamefont {C.~M.}\ \bibnamefont {Biwer}},\ }\bibfield  {title} {\bibinfo {title} {{Tidal Deformabilities and Radii of Neutron Stars from the Observation of GW170817}},\ }\href {https://doi.org/10.1103/PhysRevLett.121.091102} {\bibfield  {journal} {\bibinfo  {journal} {Phys. Rev. Lett.}\ }\textbf {\bibinfo {volume} {121}},\ \bibinfo {pages} {091102} (\bibinfo {year} {2018})},\ \bibinfo {note} {[Erratum: Phys.Rev.Lett. 121, 259902 (2018)]},\ \Eprint {https://arxiv.org/abs/1804.08583} {arXiv:1804.08583 [astro-ph.HE]} \BibitemShut {NoStop}%
\bibitem [{\citenamefont {Most}\ \emph {et~al.}(2018)\citenamefont {Most}, \citenamefont {Weih}, \citenamefont {Rezzolla},\ and\ \citenamefont {Schaffner-Bielich}}]{Most:2018hfd}%
  \BibitemOpen
  \bibfield  {author} {\bibinfo {author} {\bibfnamefont {E.~R.}\ \bibnamefont {Most}}, \bibinfo {author} {\bibfnamefont {L.~R.}\ \bibnamefont {Weih}}, \bibinfo {author} {\bibfnamefont {L.}~\bibnamefont {Rezzolla}},\ and\ \bibinfo {author} {\bibfnamefont {J.}~\bibnamefont {Schaffner-Bielich}},\ }\bibfield  {title} {\bibinfo {title} {{New constraints on radii and tidal deformabilities of neutron stars from GW170817}},\ }\href {https://doi.org/10.1103/PhysRevLett.120.261103} {\bibfield  {journal} {\bibinfo  {journal} {Phys. Rev. Lett.}\ }\textbf {\bibinfo {volume} {120}},\ \bibinfo {pages} {261103} (\bibinfo {year} {2018})},\ \Eprint {https://arxiv.org/abs/1803.00549} {arXiv:1803.00549 [gr-qc]} \BibitemShut {NoStop}%
\bibitem [{\citenamefont {Tews}\ \emph {et~al.}(2018)\citenamefont {Tews}, \citenamefont {Margueron},\ and\ \citenamefont {Reddy}}]{Tews:2018iwm}%
  \BibitemOpen
  \bibfield  {author} {\bibinfo {author} {\bibfnamefont {I.}~\bibnamefont {Tews}}, \bibinfo {author} {\bibfnamefont {J.}~\bibnamefont {Margueron}},\ and\ \bibinfo {author} {\bibfnamefont {S.}~\bibnamefont {Reddy}},\ }\bibfield  {title} {\bibinfo {title} {{Critical examination of constraints on the equation of state of dense matter obtained from GW170817}},\ }\href {https://doi.org/10.1103/PhysRevC.98.045804} {\bibfield  {journal} {\bibinfo  {journal} {Phys. Rev. C}\ }\textbf {\bibinfo {volume} {98}},\ \bibinfo {pages} {045804} (\bibinfo {year} {2018})},\ \Eprint {https://arxiv.org/abs/1804.02783} {arXiv:1804.02783 [nucl-th]} \BibitemShut {NoStop}%
\bibitem [{\citenamefont {Shibata}\ \emph {et~al.}(2019)\citenamefont {Shibata}, \citenamefont {Zhou}, \citenamefont {Kiuchi},\ and\ \citenamefont {Fujibayashi}}]{Shibata:2019ctb}%
  \BibitemOpen
  \bibfield  {author} {\bibinfo {author} {\bibfnamefont {M.}~\bibnamefont {Shibata}}, \bibinfo {author} {\bibfnamefont {E.}~\bibnamefont {Zhou}}, \bibinfo {author} {\bibfnamefont {K.}~\bibnamefont {Kiuchi}},\ and\ \bibinfo {author} {\bibfnamefont {S.}~\bibnamefont {Fujibayashi}},\ }\bibfield  {title} {\bibinfo {title} {{Constraint on the maximum mass of neutron stars using GW170817 event}},\ }\href {https://doi.org/10.1103/PhysRevD.100.023015} {\bibfield  {journal} {\bibinfo  {journal} {Phys. Rev. D}\ }\textbf {\bibinfo {volume} {100}},\ \bibinfo {pages} {023015} (\bibinfo {year} {2019})},\ \Eprint {https://arxiv.org/abs/1905.03656} {arXiv:1905.03656 [astro-ph.HE]} \BibitemShut {NoStop}%
\bibitem [{\citenamefont {Capano}\ \emph {et~al.}(2020)\citenamefont {Capano}, \citenamefont {Tews}, \citenamefont {Brown}, \citenamefont {Margalit}, \citenamefont {De}, \citenamefont {Kumar}, \citenamefont {Brown}, \citenamefont {Krishnan},\ and\ \citenamefont {Reddy}}]{Capano:2019eae}%
  \BibitemOpen
  \bibfield  {author} {\bibinfo {author} {\bibfnamefont {C.~D.}\ \bibnamefont {Capano}}, \bibinfo {author} {\bibfnamefont {I.}~\bibnamefont {Tews}}, \bibinfo {author} {\bibfnamefont {S.~M.}\ \bibnamefont {Brown}}, \bibinfo {author} {\bibfnamefont {B.}~\bibnamefont {Margalit}}, \bibinfo {author} {\bibfnamefont {S.}~\bibnamefont {De}}, \bibinfo {author} {\bibfnamefont {S.}~\bibnamefont {Kumar}}, \bibinfo {author} {\bibfnamefont {D.~A.}\ \bibnamefont {Brown}}, \bibinfo {author} {\bibfnamefont {B.}~\bibnamefont {Krishnan}},\ and\ \bibinfo {author} {\bibfnamefont {S.}~\bibnamefont {Reddy}},\ }\bibfield  {title} {\bibinfo {title} {{Stringent constraints on neutron-star radii from multimessenger observations and nuclear theory}},\ }\href {https://doi.org/10.1038/s41550-020-1014-6} {\bibfield  {journal} {\bibinfo  {journal} {Nature Astron.}\ }\textbf {\bibinfo {volume} {4}},\ \bibinfo {pages} {625} (\bibinfo {year} {2020})},\ \Eprint {https://arxiv.org/abs/1908.10352} {arXiv:1908.10352 [astro-ph.HE]} \BibitemShut
  {NoStop}%
\bibitem [{\citenamefont {Landry}\ \emph {et~al.}(2020)\citenamefont {Landry}, \citenamefont {Essick},\ and\ \citenamefont {Chatziioannou}}]{Landry:2020vaw}%
  \BibitemOpen
  \bibfield  {author} {\bibinfo {author} {\bibfnamefont {P.}~\bibnamefont {Landry}}, \bibinfo {author} {\bibfnamefont {R.}~\bibnamefont {Essick}},\ and\ \bibinfo {author} {\bibfnamefont {K.}~\bibnamefont {Chatziioannou}},\ }\bibfield  {title} {\bibinfo {title} {{Nonparametric constraints on neutron star matter with existing and upcoming gravitational wave and pulsar observations}},\ }\href {https://doi.org/10.1103/PhysRevD.101.123007} {\bibfield  {journal} {\bibinfo  {journal} {Phys. Rev. D}\ }\textbf {\bibinfo {volume} {101}},\ \bibinfo {pages} {123007} (\bibinfo {year} {2020})},\ \Eprint {https://arxiv.org/abs/2003.04880} {arXiv:2003.04880 [astro-ph.HE]} \BibitemShut {NoStop}%
\bibitem [{\citenamefont {Koehn}\ \emph {et~al.}(2024)\citenamefont {Koehn} \emph {et~al.}}]{Koehn:2024set}%
  \BibitemOpen
  \bibfield  {author} {\bibinfo {author} {\bibfnamefont {H.}~\bibnamefont {Koehn}} \emph {et~al.},\ }\bibfield  {title} {\bibinfo {title} {{An overview of existing and new nuclear and astrophysical constraints on the equation of state of neutron-rich dense matter}},\ }\href@noop {} {\bibfield  {journal} {\bibinfo  {journal} {{arXiv:2402.04172}}\ } (\bibinfo {year} {2024})},\ \Eprint {https://arxiv.org/abs/2402.04172} {arXiv:2402.04172 [astro-ph.HE]} \BibitemShut {NoStop}%
\bibitem [{\citenamefont {Papenfort}\ \emph {et~al.}(2021)\citenamefont {Papenfort}, \citenamefont {Tootle}, \citenamefont {Grandcl\'ement}, \citenamefont {Most},\ and\ \citenamefont {Rezzolla}}]{Papenfort:2021hod}%
  \BibitemOpen
  \bibfield  {author} {\bibinfo {author} {\bibfnamefont {L.~J.}\ \bibnamefont {Papenfort}}, \bibinfo {author} {\bibfnamefont {S.~D.}\ \bibnamefont {Tootle}}, \bibinfo {author} {\bibfnamefont {P.}~\bibnamefont {Grandcl\'ement}}, \bibinfo {author} {\bibfnamefont {E.~R.}\ \bibnamefont {Most}},\ and\ \bibinfo {author} {\bibfnamefont {L.}~\bibnamefont {Rezzolla}},\ }\bibfield  {title} {\bibinfo {title} {{New public code for initial data of unequal-mass, spinning compact-object binaries}},\ }\href {https://doi.org/10.1103/PhysRevD.104.024057} {\bibfield  {journal} {\bibinfo  {journal} {Phys. Rev. D}\ }\textbf {\bibinfo {volume} {104}},\ \bibinfo {pages} {024057} (\bibinfo {year} {2021})},\ \Eprint {https://arxiv.org/abs/2103.09911} {arXiv:2103.09911 [gr-qc]} \BibitemShut {NoStop}%
\bibitem [{\citenamefont {Grandclement}(2010)}]{Grandclement:2009ju}%
  \BibitemOpen
  \bibfield  {author} {\bibinfo {author} {\bibfnamefont {P.}~\bibnamefont {Grandclement}},\ }\bibfield  {title} {\bibinfo {title} {{Kadath: A Spectral solver for theoretical physics}},\ }\href {https://doi.org/10.1016/j.jcp.2010.01.005} {\bibfield  {journal} {\bibinfo  {journal} {J. Comput. Phys.}\ }\textbf {\bibinfo {volume} {229}},\ \bibinfo {pages} {3334} (\bibinfo {year} {2010})},\ \Eprint {https://arxiv.org/abs/0909.1228} {arXiv:0909.1228 [gr-qc]} \BibitemShut {NoStop}%
\bibitem [{\citenamefont {Tichy}(2009)}]{Tichy:2009yr}%
  \BibitemOpen
  \bibfield  {author} {\bibinfo {author} {\bibfnamefont {W.}~\bibnamefont {Tichy}},\ }\bibfield  {title} {\bibinfo {title} {{A New numerical method to construct binary neutron star initial data}},\ }\href {https://doi.org/10.1088/0264-9381/26/17/175018} {\bibfield  {journal} {\bibinfo  {journal} {Class. Quant. Grav.}\ }\textbf {\bibinfo {volume} {26}},\ \bibinfo {pages} {175018} (\bibinfo {year} {2009})},\ \Eprint {https://arxiv.org/abs/0908.0620} {arXiv:0908.0620 [gr-qc]} \BibitemShut {NoStop}%
\bibitem [{\citenamefont {Tichy}(2011)}]{Tichy:2011gw}%
  \BibitemOpen
  \bibfield  {author} {\bibinfo {author} {\bibfnamefont {W.}~\bibnamefont {Tichy}},\ }\bibfield  {title} {\bibinfo {title} {{Initial data for binary neutron stars with arbitrary spins}},\ }\href {https://doi.org/10.1103/PhysRevD.84.024041} {\bibfield  {journal} {\bibinfo  {journal} {Phys. Rev. D}\ }\textbf {\bibinfo {volume} {84}},\ \bibinfo {pages} {024041} (\bibinfo {year} {2011})},\ \Eprint {https://arxiv.org/abs/1107.1440} {arXiv:1107.1440 [gr-qc]} \BibitemShut {NoStop}%
\bibitem [{\citenamefont {Tichy}(2012)}]{Tichy:2012rp}%
  \BibitemOpen
  \bibfield  {author} {\bibinfo {author} {\bibfnamefont {W.}~\bibnamefont {Tichy}},\ }\bibfield  {title} {\bibinfo {title} {{Constructing quasi-equilibrium initial data for binary neutron stars with arbitrary spins}},\ }\href {https://doi.org/10.1103/PhysRevD.86.064024} {\bibfield  {journal} {\bibinfo  {journal} {Phys. Rev. D}\ }\textbf {\bibinfo {volume} {86}},\ \bibinfo {pages} {064024} (\bibinfo {year} {2012})},\ \Eprint {https://arxiv.org/abs/1209.5336} {arXiv:1209.5336 [gr-qc]} \BibitemShut {NoStop}%
\bibitem [{\citenamefont {Tichy}(2017)}]{Tichy:2016vmv}%
  \BibitemOpen
  \bibfield  {author} {\bibinfo {author} {\bibfnamefont {W.}~\bibnamefont {Tichy}},\ }\bibfield  {title} {\bibinfo {title} {{The initial value problem as it relates to numerical relativity}},\ }\href {https://doi.org/10.1088/1361-6633/80/2/026901} {\bibfield  {journal} {\bibinfo  {journal} {Rept. Prog. Phys.}\ }\textbf {\bibinfo {volume} {80}},\ \bibinfo {pages} {026901} (\bibinfo {year} {2017})},\ \Eprint {https://arxiv.org/abs/1610.03805} {arXiv:1610.03805 [gr-qc]} \BibitemShut {NoStop}%
\bibitem [{\citenamefont {Pfeiffer}\ and\ \citenamefont {York}(2003)}]{Pfeiffer:2002iy}%
  \BibitemOpen
  \bibfield  {author} {\bibinfo {author} {\bibfnamefont {H.~P.}\ \bibnamefont {Pfeiffer}}\ and\ \bibinfo {author} {\bibfnamefont {J.~W.}\ \bibnamefont {York}, \bibfnamefont {Jr.}},\ }\bibfield  {title} {\bibinfo {title} {{Extrinsic curvature and the Einstein constraints}},\ }\href {https://doi.org/10.1103/PhysRevD.67.044022} {\bibfield  {journal} {\bibinfo  {journal} {Phys. Rev. D}\ }\textbf {\bibinfo {volume} {67}},\ \bibinfo {pages} {044022} (\bibinfo {year} {2003})},\ \Eprint {https://arxiv.org/abs/gr-qc/0207095} {arXiv:gr-qc/0207095} \BibitemShut {NoStop}%
\bibitem [{\citenamefont {Pfeiffer}\ and\ \citenamefont {York}(2005)}]{Pfeiffer:2005jf}%
  \BibitemOpen
  \bibfield  {author} {\bibinfo {author} {\bibfnamefont {H.~P.}\ \bibnamefont {Pfeiffer}}\ and\ \bibinfo {author} {\bibfnamefont {J.~W.}\ \bibnamefont {York}, \bibfnamefont {Jr.}},\ }\bibfield  {title} {\bibinfo {title} {{Uniqueness and non-uniqueness in the Einstein constraints}},\ }\href {https://doi.org/10.1103/PhysRevLett.95.091101} {\bibfield  {journal} {\bibinfo  {journal} {Phys. Rev. Lett.}\ }\textbf {\bibinfo {volume} {95}},\ \bibinfo {pages} {091101} (\bibinfo {year} {2005})},\ \Eprint {https://arxiv.org/abs/gr-qc/0504142} {arXiv:gr-qc/0504142} \BibitemShut {NoStop}%
\bibitem [{\citenamefont {Pfeiffer}\ \emph {et~al.}(2007)\citenamefont {Pfeiffer}, \citenamefont {Brown}, \citenamefont {Kidder}, \citenamefont {Lindblom}, \citenamefont {Lovelace},\ and\ \citenamefont {Scheel}}]{Pfeiffer:2007yz}%
  \BibitemOpen
  \bibfield  {author} {\bibinfo {author} {\bibfnamefont {H.~P.}\ \bibnamefont {Pfeiffer}}, \bibinfo {author} {\bibfnamefont {D.~A.}\ \bibnamefont {Brown}}, \bibinfo {author} {\bibfnamefont {L.~E.}\ \bibnamefont {Kidder}}, \bibinfo {author} {\bibfnamefont {L.}~\bibnamefont {Lindblom}}, \bibinfo {author} {\bibfnamefont {G.}~\bibnamefont {Lovelace}},\ and\ \bibinfo {author} {\bibfnamefont {M.~A.}\ \bibnamefont {Scheel}},\ }\bibfield  {title} {\bibinfo {title} {{Reducing orbital eccentricity in binary black hole simulations}},\ }\href {https://doi.org/10.1088/0264-9381/24/12/S06} {\bibfield  {journal} {\bibinfo  {journal} {Class. Quant. Grav.}\ }\textbf {\bibinfo {volume} {24}},\ \bibinfo {pages} {S59} (\bibinfo {year} {2007})},\ \Eprint {https://arxiv.org/abs/gr-qc/0702106} {arXiv:gr-qc/0702106} \BibitemShut {NoStop}%
\bibitem [{\citenamefont {Boyle}\ \emph {et~al.}(2007)\citenamefont {Boyle}, \citenamefont {Brown}, \citenamefont {Kidder}, \citenamefont {Mroue}, \citenamefont {Pfeiffer}, \citenamefont {Scheel}, \citenamefont {Cook},\ and\ \citenamefont {Teukolsky}}]{Boyle:2007ft}%
  \BibitemOpen
  \bibfield  {author} {\bibinfo {author} {\bibfnamefont {M.}~\bibnamefont {Boyle}}, \bibinfo {author} {\bibfnamefont {D.~A.}\ \bibnamefont {Brown}}, \bibinfo {author} {\bibfnamefont {L.~E.}\ \bibnamefont {Kidder}}, \bibinfo {author} {\bibfnamefont {A.~H.}\ \bibnamefont {Mroue}}, \bibinfo {author} {\bibfnamefont {H.~P.}\ \bibnamefont {Pfeiffer}}, \bibinfo {author} {\bibfnamefont {M.~A.}\ \bibnamefont {Scheel}}, \bibinfo {author} {\bibfnamefont {G.~B.}\ \bibnamefont {Cook}},\ and\ \bibinfo {author} {\bibfnamefont {S.~A.}\ \bibnamefont {Teukolsky}},\ }\bibfield  {title} {\bibinfo {title} {{High-accuracy comparison of numerical relativity simulations with post-Newtonian expansions}},\ }\href {https://doi.org/10.1103/PhysRevD.76.124038} {\bibfield  {journal} {\bibinfo  {journal} {Phys. Rev. D}\ }\textbf {\bibinfo {volume} {76}},\ \bibinfo {pages} {124038} (\bibinfo {year} {2007})},\ \Eprint {https://arxiv.org/abs/0710.0158} {arXiv:0710.0158 [gr-qc]} \BibitemShut {NoStop}%
\bibitem [{\citenamefont {Buonanno}\ \emph {et~al.}(2011)\citenamefont {Buonanno}, \citenamefont {Kidder}, \citenamefont {Mroue}, \citenamefont {Pfeiffer},\ and\ \citenamefont {Taracchini}}]{Buonanno:2010yk}%
  \BibitemOpen
  \bibfield  {author} {\bibinfo {author} {\bibfnamefont {A.}~\bibnamefont {Buonanno}}, \bibinfo {author} {\bibfnamefont {L.~E.}\ \bibnamefont {Kidder}}, \bibinfo {author} {\bibfnamefont {A.~H.}\ \bibnamefont {Mroue}}, \bibinfo {author} {\bibfnamefont {H.~P.}\ \bibnamefont {Pfeiffer}},\ and\ \bibinfo {author} {\bibfnamefont {A.}~\bibnamefont {Taracchini}},\ }\bibfield  {title} {\bibinfo {title} {{Reducing orbital eccentricity of precessing black-hole binaries}},\ }\href {https://doi.org/10.1103/PhysRevD.83.104034} {\bibfield  {journal} {\bibinfo  {journal} {Phys. Rev. D}\ }\textbf {\bibinfo {volume} {83}},\ \bibinfo {pages} {104034} (\bibinfo {year} {2011})},\ \Eprint {https://arxiv.org/abs/1012.1549} {arXiv:1012.1549 [gr-qc]} \BibitemShut {NoStop}%
\bibitem [{\citenamefont {Kyutoku}\ \emph {et~al.}(2014)\citenamefont {Kyutoku}, \citenamefont {Shibata},\ and\ \citenamefont {Taniguchi}}]{Kyutoku:2014yba}%
  \BibitemOpen
  \bibfield  {author} {\bibinfo {author} {\bibfnamefont {K.}~\bibnamefont {Kyutoku}}, \bibinfo {author} {\bibfnamefont {M.}~\bibnamefont {Shibata}},\ and\ \bibinfo {author} {\bibfnamefont {K.}~\bibnamefont {Taniguchi}},\ }\bibfield  {title} {\bibinfo {title} {{Reducing orbital eccentricity in initial data of binary neutron stars}},\ }\href {https://doi.org/10.1103/PhysRevD.90.064006} {\bibfield  {journal} {\bibinfo  {journal} {Phys. Rev. D}\ }\textbf {\bibinfo {volume} {90}},\ \bibinfo {pages} {064006} (\bibinfo {year} {2014})},\ \Eprint {https://arxiv.org/abs/1405.6207} {arXiv:1405.6207 [gr-qc]} \BibitemShut {NoStop}%
\bibitem [{\citenamefont {Moldenhauer}\ \emph {et~al.}(2014)\citenamefont {Moldenhauer}, \citenamefont {Markakis}, \citenamefont {Johnson-McDaniel}, \citenamefont {Tichy},\ and\ \citenamefont {Br\"ugmann}}]{Moldenhauer:2014yaa}%
  \BibitemOpen
  \bibfield  {author} {\bibinfo {author} {\bibfnamefont {N.}~\bibnamefont {Moldenhauer}}, \bibinfo {author} {\bibfnamefont {C.~M.}\ \bibnamefont {Markakis}}, \bibinfo {author} {\bibfnamefont {N.~K.}\ \bibnamefont {Johnson-McDaniel}}, \bibinfo {author} {\bibfnamefont {W.}~\bibnamefont {Tichy}},\ and\ \bibinfo {author} {\bibfnamefont {B.}~\bibnamefont {Br\"ugmann}},\ }\bibfield  {title} {\bibinfo {title} {{Initial data for binary neutron stars with adjustable eccentricity}},\ }\href {https://doi.org/10.1103/PhysRevD.90.084043} {\bibfield  {journal} {\bibinfo  {journal} {Phys. Rev. D}\ }\textbf {\bibinfo {volume} {90}},\ \bibinfo {pages} {084043} (\bibinfo {year} {2014})},\ \Eprint {https://arxiv.org/abs/1408.4136} {arXiv:1408.4136 [gr-qc]} \BibitemShut {NoStop}%
\bibitem [{\citenamefont {Yamamoto}\ \emph {et~al.}(2008)\citenamefont {Yamamoto}, \citenamefont {Shibata},\ and\ \citenamefont {Taniguchi}}]{Yamamoto:2008js}%
  \BibitemOpen
  \bibfield  {author} {\bibinfo {author} {\bibfnamefont {T.}~\bibnamefont {Yamamoto}}, \bibinfo {author} {\bibfnamefont {M.}~\bibnamefont {Shibata}},\ and\ \bibinfo {author} {\bibfnamefont {K.}~\bibnamefont {Taniguchi}},\ }\bibfield  {title} {\bibinfo {title} {{Simulating coalescing compact binaries by a new code SACRA}},\ }\href {https://doi.org/10.1103/PhysRevD.78.064054} {\bibfield  {journal} {\bibinfo  {journal} {Phys. Rev. D}\ }\textbf {\bibinfo {volume} {78}},\ \bibinfo {pages} {064054} (\bibinfo {year} {2008})},\ \Eprint {https://arxiv.org/abs/0806.4007} {arXiv:0806.4007 [gr-qc]} \BibitemShut {NoStop}%
\bibitem [{\citenamefont {Bruegmann}\ \emph {et~al.}(2008)\citenamefont {Bruegmann}, \citenamefont {Gonzalez}, \citenamefont {Hannam}, \citenamefont {Husa}, \citenamefont {Sperhake},\ and\ \citenamefont {Tichy}}]{Bruegmann:2006ulg}%
  \BibitemOpen
  \bibfield  {author} {\bibinfo {author} {\bibfnamefont {B.}~\bibnamefont {Bruegmann}}, \bibinfo {author} {\bibfnamefont {J.~A.}\ \bibnamefont {Gonzalez}}, \bibinfo {author} {\bibfnamefont {M.}~\bibnamefont {Hannam}}, \bibinfo {author} {\bibfnamefont {S.}~\bibnamefont {Husa}}, \bibinfo {author} {\bibfnamefont {U.}~\bibnamefont {Sperhake}},\ and\ \bibinfo {author} {\bibfnamefont {W.}~\bibnamefont {Tichy}},\ }\bibfield  {title} {\bibinfo {title} {{Calibration of Moving Puncture Simulations}},\ }\href {https://doi.org/10.1103/PhysRevD.77.024027} {\bibfield  {journal} {\bibinfo  {journal} {Phys. Rev. D}\ }\textbf {\bibinfo {volume} {77}},\ \bibinfo {pages} {024027} (\bibinfo {year} {2008})},\ \Eprint {https://arxiv.org/abs/gr-qc/0610128} {arXiv:gr-qc/0610128} \BibitemShut {NoStop}%
\bibitem [{\citenamefont {Thierfelder}\ \emph {et~al.}(2011)\citenamefont {Thierfelder}, \citenamefont {Bernuzzi},\ and\ \citenamefont {Bruegmann}}]{Thierfelder:2011yi}%
  \BibitemOpen
  \bibfield  {author} {\bibinfo {author} {\bibfnamefont {M.}~\bibnamefont {Thierfelder}}, \bibinfo {author} {\bibfnamefont {S.}~\bibnamefont {Bernuzzi}},\ and\ \bibinfo {author} {\bibfnamefont {B.}~\bibnamefont {Bruegmann}},\ }\bibfield  {title} {\bibinfo {title} {{Numerical relativity simulations of binary neutron stars}},\ }\href {https://doi.org/10.1103/PhysRevD.84.044012} {\bibfield  {journal} {\bibinfo  {journal} {Phys. Rev. D}\ }\textbf {\bibinfo {volume} {84}},\ \bibinfo {pages} {044012} (\bibinfo {year} {2011})},\ \Eprint {https://arxiv.org/abs/1104.4751} {arXiv:1104.4751 [gr-qc]} \BibitemShut {NoStop}%
\bibitem [{\citenamefont {Dietrich}\ \emph {et~al.}(2015{\natexlab{b}})\citenamefont {Dietrich}, \citenamefont {Bernuzzi}, \citenamefont {Ujevic},\ and\ \citenamefont {Br\"ugmann}}]{Dietrich:2015iva}%
  \BibitemOpen
  \bibfield  {author} {\bibinfo {author} {\bibfnamefont {T.}~\bibnamefont {Dietrich}}, \bibinfo {author} {\bibfnamefont {S.}~\bibnamefont {Bernuzzi}}, \bibinfo {author} {\bibfnamefont {M.}~\bibnamefont {Ujevic}},\ and\ \bibinfo {author} {\bibfnamefont {B.}~\bibnamefont {Br\"ugmann}},\ }\bibfield  {title} {\bibinfo {title} {{Numerical relativity simulations of neutron star merger remnants using conservative mesh refinement}},\ }\href {https://doi.org/10.1103/PhysRevD.91.124041} {\bibfield  {journal} {\bibinfo  {journal} {Phys. Rev. D}\ }\textbf {\bibinfo {volume} {91}},\ \bibinfo {pages} {124041} (\bibinfo {year} {2015}{\natexlab{b}})},\ \Eprint {https://arxiv.org/abs/1504.01266} {arXiv:1504.01266 [gr-qc]} \BibitemShut {NoStop}%
\bibitem [{\citenamefont {Shibata}\ and\ \citenamefont {Nakamura}(1995)}]{Shibata:1995we}%
  \BibitemOpen
  \bibfield  {author} {\bibinfo {author} {\bibfnamefont {M.}~\bibnamefont {Shibata}}\ and\ \bibinfo {author} {\bibfnamefont {T.}~\bibnamefont {Nakamura}},\ }\bibfield  {title} {\bibinfo {title} {{Evolution of three-dimensional gravitational waves: Harmonic slicing case}},\ }\href {https://doi.org/10.1103/PhysRevD.52.5428} {\bibfield  {journal} {\bibinfo  {journal} {Phys. Rev. D}\ }\textbf {\bibinfo {volume} {52}},\ \bibinfo {pages} {5428} (\bibinfo {year} {1995})}\BibitemShut {NoStop}%
\bibitem [{\citenamefont {Baumgarte}\ and\ \citenamefont {Shapiro}(1998)}]{Baumgarte:1998te}%
  \BibitemOpen
  \bibfield  {author} {\bibinfo {author} {\bibfnamefont {T.~W.}\ \bibnamefont {Baumgarte}}\ and\ \bibinfo {author} {\bibfnamefont {S.~L.}\ \bibnamefont {Shapiro}},\ }\bibfield  {title} {\bibinfo {title} {{On the numerical integration of Einstein's field equations}},\ }\href {https://doi.org/10.1103/PhysRevD.59.024007} {\bibfield  {journal} {\bibinfo  {journal} {Phys. Rev. D}\ }\textbf {\bibinfo {volume} {59}},\ \bibinfo {pages} {024007} (\bibinfo {year} {1998})},\ \Eprint {https://arxiv.org/abs/gr-qc/9810065} {arXiv:gr-qc/9810065} \BibitemShut {NoStop}%
\bibitem [{\citenamefont {Campanelli}\ \emph {et~al.}(2006)\citenamefont {Campanelli}, \citenamefont {Lousto}, \citenamefont {Marronetti},\ and\ \citenamefont {Zlochower}}]{Campanelli:2005dd}%
  \BibitemOpen
  \bibfield  {author} {\bibinfo {author} {\bibfnamefont {M.}~\bibnamefont {Campanelli}}, \bibinfo {author} {\bibfnamefont {C.~O.}\ \bibnamefont {Lousto}}, \bibinfo {author} {\bibfnamefont {P.}~\bibnamefont {Marronetti}},\ and\ \bibinfo {author} {\bibfnamefont {Y.}~\bibnamefont {Zlochower}},\ }\bibfield  {title} {\bibinfo {title} {{Accurate evolutions of orbiting black-hole binaries without excision}},\ }\href {https://doi.org/10.1103/PhysRevLett.96.111101} {\bibfield  {journal} {\bibinfo  {journal} {Phys. Rev. Lett.}\ }\textbf {\bibinfo {volume} {96}},\ \bibinfo {pages} {111101} (\bibinfo {year} {2006})},\ \Eprint {https://arxiv.org/abs/gr-qc/0511048} {arXiv:gr-qc/0511048} \BibitemShut {NoStop}%
\bibitem [{\citenamefont {Baker}\ \emph {et~al.}(2006)\citenamefont {Baker}, \citenamefont {Centrella}, \citenamefont {Choi}, \citenamefont {Koppitz},\ and\ \citenamefont {van Meter}}]{Baker:2005vv}%
  \BibitemOpen
  \bibfield  {author} {\bibinfo {author} {\bibfnamefont {J.~G.}\ \bibnamefont {Baker}}, \bibinfo {author} {\bibfnamefont {J.}~\bibnamefont {Centrella}}, \bibinfo {author} {\bibfnamefont {D.-I.}\ \bibnamefont {Choi}}, \bibinfo {author} {\bibfnamefont {M.}~\bibnamefont {Koppitz}},\ and\ \bibinfo {author} {\bibfnamefont {J.}~\bibnamefont {van Meter}},\ }\bibfield  {title} {\bibinfo {title} {{Gravitational wave extraction from an inspiraling configuration of merging black holes}},\ }\href {https://doi.org/10.1103/PhysRevLett.96.111102} {\bibfield  {journal} {\bibinfo  {journal} {Phys. Rev. Lett.}\ }\textbf {\bibinfo {volume} {96}},\ \bibinfo {pages} {111102} (\bibinfo {year} {2006})},\ \Eprint {https://arxiv.org/abs/gr-qc/0511103} {arXiv:gr-qc/0511103} \BibitemShut {NoStop}%
\bibitem [{\citenamefont {Hilditch}\ \emph {et~al.}(2013)\citenamefont {Hilditch}, \citenamefont {Bernuzzi}, \citenamefont {Thierfelder}, \citenamefont {Cao}, \citenamefont {Tichy},\ and\ \citenamefont {Bruegmann}}]{Hilditch:2012fp}%
  \BibitemOpen
  \bibfield  {author} {\bibinfo {author} {\bibfnamefont {D.}~\bibnamefont {Hilditch}}, \bibinfo {author} {\bibfnamefont {S.}~\bibnamefont {Bernuzzi}}, \bibinfo {author} {\bibfnamefont {M.}~\bibnamefont {Thierfelder}}, \bibinfo {author} {\bibfnamefont {Z.}~\bibnamefont {Cao}}, \bibinfo {author} {\bibfnamefont {W.}~\bibnamefont {Tichy}},\ and\ \bibinfo {author} {\bibfnamefont {B.}~\bibnamefont {Bruegmann}},\ }\bibfield  {title} {\bibinfo {title} {{Compact binary evolutions with the Z4c formulation}},\ }\href {https://doi.org/10.1103/PhysRevD.88.084057} {\bibfield  {journal} {\bibinfo  {journal} {Phys. Rev. D}\ }\textbf {\bibinfo {volume} {88}},\ \bibinfo {pages} {084057} (\bibinfo {year} {2013})},\ \Eprint {https://arxiv.org/abs/1212.2901} {arXiv:1212.2901 [gr-qc]} \BibitemShut {NoStop}%
\bibitem [{\citenamefont {Alcubierre}\ \emph {et~al.}(2003)\citenamefont {Alcubierre}, \citenamefont {Bruegmann}, \citenamefont {Diener}, \citenamefont {Koppitz}, \citenamefont {Pollney}, \citenamefont {Seidel},\ and\ \citenamefont {Takahashi}}]{Alcubierre:2002kk}%
  \BibitemOpen
  \bibfield  {author} {\bibinfo {author} {\bibfnamefont {M.}~\bibnamefont {Alcubierre}}, \bibinfo {author} {\bibfnamefont {B.}~\bibnamefont {Bruegmann}}, \bibinfo {author} {\bibfnamefont {P.}~\bibnamefont {Diener}}, \bibinfo {author} {\bibfnamefont {M.}~\bibnamefont {Koppitz}}, \bibinfo {author} {\bibfnamefont {D.}~\bibnamefont {Pollney}}, \bibinfo {author} {\bibfnamefont {E.}~\bibnamefont {Seidel}},\ and\ \bibinfo {author} {\bibfnamefont {R.}~\bibnamefont {Takahashi}},\ }\bibfield  {title} {\bibinfo {title} {{Gauge conditions for long term numerical black hole evolutions without excision}},\ }\href {https://doi.org/10.1103/PhysRevD.67.084023} {\bibfield  {journal} {\bibinfo  {journal} {Phys. Rev. D}\ }\textbf {\bibinfo {volume} {67}},\ \bibinfo {pages} {084023} (\bibinfo {year} {2003})},\ \Eprint {https://arxiv.org/abs/gr-qc/0206072} {arXiv:gr-qc/0206072} \BibitemShut {NoStop}%
\bibitem [{\citenamefont {Shibata}\ \emph {et~al.}(2005)\citenamefont {Shibata}, \citenamefont {Taniguchi},\ and\ \citenamefont {Uryu}}]{Shibata:2005ss}%
  \BibitemOpen
  \bibfield  {author} {\bibinfo {author} {\bibfnamefont {M.}~\bibnamefont {Shibata}}, \bibinfo {author} {\bibfnamefont {K.}~\bibnamefont {Taniguchi}},\ and\ \bibinfo {author} {\bibfnamefont {K.}~\bibnamefont {Uryu}},\ }\bibfield  {title} {\bibinfo {title} {{Merger of binary neutron stars with realistic equations of state in full general relativity}},\ }\href {https://doi.org/10.1103/PhysRevD.71.084021} {\bibfield  {journal} {\bibinfo  {journal} {Phys. Rev. D}\ }\textbf {\bibinfo {volume} {71}},\ \bibinfo {pages} {084021} (\bibinfo {year} {2005})},\ \Eprint {https://arxiv.org/abs/gr-qc/0503119} {arXiv:gr-qc/0503119} \BibitemShut {NoStop}%
\bibitem [{\citenamefont {Bauswein}\ \emph {et~al.}(2010)\citenamefont {Bauswein}, \citenamefont {Janka},\ and\ \citenamefont {Oechslin}}]{Bauswein:2010dn}%
  \BibitemOpen
  \bibfield  {author} {\bibinfo {author} {\bibfnamefont {A.}~\bibnamefont {Bauswein}}, \bibinfo {author} {\bibfnamefont {H.~T.}\ \bibnamefont {Janka}},\ and\ \bibinfo {author} {\bibfnamefont {R.}~\bibnamefont {Oechslin}},\ }\bibfield  {title} {\bibinfo {title} {{Testing Approximations of Thermal Effects in Neutron Star Merger Simulations}},\ }\href {https://doi.org/10.1103/PhysRevD.82.084043} {\bibfield  {journal} {\bibinfo  {journal} {Phys. Rev. D}\ }\textbf {\bibinfo {volume} {82}},\ \bibinfo {pages} {084043} (\bibinfo {year} {2010})},\ \Eprint {https://arxiv.org/abs/1006.3315} {arXiv:1006.3315 [astro-ph.SR]} \BibitemShut {NoStop}%
\bibitem [{\citenamefont {Kurganov}\ and\ \citenamefont {Tadmor}(2000)}]{Kurganov:2000ovy}%
  \BibitemOpen
  \bibfield  {author} {\bibinfo {author} {\bibfnamefont {A.}~\bibnamefont {Kurganov}}\ and\ \bibinfo {author} {\bibfnamefont {E.}~\bibnamefont {Tadmor}},\ }\bibfield  {title} {\bibinfo {title} {{New High-Resolution Central Schemes for Nonlinear Conservation Laws and Convection\textendash{}Diffusion Equations}},\ }\href {https://doi.org/10.1006/jcph.2000.6459} {\bibfield  {journal} {\bibinfo  {journal} {J. Comput. Phys.}\ }\textbf {\bibinfo {volume} {160}},\ \bibinfo {pages} {241} (\bibinfo {year} {2000})}\BibitemShut {NoStop}%
\bibitem [{\citenamefont {Amiram}\ \emph {et~al.}(2006)\citenamefont {Amiram}, \citenamefont {Peter~D.},\ and\ \citenamefont {Bram~van}}]{Amiram:2006zjz}%
  \BibitemOpen
  \bibfield  {author} {\bibinfo {author} {\bibfnamefont {H.}~\bibnamefont {Amiram}}, \bibinfo {author} {\bibfnamefont {L.}~\bibnamefont {Peter~D.}},\ and\ \bibinfo {author} {\bibfnamefont {L.}~\bibnamefont {Bram~van}},\ }\bibfield  {title} {\bibinfo {title} {{On Upstream Differencing and Godunov-Type Schemes for Hyperbolic Conservation Laws}},\ }\href {https://doi.org/10.1137/1025002} {\bibfield  {journal} {\bibinfo  {journal} {SIAM Rev.}\ }\textbf {\bibinfo {volume} {25}},\ \bibinfo {pages} {35} (\bibinfo {year} {2006})}\BibitemShut {NoStop}%
\bibitem [{\citenamefont {Borges}\ \emph {et~al.}(2008)\citenamefont {Borges}, \citenamefont {Carmona}, \citenamefont {Costa},\ and\ \citenamefont {Don}}]{BORGES20083191}%
  \BibitemOpen
  \bibfield  {author} {\bibinfo {author} {\bibfnamefont {R.}~\bibnamefont {Borges}}, \bibinfo {author} {\bibfnamefont {M.}~\bibnamefont {Carmona}}, \bibinfo {author} {\bibfnamefont {B.}~\bibnamefont {Costa}},\ and\ \bibinfo {author} {\bibfnamefont {W.~S.}\ \bibnamefont {Don}},\ }\bibfield  {title} {\bibinfo {title} {An improved weighted essentially non-oscillatory scheme for hyperbolic conservation laws},\ }\href {https://doi.org/https://doi.org/10.1016/j.jcp.2007.11.038} {\bibfield  {journal} {\bibinfo  {journal} {Journal of Computational Physics}\ }\textbf {\bibinfo {volume} {227}},\ \bibinfo {pages} {3191} (\bibinfo {year} {2008})}\BibitemShut {NoStop}%
\bibitem [{\citenamefont {Bernuzzi}\ and\ \citenamefont {Dietrich}(2016)}]{Bernuzzi:2016pie}%
  \BibitemOpen
  \bibfield  {author} {\bibinfo {author} {\bibfnamefont {S.}~\bibnamefont {Bernuzzi}}\ and\ \bibinfo {author} {\bibfnamefont {T.}~\bibnamefont {Dietrich}},\ }\bibfield  {title} {\bibinfo {title} {{Gravitational waveforms from binary neutron star mergers with high-order weighted-essentially-nonoscillatory schemes in numerical relativity}},\ }\href {https://doi.org/10.1103/PhysRevD.94.064062} {\bibfield  {journal} {\bibinfo  {journal} {Phys. Rev. D}\ }\textbf {\bibinfo {volume} {94}},\ \bibinfo {pages} {064062} (\bibinfo {year} {2016})},\ \Eprint {https://arxiv.org/abs/1604.07999} {arXiv:1604.07999 [gr-qc]} \BibitemShut {NoStop}%
\bibitem [{\citenamefont {Taniguchi}\ and\ \citenamefont {Shibata}(2010)}]{Taniguchi:2010kj}%
  \BibitemOpen
  \bibfield  {author} {\bibinfo {author} {\bibfnamefont {K.}~\bibnamefont {Taniguchi}}\ and\ \bibinfo {author} {\bibfnamefont {M.}~\bibnamefont {Shibata}},\ }\bibfield  {title} {\bibinfo {title} {{Binary Neutron Stars in Quasi-equilibrium}},\ }\href {https://doi.org/10.1088/0067-0049/188/1/187} {\bibfield  {journal} {\bibinfo  {journal} {Astrophys. J. Suppl.}\ }\textbf {\bibinfo {volume} {188}},\ \bibinfo {pages} {187} (\bibinfo {year} {2010})},\ \Eprint {https://arxiv.org/abs/1005.0958} {arXiv:1005.0958 [astro-ph.SR]} \BibitemShut {NoStop}%
\bibitem [{\citenamefont {Lai}\ \emph {et~al.}(1994)\citenamefont {Lai}, \citenamefont {Rasio},\ and\ \citenamefont {Shapiro}}]{Lai:1993pa}%
  \BibitemOpen
  \bibfield  {author} {\bibinfo {author} {\bibfnamefont {D.}~\bibnamefont {Lai}}, \bibinfo {author} {\bibfnamefont {F.~A.}\ \bibnamefont {Rasio}},\ and\ \bibinfo {author} {\bibfnamefont {S.~L.}\ \bibnamefont {Shapiro}},\ }\bibfield  {title} {\bibinfo {title} {{Hydrodynamic instability and coalescence of binary neutron stars}},\ }\href {https://doi.org/10.1086/173606} {\bibfield  {journal} {\bibinfo  {journal} {Astrophys. J.}\ }\textbf {\bibinfo {volume} {420}},\ \bibinfo {pages} {811} (\bibinfo {year} {1994})},\ \Eprint {https://arxiv.org/abs/astro-ph/9304027} {arXiv:astro-ph/9304027} \BibitemShut {NoStop}%
\bibitem [{\citenamefont {Taniguchi}\ and\ \citenamefont {Nakamura}(1996)}]{Taniguchi:1996bu}%
  \BibitemOpen
  \bibfield  {author} {\bibinfo {author} {\bibfnamefont {K.}~\bibnamefont {Taniguchi}}\ and\ \bibinfo {author} {\bibfnamefont {T.}~\bibnamefont {Nakamura}},\ }\bibfield  {title} {\bibinfo {title} {{Innermost stable circular orbit of coalescing neutron star - black hole binary: Generalized pseudoNewtonian potential approach}},\ }\href {https://doi.org/10.1143/PTP.96.693} {\bibfield  {journal} {\bibinfo  {journal} {Prog. Theor. Phys.}\ }\textbf {\bibinfo {volume} {96}},\ \bibinfo {pages} {693} (\bibinfo {year} {1996})},\ \Eprint {https://arxiv.org/abs/astro-ph/9609009} {arXiv:astro-ph/9609009} \BibitemShut {NoStop}%
\bibitem [{\citenamefont {Shibata}\ and\ \citenamefont {Taniguchi}(1997)}]{Shibata:1997xn}%
  \BibitemOpen
  \bibfield  {author} {\bibinfo {author} {\bibfnamefont {M.}~\bibnamefont {Shibata}}\ and\ \bibinfo {author} {\bibfnamefont {K.}~\bibnamefont {Taniguchi}},\ }\bibfield  {title} {\bibinfo {title} {{Solving the Darwin problem in the first postNewtonian approximation of general relativity: Compressible model}},\ }\href {https://doi.org/10.1103/PhysRevD.56.811} {\bibfield  {journal} {\bibinfo  {journal} {Phys. Rev. D}\ }\textbf {\bibinfo {volume} {56}},\ \bibinfo {pages} {811} (\bibinfo {year} {1997})},\ \Eprint {https://arxiv.org/abs/gr-qc/9705028} {arXiv:gr-qc/9705028} \BibitemShut {NoStop}%
\bibitem [{\citenamefont {Bernuzzi}\ \emph {et~al.}(2012{\natexlab{b}})\citenamefont {Bernuzzi}, \citenamefont {Thierfelder},\ and\ \citenamefont {Bruegmann}}]{Bernuzzi:2011aq}%
  \BibitemOpen
  \bibfield  {author} {\bibinfo {author} {\bibfnamefont {S.}~\bibnamefont {Bernuzzi}}, \bibinfo {author} {\bibfnamefont {M.}~\bibnamefont {Thierfelder}},\ and\ \bibinfo {author} {\bibfnamefont {B.}~\bibnamefont {Bruegmann}},\ }\bibfield  {title} {\bibinfo {title} {{Accuracy of numerical relativity waveforms from binary neutron star mergers and their comparison with post-Newtonian waveforms}},\ }\href {https://doi.org/10.1103/PhysRevD.85.104030} {\bibfield  {journal} {\bibinfo  {journal} {Phys. Rev. D}\ }\textbf {\bibinfo {volume} {85}},\ \bibinfo {pages} {104030} (\bibinfo {year} {2012}{\natexlab{b}})},\ \Eprint {https://arxiv.org/abs/1109.3611} {arXiv:1109.3611 [gr-qc]} \BibitemShut {NoStop}%
\bibitem [{\citenamefont {Hotokezaka}\ \emph {et~al.}(2015)\citenamefont {Hotokezaka}, \citenamefont {Kyutoku}, \citenamefont {Okawa},\ and\ \citenamefont {Shibata}}]{Hotokezaka:2015xka}%
  \BibitemOpen
  \bibfield  {author} {\bibinfo {author} {\bibfnamefont {K.}~\bibnamefont {Hotokezaka}}, \bibinfo {author} {\bibfnamefont {K.}~\bibnamefont {Kyutoku}}, \bibinfo {author} {\bibfnamefont {H.}~\bibnamefont {Okawa}},\ and\ \bibinfo {author} {\bibfnamefont {M.}~\bibnamefont {Shibata}},\ }\bibfield  {title} {\bibinfo {title} {{Exploring tidal effects of coalescing binary neutron stars in numerical relativity. II. Long-term simulations}},\ }\href {https://doi.org/10.1103/PhysRevD.91.064060} {\bibfield  {journal} {\bibinfo  {journal} {Phys. Rev. D}\ }\textbf {\bibinfo {volume} {91}},\ \bibinfo {pages} {064060} (\bibinfo {year} {2015})},\ \Eprint {https://arxiv.org/abs/1502.03457} {arXiv:1502.03457 [gr-qc]} \BibitemShut {NoStop}%
\bibitem [{\citenamefont {Boyle}\ and\ \citenamefont {Mroue}(2009)}]{Boyle:2009vi}%
  \BibitemOpen
  \bibfield  {author} {\bibinfo {author} {\bibfnamefont {M.}~\bibnamefont {Boyle}}\ and\ \bibinfo {author} {\bibfnamefont {A.~H.}\ \bibnamefont {Mroue}},\ }\bibfield  {title} {\bibinfo {title} {{Extrapolating gravitational-wave data from numerical simulations}},\ }\href {https://doi.org/10.1103/PhysRevD.80.124045} {\bibfield  {journal} {\bibinfo  {journal} {Phys. Rev. D}\ }\textbf {\bibinfo {volume} {80}},\ \bibinfo {pages} {124045} (\bibinfo {year} {2009})},\ \Eprint {https://arxiv.org/abs/0905.3177} {arXiv:0905.3177 [gr-qc]} \BibitemShut {NoStop}%
\bibitem [{\citenamefont {Pollney}\ \emph {et~al.}(2009)\citenamefont {Pollney}, \citenamefont {Reisswig}, \citenamefont {Dorband}, \citenamefont {Schnetter},\ and\ \citenamefont {Diener}}]{Pollney:2009ut}%
  \BibitemOpen
  \bibfield  {author} {\bibinfo {author} {\bibfnamefont {D.}~\bibnamefont {Pollney}}, \bibinfo {author} {\bibfnamefont {C.}~\bibnamefont {Reisswig}}, \bibinfo {author} {\bibfnamefont {N.}~\bibnamefont {Dorband}}, \bibinfo {author} {\bibfnamefont {E.}~\bibnamefont {Schnetter}},\ and\ \bibinfo {author} {\bibfnamefont {P.}~\bibnamefont {Diener}},\ }\bibfield  {title} {\bibinfo {title} {{The Asymptotic Falloff of Local Waveform Measurements in Numerical Relativity}},\ }\href {https://doi.org/10.1103/PhysRevD.80.121502} {\bibfield  {journal} {\bibinfo  {journal} {Phys. Rev. D}\ }\textbf {\bibinfo {volume} {80}},\ \bibinfo {pages} {121502} (\bibinfo {year} {2009})},\ \Eprint {https://arxiv.org/abs/0910.3656} {arXiv:0910.3656 [gr-qc]} \BibitemShut {NoStop}%
\bibitem [{\citenamefont {Iozzo}\ \emph {et~al.}(2021)\citenamefont {Iozzo}, \citenamefont {Boyle}, \citenamefont {Deppe}, \citenamefont {Moxon}, \citenamefont {Scheel}, \citenamefont {Kidder}, \citenamefont {Pfeiffer},\ and\ \citenamefont {Teukolsky}}]{Iozzo:2020jcu}%
  \BibitemOpen
  \bibfield  {author} {\bibinfo {author} {\bibfnamefont {D.~A.~B.}\ \bibnamefont {Iozzo}}, \bibinfo {author} {\bibfnamefont {M.}~\bibnamefont {Boyle}}, \bibinfo {author} {\bibfnamefont {N.}~\bibnamefont {Deppe}}, \bibinfo {author} {\bibfnamefont {J.}~\bibnamefont {Moxon}}, \bibinfo {author} {\bibfnamefont {M.~A.}\ \bibnamefont {Scheel}}, \bibinfo {author} {\bibfnamefont {L.~E.}\ \bibnamefont {Kidder}}, \bibinfo {author} {\bibfnamefont {H.~P.}\ \bibnamefont {Pfeiffer}},\ and\ \bibinfo {author} {\bibfnamefont {S.~A.}\ \bibnamefont {Teukolsky}},\ }\bibfield  {title} {\bibinfo {title} {{Extending gravitational wave extraction using Weyl characteristic fields}},\ }\href {https://doi.org/10.1103/PhysRevD.103.024039} {\bibfield  {journal} {\bibinfo  {journal} {Phys. Rev. D}\ }\textbf {\bibinfo {volume} {103}},\ \bibinfo {pages} {024039} (\bibinfo {year} {2021})},\ \Eprint {https://arxiv.org/abs/2010.15200} {arXiv:2010.15200 [gr-qc]} \BibitemShut {NoStop}%
\bibitem [{\citenamefont {Boyle}\ \emph {et~al.}(2024)\citenamefont {Boyle}, \citenamefont {Iozzo}, \citenamefont {Stein}, \citenamefont {Khairnar}, \citenamefont {Rüter}, \citenamefont {Scheel}, \citenamefont {Varma},\ and\ \citenamefont {Mitman}}]{scri}%
  \BibitemOpen
  \bibfield  {author} {\bibinfo {author} {\bibfnamefont {M.}~\bibnamefont {Boyle}}, \bibinfo {author} {\bibfnamefont {D.}~\bibnamefont {Iozzo}}, \bibinfo {author} {\bibfnamefont {L.}~\bibnamefont {Stein}}, \bibinfo {author} {\bibfnamefont {A.}~\bibnamefont {Khairnar}}, \bibinfo {author} {\bibfnamefont {H.}~\bibnamefont {Rüter}}, \bibinfo {author} {\bibfnamefont {M.}~\bibnamefont {Scheel}}, \bibinfo {author} {\bibfnamefont {V.}~\bibnamefont {Varma}},\ and\ \bibinfo {author} {\bibfnamefont {K.}~\bibnamefont {Mitman}},\ }\href {https://doi.org/10.5281/zenodo.14531184} {\bibinfo {title} {scri}} (\bibinfo {year} {2024})\BibitemShut {NoStop}%
\bibitem [{\citenamefont {Lousto}\ \emph {et~al.}(2010)\citenamefont {Lousto}, \citenamefont {Nakano}, \citenamefont {Zlochower},\ and\ \citenamefont {Campanelli}}]{Lousto:2010qx}%
  \BibitemOpen
  \bibfield  {author} {\bibinfo {author} {\bibfnamefont {C.~O.}\ \bibnamefont {Lousto}}, \bibinfo {author} {\bibfnamefont {H.}~\bibnamefont {Nakano}}, \bibinfo {author} {\bibfnamefont {Y.}~\bibnamefont {Zlochower}},\ and\ \bibinfo {author} {\bibfnamefont {M.}~\bibnamefont {Campanelli}},\ }\bibfield  {title} {\bibinfo {title} {{Intermediate-mass-ratio black hole binaries: Intertwining numerical and perturbative techniques}},\ }\href {https://doi.org/10.1103/PhysRevD.82.104057} {\bibfield  {journal} {\bibinfo  {journal} {Phys. Rev. D}\ }\textbf {\bibinfo {volume} {82}},\ \bibinfo {pages} {104057} (\bibinfo {year} {2010})},\ \Eprint {https://arxiv.org/abs/1008.4360} {arXiv:1008.4360 [gr-qc]} \BibitemShut {NoStop}%
\bibitem [{\citenamefont {Nakano}\ \emph {et~al.}(2011)\citenamefont {Nakano}, \citenamefont {Zlochower}, \citenamefont {Lousto},\ and\ \citenamefont {Campanelli}}]{Nakano:2011pb}%
  \BibitemOpen
  \bibfield  {author} {\bibinfo {author} {\bibfnamefont {H.}~\bibnamefont {Nakano}}, \bibinfo {author} {\bibfnamefont {Y.}~\bibnamefont {Zlochower}}, \bibinfo {author} {\bibfnamefont {C.~O.}\ \bibnamefont {Lousto}},\ and\ \bibinfo {author} {\bibfnamefont {M.}~\bibnamefont {Campanelli}},\ }\bibfield  {title} {\bibinfo {title} {{Intermediate-mass-ratio black hole binaries II: Modeling Trajectories and Gravitational Waveforms}},\ }\href {https://doi.org/10.1103/PhysRevD.84.124006} {\bibfield  {journal} {\bibinfo  {journal} {Phys. Rev. D}\ }\textbf {\bibinfo {volume} {84}},\ \bibinfo {pages} {124006} (\bibinfo {year} {2011})},\ \Eprint {https://arxiv.org/abs/1108.4421} {arXiv:1108.4421 [gr-qc]} \BibitemShut {NoStop}%
\bibitem [{\citenamefont {Hotokezaka}\ \emph {et~al.}(2016)\citenamefont {Hotokezaka}, \citenamefont {Kyutoku}, \citenamefont {Sekiguchi},\ and\ \citenamefont {Shibata}}]{Hotokezaka:2016bzh}%
  \BibitemOpen
  \bibfield  {author} {\bibinfo {author} {\bibfnamefont {K.}~\bibnamefont {Hotokezaka}}, \bibinfo {author} {\bibfnamefont {K.}~\bibnamefont {Kyutoku}}, \bibinfo {author} {\bibfnamefont {Y.-i.}\ \bibnamefont {Sekiguchi}},\ and\ \bibinfo {author} {\bibfnamefont {M.}~\bibnamefont {Shibata}},\ }\bibfield  {title} {\bibinfo {title} {{Measurability of the tidal deformability by gravitational waves from coalescing binary neutron stars}},\ }\href {https://doi.org/10.1103/PhysRevD.93.064082} {\bibfield  {journal} {\bibinfo  {journal} {Phys. Rev. D}\ }\textbf {\bibinfo {volume} {93}},\ \bibinfo {pages} {064082} (\bibinfo {year} {2016})},\ \Eprint {https://arxiv.org/abs/1603.01286} {arXiv:1603.01286 [gr-qc]} \BibitemShut {NoStop}%
\bibitem [{\citenamefont {Reisswig}\ and\ \citenamefont {Pollney}(2011)}]{Reisswig:2010di}%
  \BibitemOpen
  \bibfield  {author} {\bibinfo {author} {\bibfnamefont {C.}~\bibnamefont {Reisswig}}\ and\ \bibinfo {author} {\bibfnamefont {D.}~\bibnamefont {Pollney}},\ }\bibfield  {title} {\bibinfo {title} {{Notes on the integration of numerical relativity waveforms}},\ }\href {https://doi.org/10.1088/0264-9381/28/19/195015} {\bibfield  {journal} {\bibinfo  {journal} {Class. Quant. Grav.}\ }\textbf {\bibinfo {volume} {28}},\ \bibinfo {pages} {195015} (\bibinfo {year} {2011})},\ \Eprint {https://arxiv.org/abs/1006.1632} {arXiv:1006.1632 [gr-qc]} \BibitemShut {NoStop}%
\bibitem [{\citenamefont {Dietrich}(2016)}]{Dietrich2016BinaryNS}%
  \BibitemOpen
  \bibfield  {author} {\bibinfo {author} {\bibfnamefont {T.}~\bibnamefont {Dietrich}},\ }\bibfield  {title} {\bibinfo {title} {Binary neutron star merger simulations}\ }(\bibinfo {year} {2016})\BibitemShut {NoStop}%
\bibitem [{\citenamefont {Lackey}\ \emph {et~al.}(2019)\citenamefont {Lackey}, \citenamefont {P\"urrer}, \citenamefont {Taracchini},\ and\ \citenamefont {Marsat}}]{Lackey:2018zvw}%
  \BibitemOpen
  \bibfield  {author} {\bibinfo {author} {\bibfnamefont {B.~D.}\ \bibnamefont {Lackey}}, \bibinfo {author} {\bibfnamefont {M.}~\bibnamefont {P\"urrer}}, \bibinfo {author} {\bibfnamefont {A.}~\bibnamefont {Taracchini}},\ and\ \bibinfo {author} {\bibfnamefont {S.}~\bibnamefont {Marsat}},\ }\bibfield  {title} {\bibinfo {title} {{Surrogate model for an aligned-spin effective one body waveform model of binary neutron star inspirals using Gaussian process regression}},\ }\href {https://doi.org/10.1103/PhysRevD.100.024002} {\bibfield  {journal} {\bibinfo  {journal} {Phys. Rev. D}\ }\textbf {\bibinfo {volume} {100}},\ \bibinfo {pages} {024002} (\bibinfo {year} {2019})},\ \Eprint {https://arxiv.org/abs/1812.08643} {arXiv:1812.08643 [gr-qc]} \BibitemShut {NoStop}%
\bibitem [{\citenamefont {{LIGO Scientific Collaboration}}\ \emph {et~al.}(2018)\citenamefont {{LIGO Scientific Collaboration}}, \citenamefont {{Virgo Collaboration}},\ and\ \citenamefont {{KAGRA Collaboration}}}]{lalsuite}%
  \BibitemOpen
  \bibfield  {author} {\bibinfo {author} {\bibnamefont {{LIGO Scientific Collaboration}}}, \bibinfo {author} {\bibnamefont {{Virgo Collaboration}}},\ and\ \bibinfo {author} {\bibnamefont {{KAGRA Collaboration}}},\ }\href {https://doi.org/10.7935/GT1W-FZ16} {\bibinfo {title} {{LVK} {A}lgorithm {L}ibrary - {LALS}uite}},\ \bibinfo {howpublished} {Free software (GPL)} (\bibinfo {year} {2018})\BibitemShut {NoStop}%
\bibitem [{\citenamefont {Nitz}\ \emph {et~al.}(2024)\citenamefont {Nitz}, \citenamefont {Harry}, \citenamefont {Brown}, \citenamefont {Biwer}, \citenamefont {Willis}, \citenamefont {Canton}, \citenamefont {Capano}, \citenamefont {Dent}, \citenamefont {Pekowsky}, \citenamefont {Davies}, \citenamefont {De}, \citenamefont {Cabero}, \citenamefont {Wu}, \citenamefont {Williamson}, \citenamefont {Machenschalk}, \citenamefont {Macleod}, \citenamefont {Pannarale}, \citenamefont {Kumar}, \citenamefont {Reyes}, \citenamefont {dfinstad}, \citenamefont {Kumar}, \citenamefont {Tápai}, \citenamefont {Singer}, \citenamefont {Kumar}, \citenamefont {veronica villa}, \citenamefont {maxtrevor}, \citenamefont {Gadre}, \citenamefont {Khan}, \citenamefont {Fairhurst},\ and\ \citenamefont {Tolley}}]{alex_nitz_2024_10473621}%
  \BibitemOpen
  \bibfield  {author} {\bibinfo {author} {\bibfnamefont {A.}~\bibnamefont {Nitz}}, \bibinfo {author} {\bibfnamefont {I.}~\bibnamefont {Harry}}, \bibinfo {author} {\bibfnamefont {D.}~\bibnamefont {Brown}}, \bibinfo {author} {\bibfnamefont {C.~M.}\ \bibnamefont {Biwer}}, \bibinfo {author} {\bibfnamefont {J.}~\bibnamefont {Willis}}, \bibinfo {author} {\bibfnamefont {T.~D.}\ \bibnamefont {Canton}}, \bibinfo {author} {\bibfnamefont {C.}~\bibnamefont {Capano}}, \bibinfo {author} {\bibfnamefont {T.}~\bibnamefont {Dent}}, \bibinfo {author} {\bibfnamefont {L.}~\bibnamefont {Pekowsky}}, \bibinfo {author} {\bibfnamefont {G.~S.~C.}\ \bibnamefont {Davies}}, \bibinfo {author} {\bibfnamefont {S.}~\bibnamefont {De}}, \bibinfo {author} {\bibfnamefont {M.}~\bibnamefont {Cabero}}, \bibinfo {author} {\bibfnamefont {S.}~\bibnamefont {Wu}}, \bibinfo {author} {\bibfnamefont {A.~R.}\ \bibnamefont {Williamson}}, \bibinfo {author} {\bibfnamefont {B.}~\bibnamefont {Machenschalk}}, \bibinfo {author} {\bibfnamefont {D.}~\bibnamefont
  {Macleod}}, \bibinfo {author} {\bibfnamefont {F.}~\bibnamefont {Pannarale}}, \bibinfo {author} {\bibfnamefont {P.}~\bibnamefont {Kumar}}, \bibinfo {author} {\bibfnamefont {S.}~\bibnamefont {Reyes}}, \bibinfo {author} {\bibnamefont {dfinstad}}, \bibinfo {author} {\bibfnamefont {S.}~\bibnamefont {Kumar}}, \bibinfo {author} {\bibfnamefont {M.}~\bibnamefont {Tápai}}, \bibinfo {author} {\bibfnamefont {L.}~\bibnamefont {Singer}}, \bibinfo {author} {\bibfnamefont {P.}~\bibnamefont {Kumar}}, \bibinfo {author} {\bibnamefont {veronica villa}}, \bibinfo {author} {\bibnamefont {maxtrevor}}, \bibinfo {author} {\bibfnamefont {B.~U.~V.}\ \bibnamefont {Gadre}}, \bibinfo {author} {\bibfnamefont {S.}~\bibnamefont {Khan}}, \bibinfo {author} {\bibfnamefont {S.}~\bibnamefont {Fairhurst}},\ and\ \bibinfo {author} {\bibfnamefont {A.}~\bibnamefont {Tolley}},\ }\href {https://doi.org/10.5281/zenodo.10473621} {\bibinfo {title} {gwastro/pycbc: v2.3.3 release of pycbc}} (\bibinfo {year} {2024})\BibitemShut {NoStop}%
\bibitem [{\citenamefont {Tagoshi}\ \emph {et~al.}(1997)\citenamefont {Tagoshi}, \citenamefont {Mano},\ and\ \citenamefont {Takasugi}}]{Tagoshi:1997jy}%
  \BibitemOpen
  \bibfield  {author} {\bibinfo {author} {\bibfnamefont {H.}~\bibnamefont {Tagoshi}}, \bibinfo {author} {\bibfnamefont {S.}~\bibnamefont {Mano}},\ and\ \bibinfo {author} {\bibfnamefont {E.}~\bibnamefont {Takasugi}},\ }\bibfield  {title} {\bibinfo {title} {{PostNewtonian expansion of gravitational waves from a particle in circular orbits around a rotating black hole: Effects of black hole absorption}},\ }\href {https://doi.org/10.1143/PTP.98.829} {\bibfield  {journal} {\bibinfo  {journal} {Prog. Theor. Phys.}\ }\textbf {\bibinfo {volume} {98}},\ \bibinfo {pages} {829} (\bibinfo {year} {1997})},\ \Eprint {https://arxiv.org/abs/gr-qc/9711072} {arXiv:gr-qc/9711072} \BibitemShut {NoStop}%
\bibitem [{\citenamefont {Alvi}(2001)}]{Alvi:2001mx}%
  \BibitemOpen
  \bibfield  {author} {\bibinfo {author} {\bibfnamefont {K.}~\bibnamefont {Alvi}},\ }\bibfield  {title} {\bibinfo {title} {{Energy and angular momentum flow into a black hole in a binary}},\ }\href {https://doi.org/10.1103/PhysRevD.64.104020} {\bibfield  {journal} {\bibinfo  {journal} {Phys. Rev. D}\ }\textbf {\bibinfo {volume} {64}},\ \bibinfo {pages} {104020} (\bibinfo {year} {2001})},\ \Eprint {https://arxiv.org/abs/gr-qc/0107080} {arXiv:gr-qc/0107080} \BibitemShut {NoStop}%
\bibitem [{\citenamefont {Porto}(2008)}]{Porto:2007qi}%
  \BibitemOpen
  \bibfield  {author} {\bibinfo {author} {\bibfnamefont {R.~A.}\ \bibnamefont {Porto}},\ }\bibfield  {title} {\bibinfo {title} {{Absorption effects due to spin in the worldline approach to black hole dynamics}},\ }\href {https://doi.org/10.1103/PhysRevD.77.064026} {\bibfield  {journal} {\bibinfo  {journal} {Phys. Rev. D}\ }\textbf {\bibinfo {volume} {77}},\ \bibinfo {pages} {064026} (\bibinfo {year} {2008})},\ \Eprint {https://arxiv.org/abs/0710.5150} {arXiv:0710.5150 [hep-th]} \BibitemShut {NoStop}%
\bibitem [{\citenamefont {Saketh}\ \emph {et~al.}(2023)\citenamefont {Saketh}, \citenamefont {Steinhoff}, \citenamefont {Vines},\ and\ \citenamefont {Buonanno}}]{Saketh:2022xjb}%
  \BibitemOpen
  \bibfield  {author} {\bibinfo {author} {\bibfnamefont {M.~V.~S.}\ \bibnamefont {Saketh}}, \bibinfo {author} {\bibfnamefont {J.}~\bibnamefont {Steinhoff}}, \bibinfo {author} {\bibfnamefont {J.}~\bibnamefont {Vines}},\ and\ \bibinfo {author} {\bibfnamefont {A.}~\bibnamefont {Buonanno}},\ }\bibfield  {title} {\bibinfo {title} {{Modeling horizon absorption in spinning binary black holes using effective worldline theory}},\ }\href {https://doi.org/10.1103/PhysRevD.107.084006} {\bibfield  {journal} {\bibinfo  {journal} {Phys. Rev. D}\ }\textbf {\bibinfo {volume} {107}},\ \bibinfo {pages} {084006} (\bibinfo {year} {2023})},\ \Eprint {https://arxiv.org/abs/2212.13095} {arXiv:2212.13095 [gr-qc]} \BibitemShut {NoStop}%
\bibitem [{\citenamefont {Baiotti}\ \emph {et~al.}(2010{\natexlab{b}})\citenamefont {Baiotti}, \citenamefont {Damour}, \citenamefont {Giacomazzo}, \citenamefont {Nagar},\ and\ \citenamefont {Rezzolla}}]{Baiotti:2010xh}%
  \BibitemOpen
  \bibfield  {author} {\bibinfo {author} {\bibfnamefont {L.}~\bibnamefont {Baiotti}}, \bibinfo {author} {\bibfnamefont {T.}~\bibnamefont {Damour}}, \bibinfo {author} {\bibfnamefont {B.}~\bibnamefont {Giacomazzo}}, \bibinfo {author} {\bibfnamefont {A.}~\bibnamefont {Nagar}},\ and\ \bibinfo {author} {\bibfnamefont {L.}~\bibnamefont {Rezzolla}},\ }\bibfield  {title} {\bibinfo {title} {{Analytic modelling of tidal effects in the relativistic inspiral of binary neutron stars}},\ }\href {https://doi.org/10.1103/PhysRevLett.105.261101} {\bibfield  {journal} {\bibinfo  {journal} {Phys. Rev. Lett.}\ }\textbf {\bibinfo {volume} {105}},\ \bibinfo {pages} {261101} (\bibinfo {year} {2010}{\natexlab{b}})},\ \Eprint {https://arxiv.org/abs/1009.0521} {arXiv:1009.0521 [gr-qc]} \BibitemShut {NoStop}%
\bibitem [{\citenamefont {Thorne}(1980)}]{Thorne:1980ru}%
  \BibitemOpen
  \bibfield  {author} {\bibinfo {author} {\bibfnamefont {K.~S.}\ \bibnamefont {Thorne}},\ }\bibfield  {title} {\bibinfo {title} {{Multipole Expansions of Gravitational Radiation}},\ }\href {https://doi.org/10.1103/RevModPhys.52.299} {\bibfield  {journal} {\bibinfo  {journal} {Rev. Mod. Phys.}\ }\textbf {\bibinfo {volume} {52}},\ \bibinfo {pages} {299} (\bibinfo {year} {1980})}\BibitemShut {NoStop}%
\bibitem [{\citenamefont {Mino}\ \emph {et~al.}(1997)\citenamefont {Mino}, \citenamefont {Sasaki}, \citenamefont {Shibata}, \citenamefont {Tagoshi},\ and\ \citenamefont {Tanaka}}]{Mino:1997bx}%
  \BibitemOpen
  \bibfield  {author} {\bibinfo {author} {\bibfnamefont {Y.}~\bibnamefont {Mino}}, \bibinfo {author} {\bibfnamefont {M.}~\bibnamefont {Sasaki}}, \bibinfo {author} {\bibfnamefont {M.}~\bibnamefont {Shibata}}, \bibinfo {author} {\bibfnamefont {H.}~\bibnamefont {Tagoshi}},\ and\ \bibinfo {author} {\bibfnamefont {T.}~\bibnamefont {Tanaka}},\ }\bibfield  {title} {\bibinfo {title} {{Black hole perturbation: Chapter 1}},\ }\href {https://doi.org/10.1143/PTPS.128.1} {\bibfield  {journal} {\bibinfo  {journal} {Prog. Theor. Phys. Suppl.}\ }\textbf {\bibinfo {volume} {128}},\ \bibinfo {pages} {1} (\bibinfo {year} {1997})},\ \Eprint {https://arxiv.org/abs/gr-qc/9712057} {arXiv:gr-qc/9712057} \BibitemShut {NoStop}%
\bibitem [{\citenamefont {Shibata}\ and\ \citenamefont {Sasaki}(1998)}]{Shibata:1998xw}%
  \BibitemOpen
  \bibfield  {author} {\bibinfo {author} {\bibfnamefont {M.}~\bibnamefont {Shibata}}\ and\ \bibinfo {author} {\bibfnamefont {M.}~\bibnamefont {Sasaki}},\ }\bibfield  {title} {\bibinfo {title} {{Innermost stable circular orbits around relativistic rotating stars}},\ }\href {https://doi.org/10.1103/PhysRevD.58.104011} {\bibfield  {journal} {\bibinfo  {journal} {Phys. Rev. D}\ }\textbf {\bibinfo {volume} {58}},\ \bibinfo {pages} {104011} (\bibinfo {year} {1998})},\ \Eprint {https://arxiv.org/abs/gr-qc/9807046} {arXiv:gr-qc/9807046} \BibitemShut {NoStop}%
\bibitem [{\citenamefont {Pappas}\ and\ \citenamefont {Apostolatos}(2014)}]{Pappas:2013naa}%
  \BibitemOpen
  \bibfield  {author} {\bibinfo {author} {\bibfnamefont {G.}~\bibnamefont {Pappas}}\ and\ \bibinfo {author} {\bibfnamefont {T.~A.}\ \bibnamefont {Apostolatos}},\ }\bibfield  {title} {\bibinfo {title} {{Effectively universal behavior of rotating neutron stars in general relativity makes them even simpler than their Newtonian counterparts}},\ }\href {https://doi.org/10.1103/PhysRevLett.112.121101} {\bibfield  {journal} {\bibinfo  {journal} {Phys. Rev. Lett.}\ }\textbf {\bibinfo {volume} {112}},\ \bibinfo {pages} {121101} (\bibinfo {year} {2014})},\ \Eprint {https://arxiv.org/abs/1311.5508} {arXiv:1311.5508 [gr-qc]} \BibitemShut {NoStop}%
\bibitem [{\citenamefont {Yagi}\ \emph {et~al.}(2014)\citenamefont {Yagi}, \citenamefont {Kyutoku}, \citenamefont {Pappas}, \citenamefont {Yunes},\ and\ \citenamefont {Apostolatos}}]{Yagi:2014bxa}%
  \BibitemOpen
  \bibfield  {author} {\bibinfo {author} {\bibfnamefont {K.}~\bibnamefont {Yagi}}, \bibinfo {author} {\bibfnamefont {K.}~\bibnamefont {Kyutoku}}, \bibinfo {author} {\bibfnamefont {G.}~\bibnamefont {Pappas}}, \bibinfo {author} {\bibfnamefont {N.}~\bibnamefont {Yunes}},\ and\ \bibinfo {author} {\bibfnamefont {T.~A.}\ \bibnamefont {Apostolatos}},\ }\bibfield  {title} {\bibinfo {title} {{Effective No-Hair Relations for Neutron Stars and Quark Stars: Relativistic Results}},\ }\href {https://doi.org/10.1103/PhysRevD.89.124013} {\bibfield  {journal} {\bibinfo  {journal} {Phys. Rev. D}\ }\textbf {\bibinfo {volume} {89}},\ \bibinfo {pages} {124013} (\bibinfo {year} {2014})},\ \Eprint {https://arxiv.org/abs/1403.6243} {arXiv:1403.6243 [gr-qc]} \BibitemShut {NoStop}%
\bibitem [{\citenamefont {Levi}\ and\ \citenamefont {Yin}(2023)}]{Levi:2022rrq}%
  \BibitemOpen
  \bibfield  {author} {\bibinfo {author} {\bibfnamefont {M.}~\bibnamefont {Levi}}\ and\ \bibinfo {author} {\bibfnamefont {Z.}~\bibnamefont {Yin}},\ }\bibfield  {title} {\bibinfo {title} {{Completing the fifth PN precision frontier via the EFT of spinning gravitating objects}},\ }\href {https://doi.org/10.1007/JHEP04(2023)079} {\bibfield  {journal} {\bibinfo  {journal} {JHEP}\ }\textbf {\bibinfo {volume} {04}},\ \bibinfo {pages} {079}},\ \Eprint {https://arxiv.org/abs/2211.14018} {arXiv:2211.14018 [hep-th]} \BibitemShut {NoStop}%
\bibitem [{\citenamefont {Bautista}\ \emph {et~al.}(2024)\citenamefont {Bautista}, \citenamefont {Khalil}, \citenamefont {Sergola}, \citenamefont {Kavanagh},\ and\ \citenamefont {Vines}}]{Bautista:2024agp}%
  \BibitemOpen
  \bibfield  {author} {\bibinfo {author} {\bibfnamefont {Y.~F.}\ \bibnamefont {Bautista}}, \bibinfo {author} {\bibfnamefont {M.}~\bibnamefont {Khalil}}, \bibinfo {author} {\bibfnamefont {M.}~\bibnamefont {Sergola}}, \bibinfo {author} {\bibfnamefont {C.}~\bibnamefont {Kavanagh}},\ and\ \bibinfo {author} {\bibfnamefont {J.}~\bibnamefont {Vines}},\ }\bibfield  {title} {\bibinfo {title} {{Post-Newtonian observables for aligned-spin binaries to sixth order in spin from gravitational self-force and Compton amplitudes}},\ }\href {https://doi.org/10.1103/PhysRevD.110.124005} {\bibfield  {journal} {\bibinfo  {journal} {Phys. Rev. D}\ }\textbf {\bibinfo {volume} {110}},\ \bibinfo {pages} {124005} (\bibinfo {year} {2024})},\ \Eprint {https://arxiv.org/abs/2408.01871} {arXiv:2408.01871 [gr-qc]} \BibitemShut {NoStop}%
\bibitem [{\citenamefont {Levi}\ and\ \citenamefont {Steinhoff}(2015)}]{Levi:2014gsa}%
  \BibitemOpen
  \bibfield  {author} {\bibinfo {author} {\bibfnamefont {M.}~\bibnamefont {Levi}}\ and\ \bibinfo {author} {\bibfnamefont {J.}~\bibnamefont {Steinhoff}},\ }\bibfield  {title} {\bibinfo {title} {{Leading order finite size effects with spins for inspiralling compact binaries}},\ }\href {https://doi.org/10.1007/JHEP06(2015)059} {\bibfield  {journal} {\bibinfo  {journal} {JHEP}\ }\textbf {\bibinfo {volume} {06}},\ \bibinfo {pages} {059}},\ \Eprint {https://arxiv.org/abs/1410.2601} {arXiv:1410.2601 [gr-qc]} \BibitemShut {NoStop}%
\bibitem [{\citenamefont {Kim}\ \emph {et~al.}(2022)\citenamefont {Kim}, \citenamefont {Levi},\ and\ \citenamefont {Yin}}]{Kim:2021rfj}%
  \BibitemOpen
  \bibfield  {author} {\bibinfo {author} {\bibfnamefont {J.-W.}\ \bibnamefont {Kim}}, \bibinfo {author} {\bibfnamefont {M.}~\bibnamefont {Levi}},\ and\ \bibinfo {author} {\bibfnamefont {Z.}~\bibnamefont {Yin}},\ }\bibfield  {title} {\bibinfo {title} {{Quadratic-in-spin interactions at fifth post-Newtonian order probe new physics}},\ }\href {https://doi.org/10.1016/j.physletb.2022.137410} {\bibfield  {journal} {\bibinfo  {journal} {Phys. Lett. B}\ }\textbf {\bibinfo {volume} {834}},\ \bibinfo {pages} {137410} (\bibinfo {year} {2022})},\ \Eprint {https://arxiv.org/abs/2112.01509} {arXiv:2112.01509 [hep-th]} \BibitemShut {NoStop}%
\bibitem [{\citenamefont {Kim}\ \emph {et~al.}(2023)\citenamefont {Kim}, \citenamefont {Levi},\ and\ \citenamefont {Yin}}]{Kim:2022bwv}%
  \BibitemOpen
  \bibfield  {author} {\bibinfo {author} {\bibfnamefont {J.-W.}\ \bibnamefont {Kim}}, \bibinfo {author} {\bibfnamefont {M.}~\bibnamefont {Levi}},\ and\ \bibinfo {author} {\bibfnamefont {Z.}~\bibnamefont {Yin}},\ }\bibfield  {title} {\bibinfo {title} {{N$^{3}$LO quadratic-in-spin interactions for generic compact binaries}},\ }\href {https://doi.org/10.1007/JHEP03(2023)098} {\bibfield  {journal} {\bibinfo  {journal} {JHEP}\ }\textbf {\bibinfo {volume} {03}},\ \bibinfo {pages} {098}},\ \Eprint {https://arxiv.org/abs/2209.09235} {arXiv:2209.09235 [hep-th]} \BibitemShut {NoStop}%
\bibitem [{\citenamefont {Marsat}(2015)}]{Marsat:2014xea}%
  \BibitemOpen
  \bibfield  {author} {\bibinfo {author} {\bibfnamefont {S.}~\bibnamefont {Marsat}},\ }\bibfield  {title} {\bibinfo {title} {{Cubic order spin effects in the dynamics and gravitational wave energy flux of compact object binaries}},\ }\href {https://doi.org/10.1088/0264-9381/32/8/085008} {\bibfield  {journal} {\bibinfo  {journal} {Class. Quant. Grav.}\ }\textbf {\bibinfo {volume} {32}},\ \bibinfo {pages} {085008} (\bibinfo {year} {2015})},\ \Eprint {https://arxiv.org/abs/1411.4118} {arXiv:1411.4118 [gr-qc]} \BibitemShut {NoStop}%
\bibitem [{\citenamefont {Gal'tsov}(1982)}]{Galtsov:1982hwm}%
  \BibitemOpen
  \bibfield  {author} {\bibinfo {author} {\bibfnamefont {D.~V.}\ \bibnamefont {Gal'tsov}},\ }\bibfield  {title} {\bibinfo {title} {{Radiation reaction in the Kerr gravitational field}},\ }\href {https://doi.org/10.1088/0305-4470/15/12/025} {\bibfield  {journal} {\bibinfo  {journal} {J. Phys. A}\ }\textbf {\bibinfo {volume} {15}},\ \bibinfo {pages} {3737} (\bibinfo {year} {1982})}\BibitemShut {NoStop}%
\bibitem [{\citenamefont {Damour}\ \emph {et~al.}(2000)\citenamefont {Damour}, \citenamefont {Iyer},\ and\ \citenamefont {Sathyaprakash}}]{Damour:2000gg}%
  \BibitemOpen
  \bibfield  {author} {\bibinfo {author} {\bibfnamefont {T.}~\bibnamefont {Damour}}, \bibinfo {author} {\bibfnamefont {B.~R.}\ \bibnamefont {Iyer}},\ and\ \bibinfo {author} {\bibfnamefont {B.~S.}\ \bibnamefont {Sathyaprakash}},\ }\bibfield  {title} {\bibinfo {title} {{Frequency domain P approximant filters for time truncated inspiral gravitational wave signals from compact binaries}},\ }\href {https://doi.org/10.1103/PhysRevD.62.084036} {\bibfield  {journal} {\bibinfo  {journal} {Phys. Rev. D}\ }\textbf {\bibinfo {volume} {62}},\ \bibinfo {pages} {084036} (\bibinfo {year} {2000})},\ \Eprint {https://arxiv.org/abs/gr-qc/0001023} {arXiv:gr-qc/0001023} \BibitemShut {NoStop}%
\bibitem [{\citenamefont {Albertini}\ \emph {et~al.}(2024)\citenamefont {Albertini}, \citenamefont {Nagar}, \citenamefont {Mathews},\ and\ \citenamefont {Lukes-Gerakopoulos}}]{Albertini:2024rrs}%
  \BibitemOpen
  \bibfield  {author} {\bibinfo {author} {\bibfnamefont {A.}~\bibnamefont {Albertini}}, \bibinfo {author} {\bibfnamefont {A.}~\bibnamefont {Nagar}}, \bibinfo {author} {\bibfnamefont {J.}~\bibnamefont {Mathews}},\ and\ \bibinfo {author} {\bibfnamefont {G.}~\bibnamefont {Lukes-Gerakopoulos}},\ }\bibfield  {title} {\bibinfo {title} {{Comparing second-order gravitational self-force and effective-one-body waveforms from inspiralling, quasicircular black hole binaries with a nonspinning primary and a spinning secondary}},\ }\href {https://doi.org/10.1103/PhysRevD.110.044034} {\bibfield  {journal} {\bibinfo  {journal} {Phys. Rev. D}\ }\textbf {\bibinfo {volume} {110}},\ \bibinfo {pages} {044034} (\bibinfo {year} {2024})},\ \Eprint {https://arxiv.org/abs/2406.04108} {arXiv:2406.04108 [gr-qc]} \BibitemShut {NoStop}%
\end{thebibliography}%

\end{document}